\documentclass[10pt,journal,final,compsoc]{IEEEtran}
\usepackage{graphicx}
\usepackage{epstopdf}
\usepackage{graphicx}
\usepackage{subfigure}

\usepackage{hhline}
\usepackage{amsmath}
\usepackage{changepage}

\usepackage{bigstrut,multirow,rotating}
\usepackage{mdwlist}
\usepackage{float}
\usepackage{amssymb}
\usepackage{multirow}
\usepackage{mathrsfs}
\usepackage{algorithmic}
\usepackage[ruled]{algorithm2e}
\usepackage{rotating}
\usepackage{balance}
\usepackage{threeparttable}
\usepackage{color}
\usepackage{colortbl}
\usepackage{makecell}
\usepackage{pdflscape}
\usepackage{siunitx}
\usepackage[pagewise]{lineno}
\usepackage{makecell}
\usepackage{stfloats}
\usepackage{verbatim}
\usepackage{setspace}
\usepackage{threeparttable}
\usepackage{supertabular,booktabs}
\usepackage{rotating}
\usepackage{supertabular}
\usepackage{pifont}
\usepackage{ragged2e}
\usepackage{xcolor,pict2e}
\usepackage{tikz}
\usepackage{lettrine}
\usepackage{hyperref}
\usepackage[hyphenbreaks]{breakurl}
\usepackage{fancyhdr}

    \usepackage{xcolor}

    \newcounter{DaveCommentCounter}
    \newcounter{CuiCommentCounter}
    \newcounter{CuiResCounter}
\newcommand{\VPP}{VPP-ART}
\newcommand{\FSCS}{FSCS-ART}
\newcommand{\Naive}{Naive-KDFC}
\newcommand{\SemiBal}{SemiBal-KDFC}
\newcommand{\LimBal}{LimBal-KDFC}
\newcommand{\KDFC}{KDFC-ART}
\newcommand{\fratio}{ART F-ratio}
\newcommand{\es}{effect size}
\newcommand{\NN}{NN}

\renewcommand{\algorithmicrequire}{\textbf{Inputs:}}  
\renewcommand{\algorithmicensure}{\textbf{Output:}} 

\renewcommand{\raggedright}{\leftskip=0pt \rightskip=0pt plus 0cm}
\begin{document}

\title{VPP-ART: An Efficient Implementation of Fixed-Size-Candidate-Set Adaptive Random Testing using Vantage Point Partitioning}
\author{Rubing Huang,~\IEEEmembership{Senior Member,~IEEE,}
        Chenhui Cui,
        Dave Towey,~\IEEEmembership{Senior Member,~IEEE,}
        \\Weifeng Sun,
        Junlong Lian

    \IEEEcompsocitemizethanks{
    \IEEEcompsocthanksitem R. Huang is with the Faculty of Information Technology, Macau University of Science and Technology, Taipa, Macau 999078, China.\protect\\
    E-mail: rbhuang@must.edu.mo.

    \IEEEcompsocthanksitem C. Cui and J. Lian are with the School of Computer Science and Communication Engineering, Jiangsu University, Zhenjiang, Jiangsu 212013, China.\protect\\
    E-mail: \{2211908012, 2211908018\}@stmail.ujs.edu.cn.

    \IEEEcompsocthanksitem D. Towey is with the School of Computer Science, University of Nottingham Ningbo China, Ningbo, Zhejiang 315100, China.\protect\\
    E-mail: dave.towey@nottingham.edu.cn.
    
    \IEEEcompsocthanksitem W. Sun is with the School of Big Data and Software Engineering, Chongqing University, Chongqing, 401331, China.\protect\\
    E-mail: weifeng.sun@cqu.edu.cn.
    }
}

\markboth{\VPP\/: An Efficient Implementation of Fixed-Size-Candidate-Set Adaptive Random Testing using Vantage Point Partitioning}
{\tiny
Shell \MakeLowercase{\textit{et al.}}: Bare Demo of IEEEtran.cls for Computer Society Journals}

\IEEEtitleabstractindextext{
\begin{abstract}
    \justifying
    \textit{Adaptive Random Testing} (ART) is an enhancement of \textit{Random Testing} (RT), and aims to improve the RT failure-detection effectiveness by distributing test cases more evenly in the input domain.
    Many ART algorithms have been proposed, with \textit{Fixed-Size-Candidate-Set} ART (\FSCS\/) being one of the most effective and popular.
    \FSCS\/ ensures high failure-detection effectiveness by selecting the next test case as the candidate farthest from previously-executed test cases.
    Although \FSCS\/ has good failure-detection effectiveness, it also faces some challenges, including heavy computational overheads.
    In this paper, we propose an enhanced version of \FSCS\/, \textit{Vantage Point Partitioning ART} (\VPP\/).
    \VPP\/ addresses the \FSCS\/ computational overhead problem using vantage point partitioning, while maintaining the failure-detection effectiveness.
    \VPP\/ partitions the input domain space using a \textit{modified Vantage Point tree} (VP-tree) and finds the approximate nearest executed test cases of a candidate test case in the partitioned sub-domains
    ---
    thereby significantly reducing the time overheads compared with the searches required for \FSCS\/.
    To enable the \FSCS\/ dynamic insertion process, we modify the traditional VP-tree to support dynamic data.
    The simulation results show that \VPP\/ has a much lower time overhead compared to \FSCS\/, but also delivers similar (or better) failure-detection effectiveness,
    especially in the higher dimensional input domains.
    According to statistical analyses, \VPP\/ can improve on the \FSCS\/ failure-detection effectiveness by approximately 50\% to 58\%.
    \VPP\/ also  compares favorably with the \KDFC\/ algorithms (a series of enhanced ART algorithms based on the KD-tree).
    Our experiments also show that \VPP\/ is more cost-effective than \FSCS\/ and \KDFC\/.
\end{abstract}

}

\maketitle
\IEEEdisplaynontitleabstractindextext
\IEEEpeerreviewmaketitle

\section{Introduction
\label{intro}}





\lettrine[lines=2]{S}{~~OFTWARE} testing is an important technique for evaluating and verifying the quality of the \textit{System Under Test} (SUT), and is an important part of the software life cycle~\cite{Anand2013a},~\cite{Orso2014}.
Software testing involves executing the software, aiming to find failures.
It can be divided into four steps:
(1) definition of test objectives;
(2) generation of test cases;
(3) execution of test cases; and
(4) examination and verification of test results.
Each test case is selected from the set of all possible inputs that constitute the \textit{input domain}.
When the output or behavior of the SUT during the execution of test case does not meet the expectation (as determined by the \textit{test oracle}~\cite{Weyuker1982,BarrHMSY2015,Jahangirova2017}), the test is considered to \textit{fail}, otherwise, it \textit{passes}.

\textit{Random Testing} (RT)~\cite{Arcuri2012} is a simple and efficient black-box testing method that generates test cases randomly within the input domain.
RT has been used in a wide variety of environments and systems, including:
in a stochastic scheduling algorithm for testing distributed systems \cite{Ozkan2018};
for testing GCC, LLVM and Intel C++ compilers \cite{Livinskii2020}; and
for GUI testing \cite{White2019}.
Research into RT enhancement has also been popular, with Dynamic Random Testing (DRT) \cite{Pei2019,Pei2021}, for example, improving the selection probability of sub-domains with high failure-detection rates.

\begin{figure*}[!b]
\graphicspath{{VPP-ART-Figures/}}
  \centering
  \subfigure[Block pattern]
  {
    \includegraphics[width=0.235\textwidth]{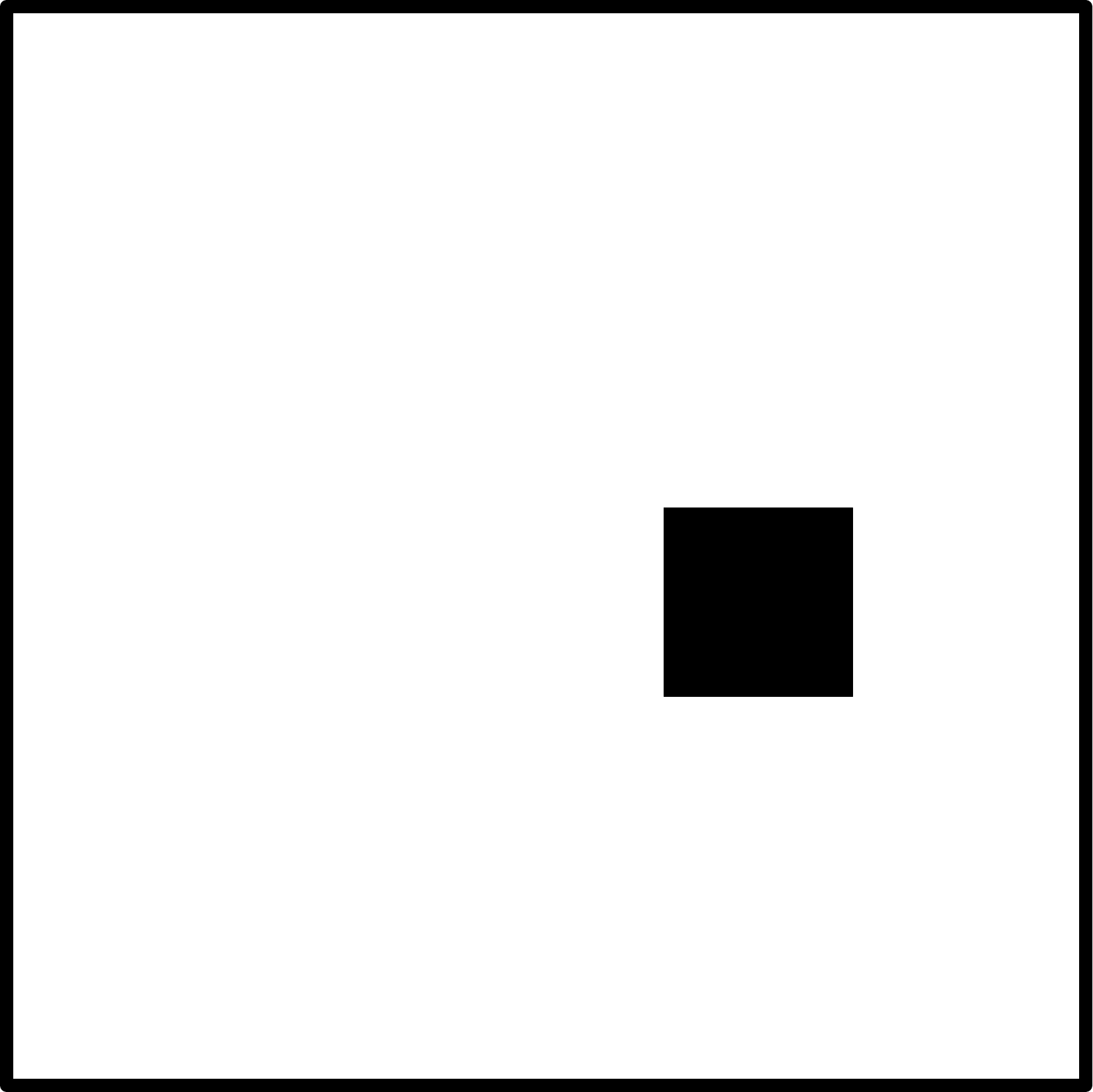}
    \label{block pattern}
  }
  \hspace{10mm}
  \subfigure[Strip pattern]
  {
    \includegraphics[width=0.235\textwidth]{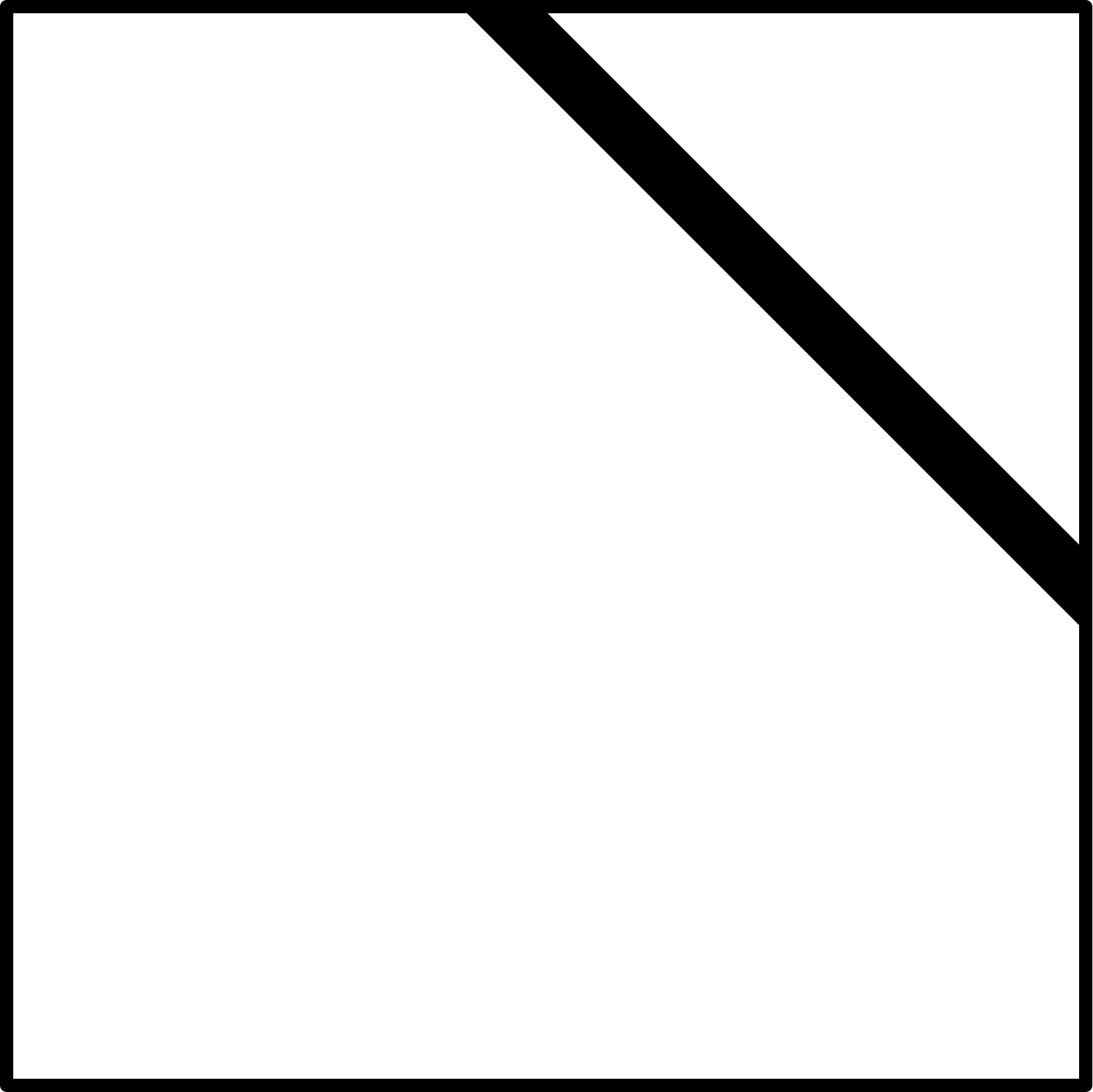}
    \label{strip pattern}
  }
  \hspace{10mm}
  \subfigure[Point pattern]
  {
    \includegraphics[width=0.235\textwidth]{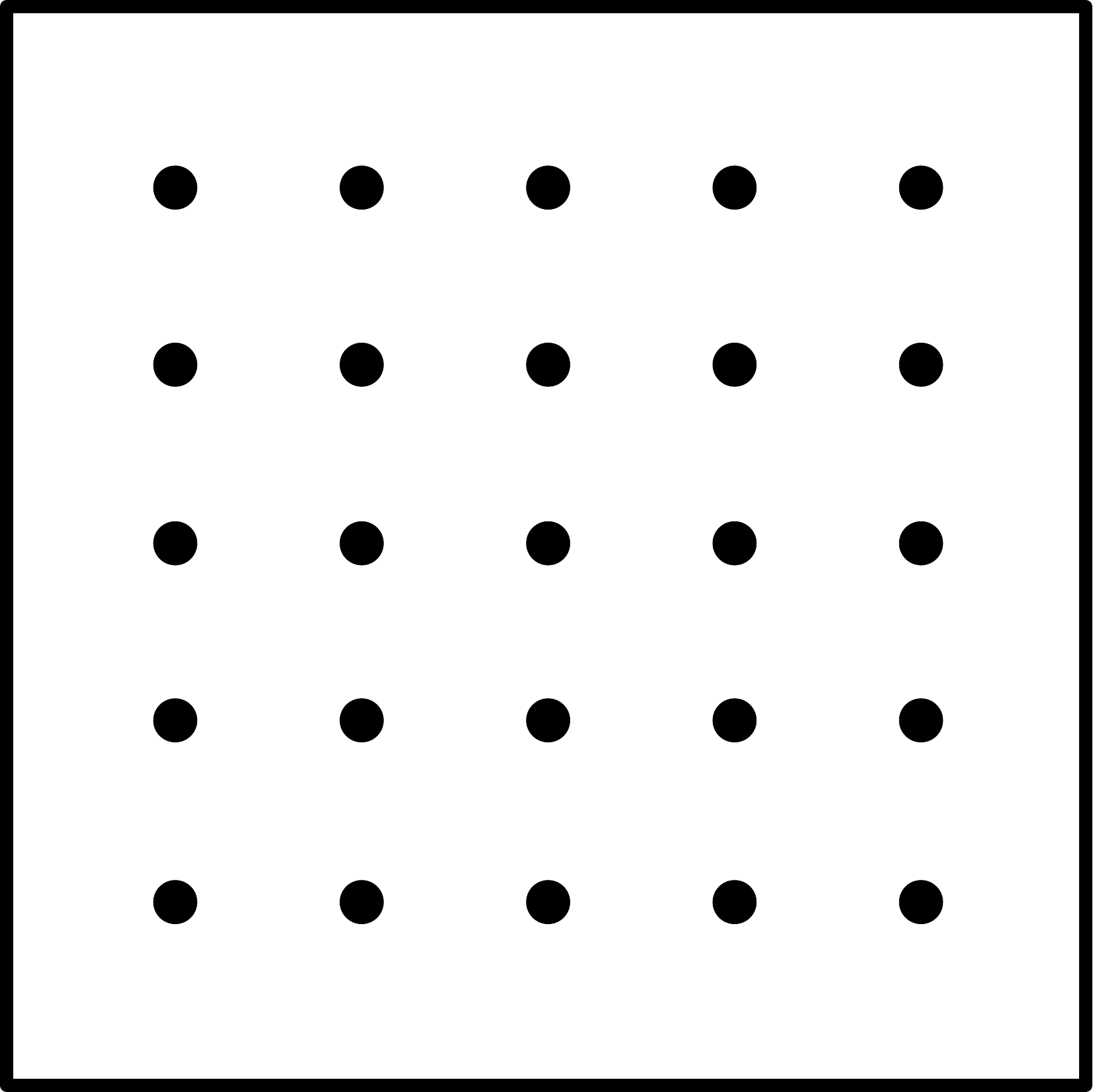}
    \label{point pattern}
  }
  \caption{An illustration of three types of failure patterns in 2-dimensional input domains.}
  \label{FIG:FP}
\end{figure*}

Because RT does not make use of any additional information beyond input parameter requirements, research is ongoing into how to improve its testing effectiveness.
\textit{Adaptive Random Testing} (ART) \cite{Huang2019} is a family of RT-based testing techniques that aims to improve on RT testing effectiveness by more evenly spreading the test cases throughout the input domain.
One of the first, and still most popular, ART implementations is  \textit{Fixed-Size-Candidate-Set} ART (\FSCS\/)~\cite{Chen2004a}.
Basically, for each next test case, \FSCS\/ randomly generates $k$ candidate test cases, calculates the distance between each candidate and each previously-executed test case (that did not reveal any failure), and selects the candidate furthest from them as the next test case to execute.
Many previous studies have demonstrated the high effectiveness of \FSCS\/ compared to RT \cite{Chen2004a,Chen2004c,Chen2006a,Chen2007,Chen2007b,Chen2013}.
However, as reported by Wu \textit{et al.}~\cite{Wu2020}, although ART enhances RT, and is comparable to combinatorial testing in $96\%$ of scenarios, it can be $3.5$ times more computationally expensive than severely-constrained combinatorial testing.

Although ART is very effective, many problems and challenges remain that need to be addressed \cite{Huang2019}.
One of these problems relates to the time required by \FSCS\/ to select test cases, which can be much greater than the execution time:
This is referred to as the \textit{high computational overhead problem}.
Many studies have investigated potential performance improvements for ART algorithms, including:
a forgetting strategy \cite{Chan2006} that reduces the number of distance calculations to previously-executed tests;
an approach, DF-\FSCS\/, that ignores executed test cases not in the line of \textit{sight} of a given candidate \cite{Mao2017};
implementations based on a $K$-Dimensional tree (KD-tree) structure, \KDFC\/~\cite{Mao2019};
a \textit{Single-Instruction-Multiple-Data} (SIMD) mechanism \cite{Ashfaq2020} to calculate all pairwise distances for a single distance calculation instruction;
ART-DC~\cite{Chow2013}, which uses a \textit{divide-and-conquer} strategy to generate the test cases from the entire input domain; and
DMART~\cite{Huang2015}, an enhancement of Mirror ART (MART)~\cite{Chen2004} based on dynamic partitioning,
that generates test cases using a specific ART algorithm in half of the sub-domains, and then \textit{mirrors} these test cases to the other half of the sub-domains to generate the remaining test cases.

As stated, a significant problem faced by \FSCS\/ is the heavy time overheads related to the large number of distance calculations required to find the nearest executed test cases for each candidate test case.
Alleviating this problem will require a better way of identifying nearest executed test cases.
In this paper,
we propose a new ART approach using \textit{Vantage Point Partitioning} (\VPP\/), to improve the efficiency of \FSCS\/.
\VPP\/ uses a modified VP-tree spatial partitioning structure to avoid redundant distance calculations, reducing the computational overheads  of \FSCS\/.
An original vantage point tree (VP-tree)~\cite{Yianilos1993,Uhlmann1991,Chiueh1994} is a special kind of spatial partitioning tree that divides the input space into hyperspheres.
Using a VP-tree, space can be divided into inner and outer regions of the hypersphere,  significantly reducing the number of computations when querying nearest neighbors (\NN\/s) for a given query point.
Therefore, vantage point partitioning addresses the need for \FSCS\/ time overhead reduction.
To evaluate \VPP\/, we conducted a series of simulations and experiments on $22$ subject programs, written in C++ and Java.

\begin{figure*}[!b]
\centering
\graphicspath{{VPP-ART-Figures/}}
    \subfigure[]
    {
        \includegraphics[width=0.235\textwidth]{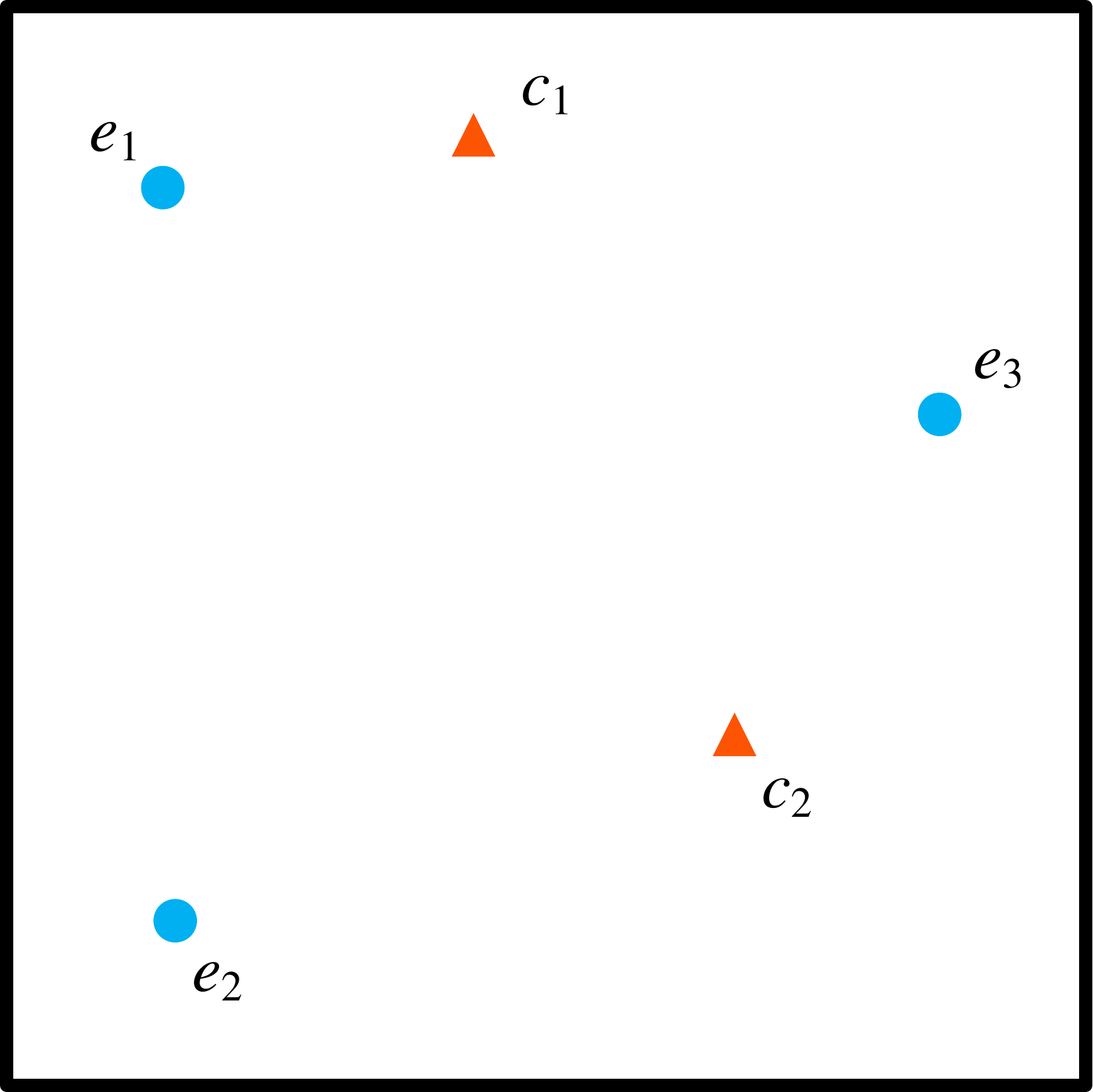}
        \label{FIG:FSCS1}
    }
    \subfigure[]
    {
        \includegraphics[width=0.235\textwidth]{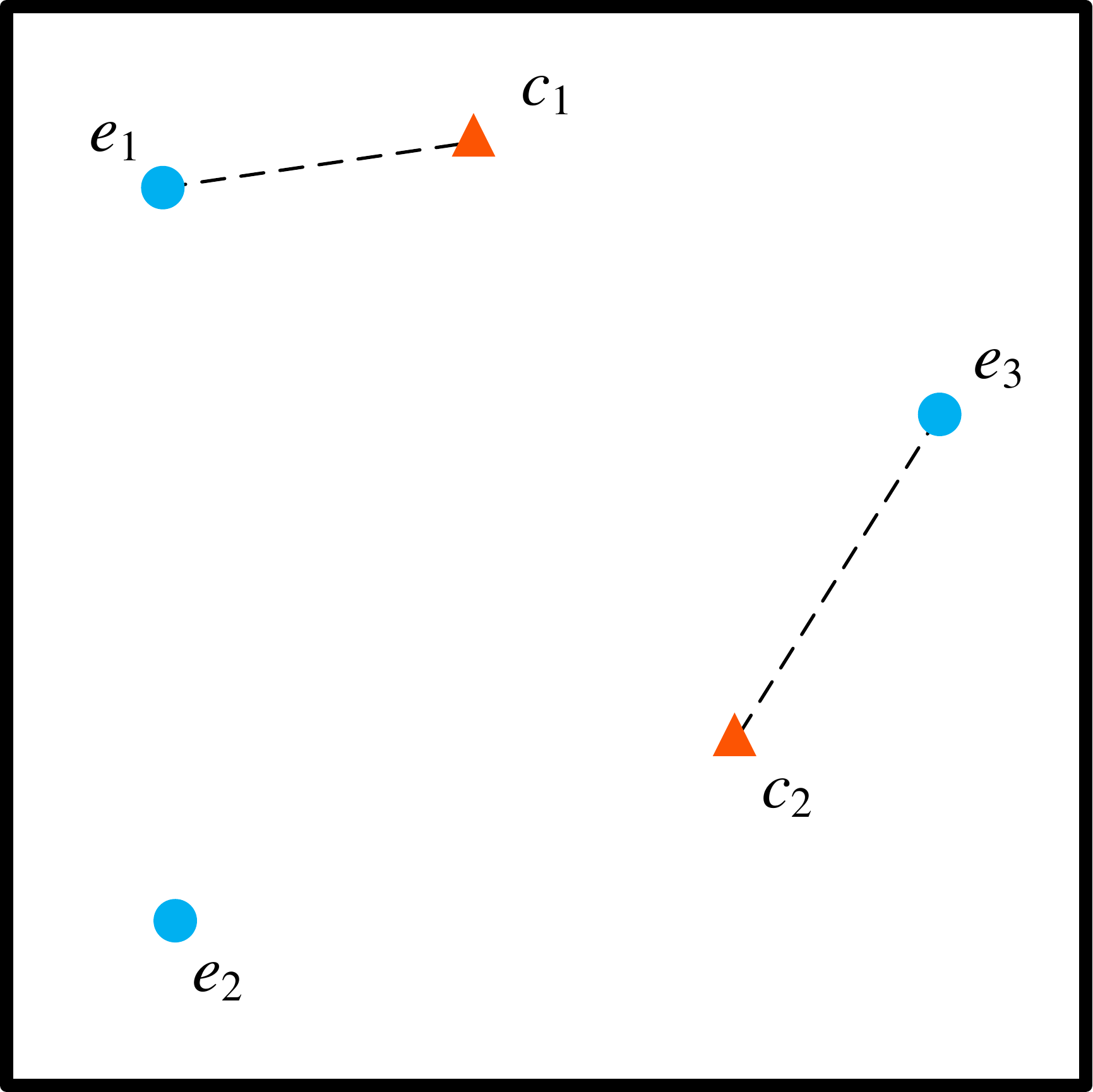}
        \label{FIG:FSCS2}
    }
    \subfigure[]
    {
        \includegraphics[width=0.235\textwidth]{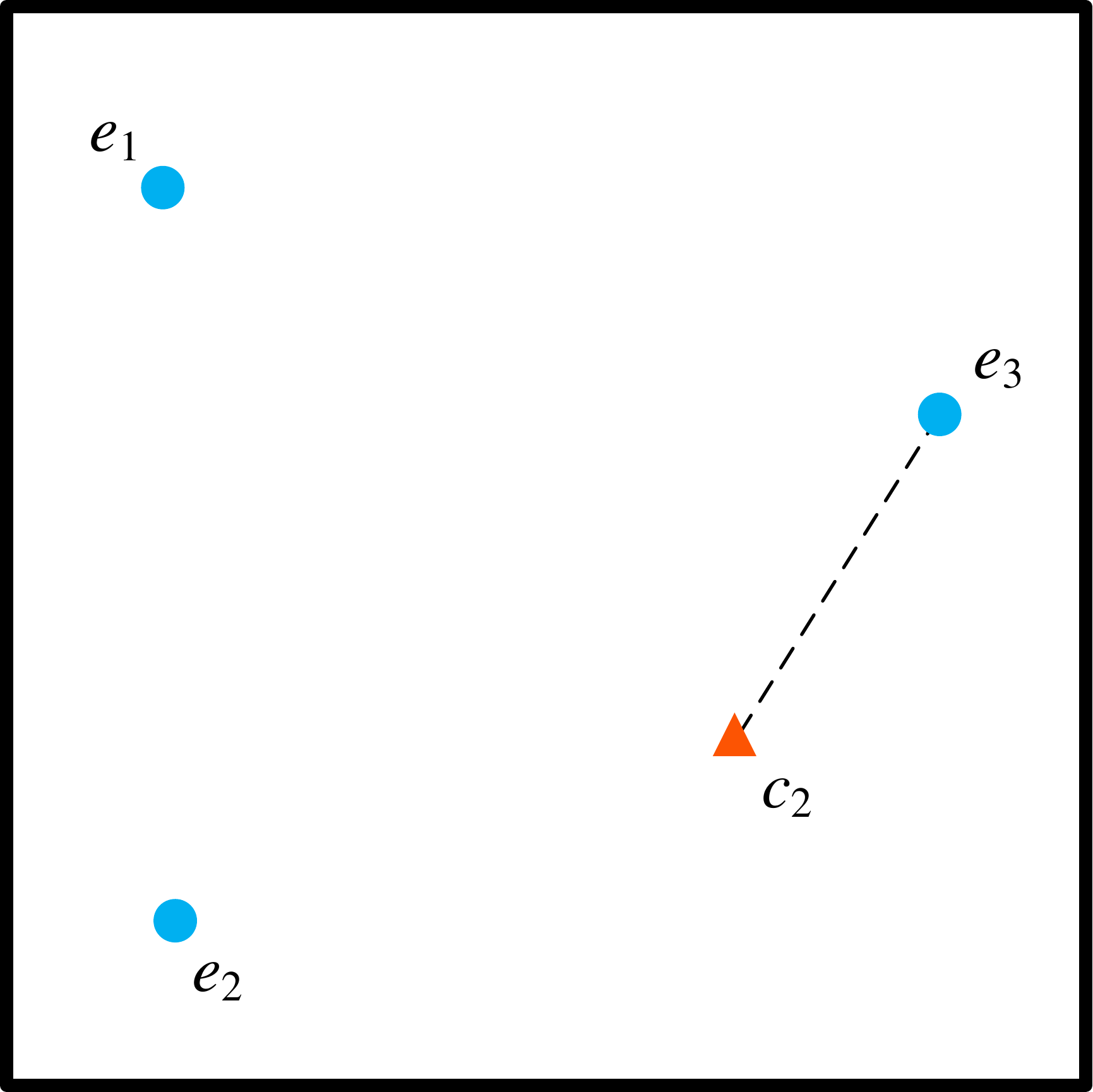}
        \label{FIG:FSCS3}
    }
    \subfigure[]
    {
        \includegraphics[width=0.235\textwidth]{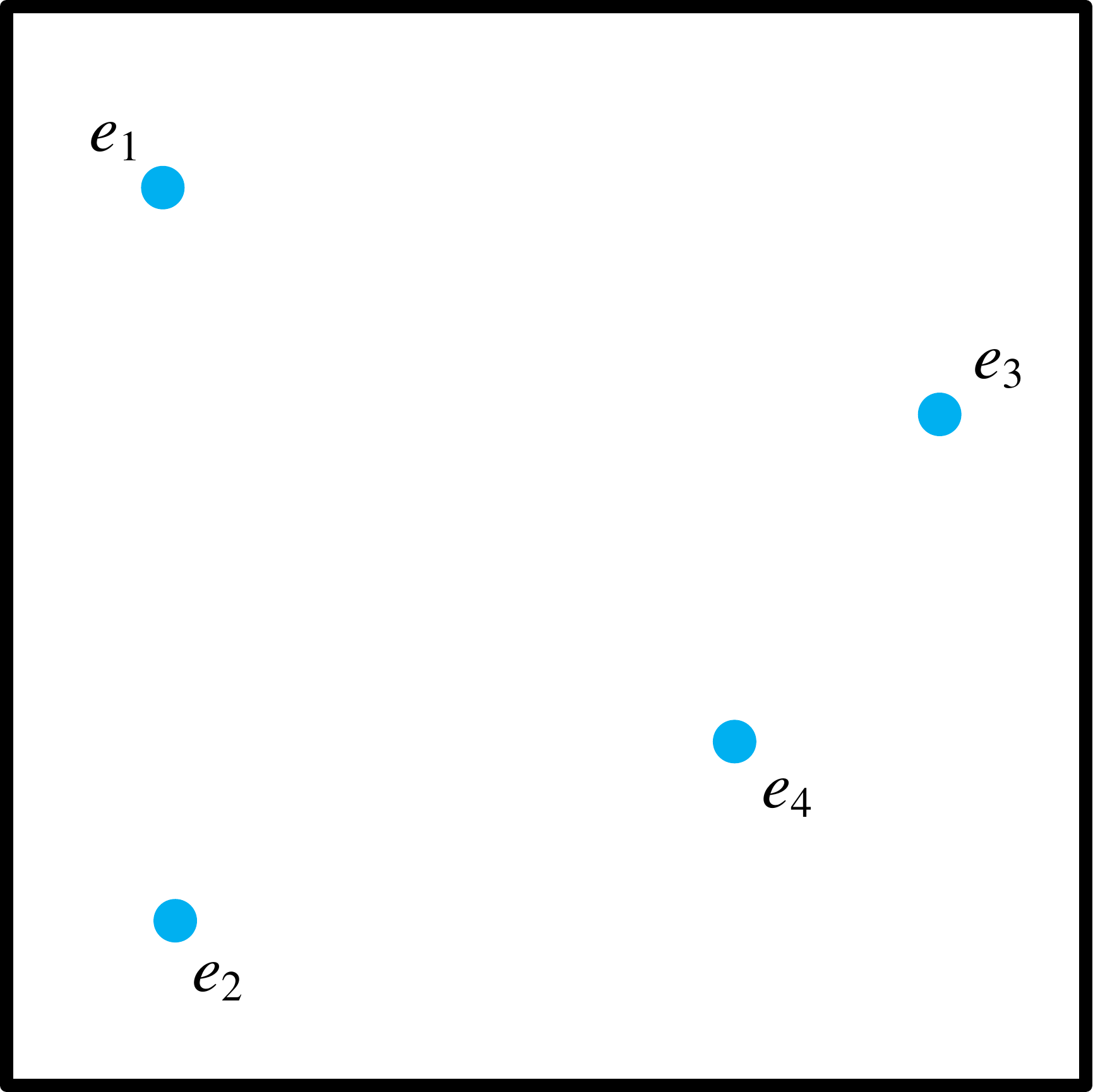}
        \label{FIG:FSCS4}
    }
    \caption{Example illustration of \FSCS\/ generating test cases in a 2-dimensional input domain (from Huang et al.~\cite{Huang2019}).}
    \label{FIG:FSCS}
\end{figure*}

A standard VP-tree is only applicable to static data
---
the points must be known before the VP-tree is constructed.
However, ART test case generation is a dynamic process:
A newly-generated test case $tc$ depends on the information of previously-executed test cases;
if no failure is found by $tc$, then $tc$ should be saved in the VP-tree.
This process requires that the tree structure support dynamic data, especially insert operations.
Because the original VP-tree structure is constructed based on distances between the vantage point and other points,
a worst-case scenario exists when a lower-level node in the tree changes, causing upper-level nodes to also (possibly) change, which may necessitate reconstruction of the entire tree.
This problem can be addressed by revising the original VP-tree structure to support dynamic data.

The main contributions of this paper are:
\begin{itemize}
  \item [1)]
      We propose an improved VP-tree structure that can support dynamic insertion.
      This tree structure can identify an approximate \NN\/ that does not differ much from the exact \NN\/, reducing the time cost of \FSCS\/.
      To the best of our knowledge, this is the first paper to propose using vantage point partitioning to address the ART time overheads problem.
  \item [2)]
      We report on simulations and experiments investigating \VPP\/, from the perspectives of testing effectiveness and efficiency.
  \item [3)]
      Compared with \FSCS\/, our approach significantly reduces computational overheads while delivering comparable, or better, failure-detection effectiveness.
      Compared with \KDFC\/ algorithms, our approach has similar or better performance, with reduced time costs in high dimensions.
\end{itemize}

The rest of this paper is organized as follows:
Section~\ref{SEC:background} introduces some background information about failure patterns, the original \FSCS\/ method, and vantage point partitioning.
Section~\ref{SEC:VPPS-ART} presents a framework to enhance \FSCS\/, and introduces \VPP\/.
Section~\ref{SEC:experimental studies} describes the simulations and experiments,
the results and analyses of which are presented in Section~\ref{SEC:results}.
Section~\ref{SEC:Threat-to-Validity} discusses the potential threats to the validity of our studies.
Related work is discussed in Section~\ref{SEC:relatedWork}.
Finally, we conclude the paper and discuss some potential future work in Section~\ref{SEC:conclusion}.

\section{Background
\label{SEC:background}}

In this section, we briefly present some background information about failure patterns and Fixed-Size-Candidate-Set ART.
We also introduce some preliminary concepts about vantage point partitioning.

\subsection{Failure Regions
\label{failure rate and failure patterns}}

The inputs to a faulty program can be divided into two distinct types:
\textit{failure-causing inputs} (those inputs which, when executed, cause a test to fail); and
\textit{non-failure-causing inputs} (inputs which do not reveal a failure).
The program's \textit{failure region} consists of the set of all its failure-causing inputs.
In software testing, knowledge of a failure region can be an extremely helpful guide for test case generation and selection.
In general, two basic features are used to describe the failure region:
the \textit{failure pattern}, which is the distributions, shape, and locations of failure-causing inputs in the input domain; and
the \textit{failure rate}, denoted $\theta$, which is the proportion of failure-causing to all possible inputs in the entire input domain.

A number of studies \cite{White1980,Ammann1988,Finelli1991,Bishop1993,Schneckenburger2007} have reported that failure-causing inputs tend to cluster into contiguous regions.
Chan \textit{et al.}~\cite{Chan1996} classified failure patterns into three types:
\textit{block pattern};
\textit{strip pattern}; and
\textit{point pattern}.
Figure~\ref{FIG:FP} shows these three main failure patterns in $2$-dimensional input domains, where the bounding boxes represent the input domain boundaries, and the shaded areas represent the failure regions (containing the failure-causing inputs).
Chan \textit{et al.}~\cite{Chan1996} also suggested that block and strip patterns are more commonly found than point patterns.

\subsection{\FSCS\/: Fixed-Size-Candidate-Set ART
\label{Original FSCS-ART Method}}

ART is a family of testing methods that improve over RT effectiveness by distributing the test cases more evenly throughout the input domain.
One of the first, and still the most popular, ART implementations is \textit{Fixed-Size-Candidate-Set} ART (\FSCS\/)~\cite{Chen2004a}.
Many studies have shown \FSCS\/ to be more effective than RT, in terms of failure-detection effectiveness~\cite{Chen2004a,Chen2004c,Chen2006a,Chen2007}, test case distribution~\cite{Chen2007b}, and code coverage~\cite{Chen2013}.

\FSCS\/~\cite{Chen2004a} makes use of the concept of {\em distance} to evaluate similarities among test cases.
It maintains two sets of test cases, the candidate set $C$ and the executed set $E$.
$C$ stores $k$ test cases that are randomly generated from the input domain, and $E$ stores the test cases that have already been executed (but without causing any failure).
Previous studies~\cite{Chen2004a} have recommended a default value of $10$ for $k$.

The \FSCS\/ test case generation process can be described as follows:
The first test case, $e_1$, is randomly generated from the input domain, and executed.
Assuming that $e_1$ does not reveal a failure, it is then stored in $E$.
From now on, each time a new test case is needed, $k$ test cases are randomly generated, and stored in the candidate set $C$.
The \textit{best} element from $C$ is selected as the next test case to be executed
---
with \FSCS\/, best is defined as being farthest from the previously-executed test cases (stored in $E$).
As testing progresses (without failures being revealed), $E$ grows larger.
Formally, given
a nonempty set of executed test cases $E$ ($\{e_1,e_2,\cdots,e_{|E|}\}$), and
a fixed number $k$ of candidate test cases in $C$ ($\{c_1,c_2,\cdots,c_k\}$), then
the requirement for selecting the next (best) test case $c_{\textit{best}}$ is:
\begin{equation}\label{EQ:2.1}
    \min_{\forall e_i \in E,c_j \in C} dist(c_j,e_i) \leq \min_{\forall e_i \in E}dist(c_{\textit{best}},e_i),
\end{equation}
where $dist(x,y)$ is the distance between test cases $x$ and $y$ (typically the \textit{Euclidean distance} for numerical input domains).

Figure~\ref{FIG:FSCS} shows an example of \FSCS\/ generating test cases in a $2$-dimensional input domain.
In the first step of the example (Figure~\ref{FIG:FSCS1}), there are
three executed test cases, $e_1$, $e_2$, $e_3$ (denoted by small dots); and
two candidate test cases, $c_1$, $c_2$ (denoted by small triangles).
In order to select the next test case, the distance between each candidate test case and each executed test case is calculated, and the \NN\/ to each candidate is identified.
As shown in Figure~\ref{FIG:FSCS2}, $e_1$ is the \NN\/ of $c_1$, and $e_3$ is the \NN\/ of $c_2$.
The candidate with the greatest distance to its \NN\/
---
$c_2$ in Figure~\ref{FIG:FSCS3}
---
is selected as the next test case, and executed.
If no termination condition is satisfied (e.g., no failure is revealed), then $c_2$ is stored in $E$ as the fourth executed test case, $e_4$, as shown in Figure~\ref{FIG:FSCS4}.
This processes continues until a termination condition is satisfied.

A challenge for the \FSCS\/ algorithm is its \textit{high computational overheads}:
Each iteration of the \FSCS\/ process requires distance calculations between each candidate test case in $C$ and all the previously-executed test cases.
The time complexity of \FSCS\/ is $O(kN^2)$, where $k$ is the size of $C$ and $N$ is the number of previously-executed test cases~\cite{Huang2019}.
In order to detect a failure in a program, \FSCS\/ could take an enormous amount of time to generate the required number of test cases.
When the program's failure rate is very small, the \FSCS\/ time cost will be very high.
A key to reducing the computational overheads is to optimize the search for candidates' nearest executed test cases.
Therefore, adoption of the highly-efficient \textit{\NN\/ Search} \cite{Sanchez2004} strategies should enable a reduction in the search time.

\subsection{Vantage Point Partitioning
\label{SEC:vp tree}}

Given a point set in a $d$-dimensional vector space, \textit{Vantage Point Partitioning} (VPP)~\cite{Yianilos1993} makes use of the relative distances between the points and a particular \textit{vantage point} to enable a very efficient \NN\/ search.

VPP can be organized into a tree structure, a \textit{Vantage Point tree} (VP-tree)~\cite{Yianilos1993,Uhlmann1991,Chiueh1994}.
The VP-tree can reduce unnecessary computations when solving NN search problems~\cite{Sanchez2004}, and has been used in various contexts, including:
computational biology~\cite{DeFreitas2016};
image processing~\cite{Markov2007};
databases~\cite{Shishibori2011,Skalak2008}; and
computer vision~\cite{Kumar2008}.
We have applied some modifications to the original VPP algorithm to enable its use with ART.
Our research is, to the best of our knowledge, the first time VP-trees have been applied to test case generation in the field of ART.

\begin{figure}[!t]
\centering
\graphicspath{{VPP-ART-Figures/}}
    \subfigure[Binary VP-tree]
    {
        \includegraphics[width=0.38\textwidth]{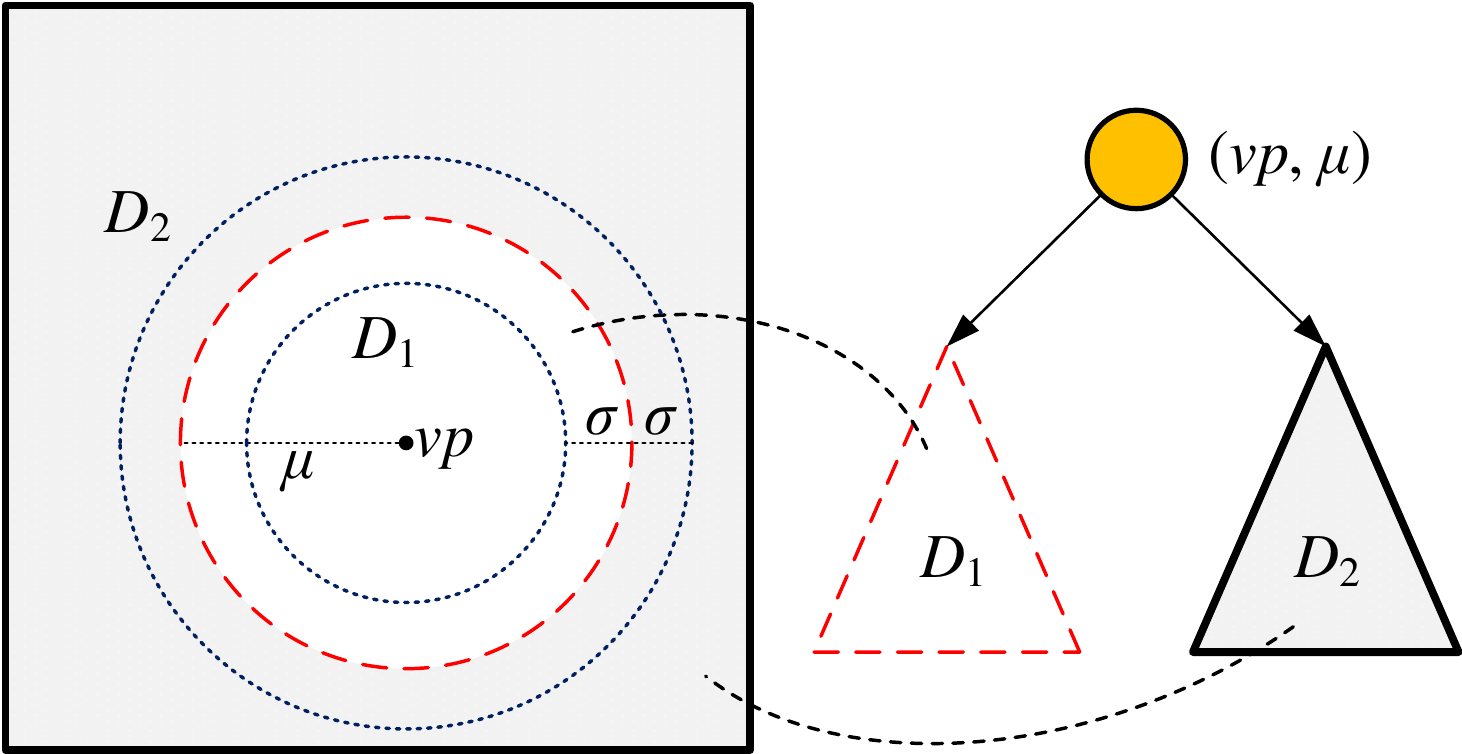}
        \label{FIG:vp-2-p1}
    }

    \subfigure[$\varepsilon$-ary VP-tree]
    {
        \includegraphics[width=0.47\textwidth]{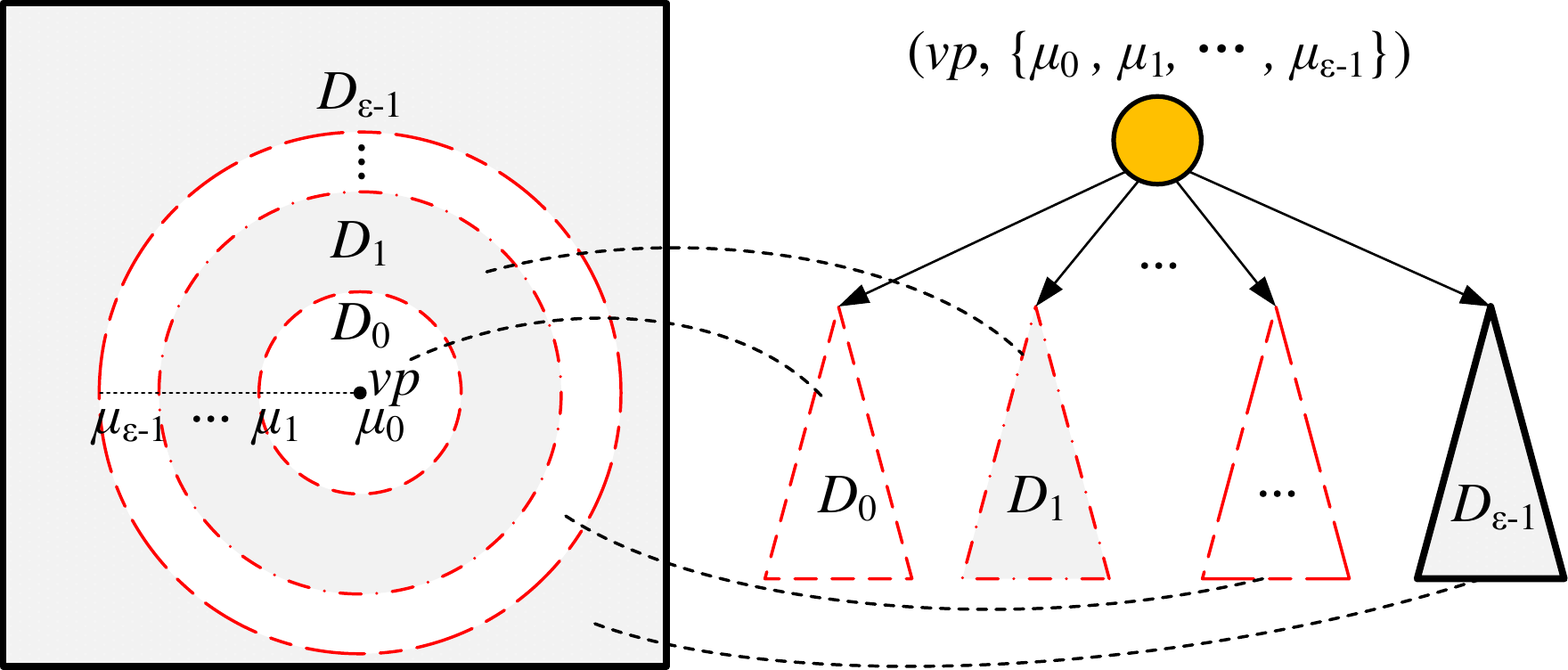}
        \label{FIG:vp-n-p1}
    }
    \caption{VP-tree partitioning strategy in a 2-dimensional input domain.}
    \label{FIG:vp}
\end{figure}

\subsubsection{VP-tree Construction Process}

To illustrate the VP-tree construction process, we use an example of binary (2-ary) partitioning.
This can easily be generalized to $\varepsilon$-ary cases, where $\varepsilon > 2$~\cite{Fu2000}.

Generally speaking, a binary VP-tree is constructed by splitting a data set into two subsets using a \textit{distance partitioning criterion} and a \textit{vantage point}.
The vantage point is stored in the root node of the binary VP-tree, and the two subsets are organized into the left and right sub-trees of the root node.
Then, the sub-trees are both processed recursively, constructing in each the next level sub-tree according to newly-selected vantage points.
This continues until each node contains only one data point, and the construction of the tree is completed.
Formally, given a set $\mathcal{D}$ of $n$ data points, a point $vp$ is randomly chosen as the vantage point.
Next, the distances between $vp$ and other points in $\mathcal{D}$ are calculated:
$\mathcal{S} =\big\{dist(p,vp)|p \in \mathcal{D}-\{vp\}\big\}$.
The entire data set can be partitioned into two subsets using the median distance value $\mu$ in $\mathcal{S}$:
As shown in Figure~\ref{FIG:vp-2-p1},
$\mathcal{D}_1$ refers to the points within a distance of $\mu$ from $vp$; and
$\mathcal{D}_2$ refers to the points that are more than a distance of $\mu$ from $vp$.


\begin{figure}[!b]
    \centering
    \graphicspath{{VPP-ART-Figures/}}
    \subfigure[Binary VP-tree partitioning]
    {
        \includegraphics[width=0.35\textwidth]{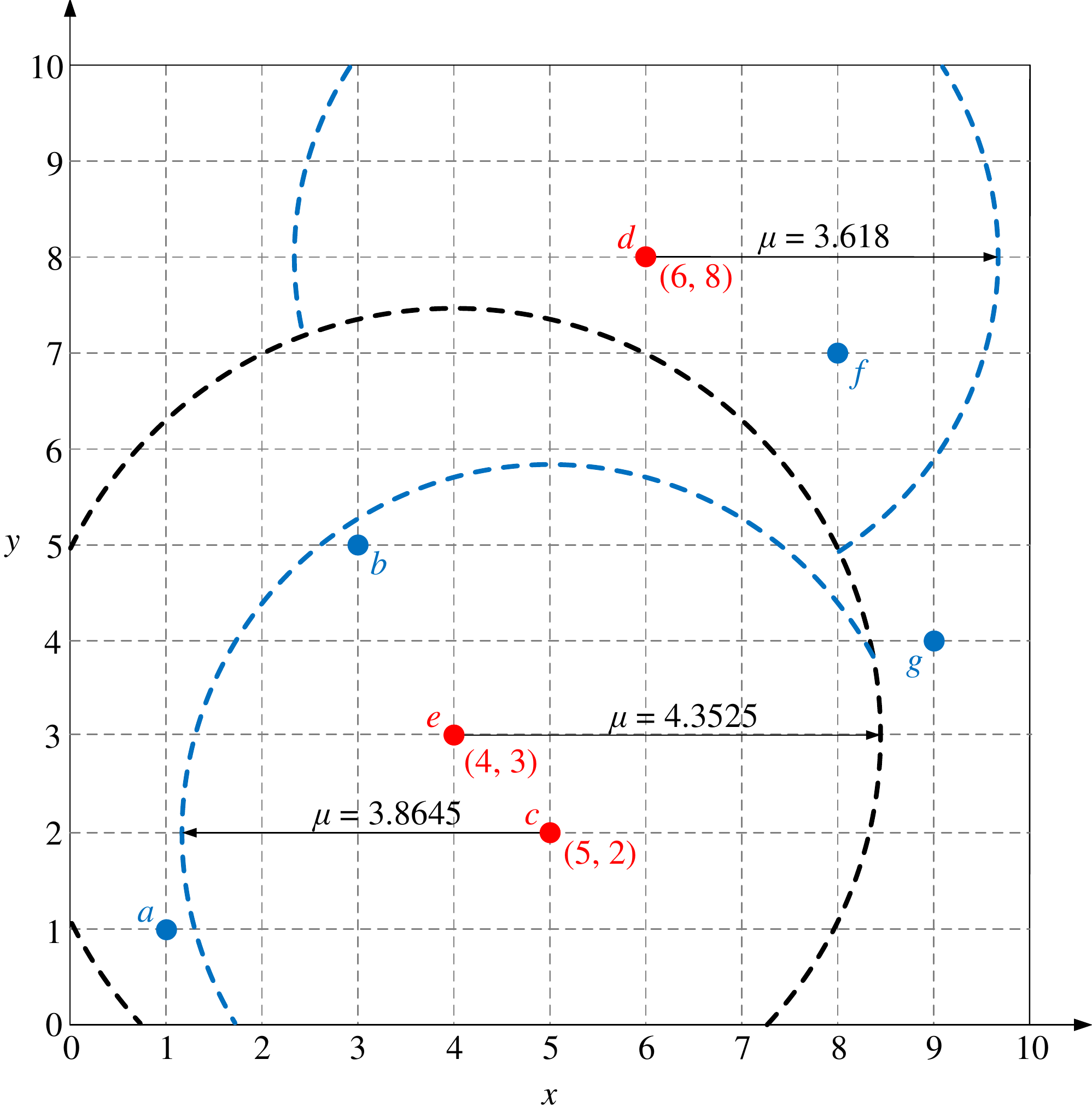}
        \label{FIG:vp-2-p}
    } 

    \subfigure[$\varepsilon$-ary VP-tree partitioning]
    {
        \includegraphics[width=0.35\textwidth]{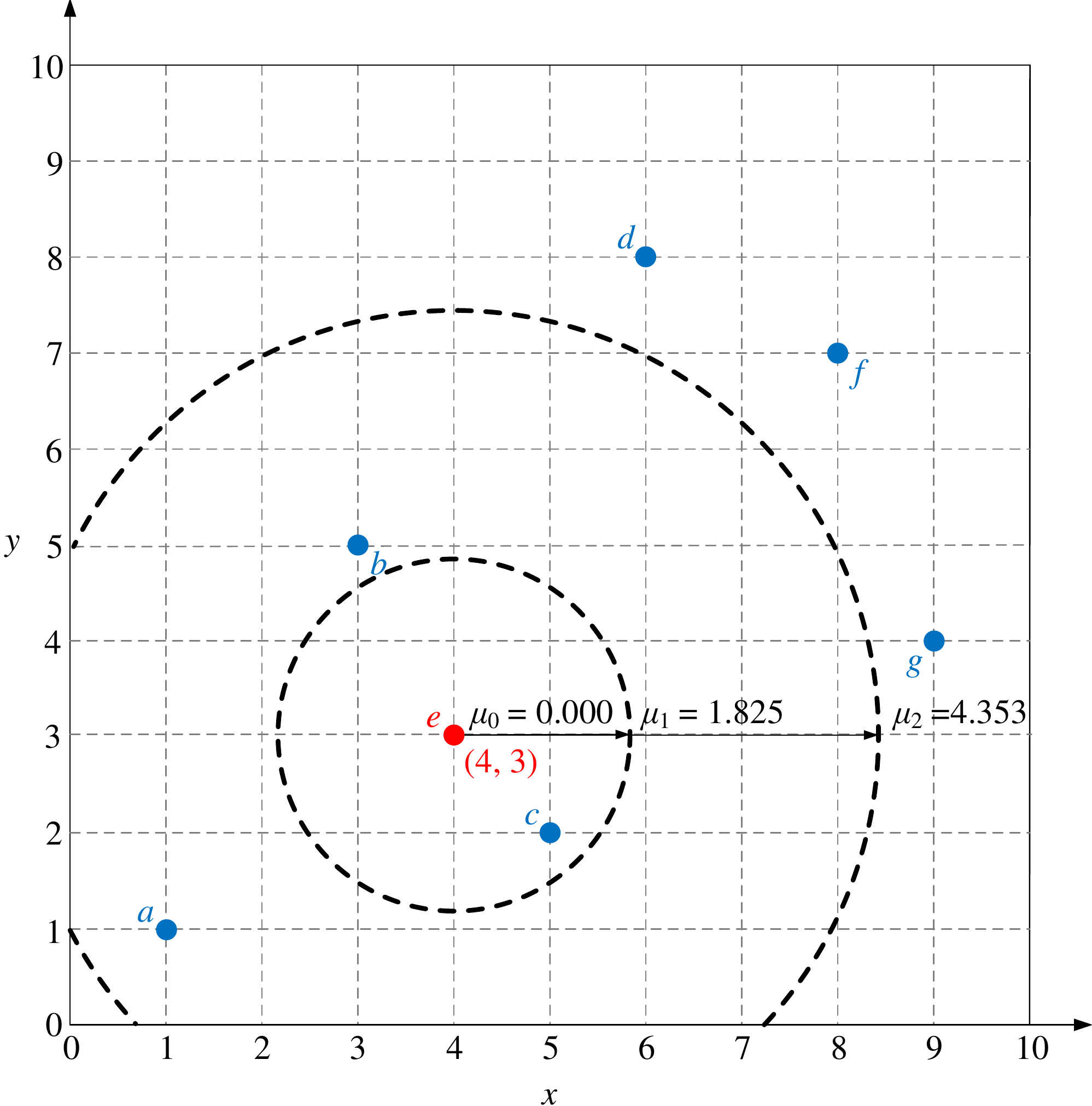}
        \label{FIG:vp-n-p}
    }
    \caption{Example comparison between binary and $\varepsilon$-ary VP-tree partitioning in a 2-dimensional input domain.}
    \label{FIG:vptree partitioning strategy}
\end{figure}

The construction process for an $\varepsilon$-ary VP-tree when $\varepsilon>2$, is similar to the binary tree case:
For a given set of points ($\mathcal{D}$), a vantage point $vp$ is again randomly chosen, and the distances between $vp$ and all points in $\mathcal{D}$ are calculated and stored in ascending order.
The differences compared with the binary case are:
(1) the data set is not partitioned into only two subsets, but into $\varepsilon$ approximately-equal subsets, and
(2) $vp$ is stored in the first subset (in this paper).
As shown in Figure~\ref{FIG:vp-n-p1},
$\mu_i$ ($i = 1,2, \cdots, \varepsilon-1$)
denote the boundary distance values that split $\mathcal{D}_{i-1}$ and $\mathcal{D}_{i}$.
Formally, for a sequence of distances, stored in ascending order, $\mathcal{S}=\big\{dist(a_j,vp)| a_j \in \mathcal{D}, j=0,1,\cdots,|\mathcal{D}|-1\big\}$,
where $|\mathcal{D}|$ is the number of elements in $\mathcal{D}$,
the boundary distance values $\mu_i$ can be calculated as follows:
\begin{equation}\label{EQ:insert1}
    \mu_i=\begin{cases}
    0,&i=0\\
    \frac{\mathcal{S}\Big(i \times \big\lfloor |\mathcal{D}| / \varepsilon \big\rfloor - 1\Big)
    + \mathcal{S}\Big(i \times \big\lfloor |\mathcal{D}| / \varepsilon \big\rfloor \Big)}{2},
    &i\in [1,\varepsilon-1]\\
    \end{cases}
\end{equation}
where
$\mathcal{S}(j)$ represents the $j$-th ordered element in $\mathcal{S}$ ($j = 0,1, \cdots, |\mathcal{D}|-1$), and
$\big\lfloor |\mathcal{D}|/ \varepsilon \big\rfloor$ indicates the rounding down integer $|\mathcal{D}| / \varepsilon$.
Therefore, for each point $p$ in $\mathcal{D}_i$, the value of $dist(p, vp)$ is between $\mu_i$ and $\mu_{i+1}$.

\begin{figure}[!t]
    \centering
    \graphicspath{{VPP-ART-Figures/}}
    \subfigure[Binary VP-tree structure]
    {
        \includegraphics[width=0.28\textwidth]{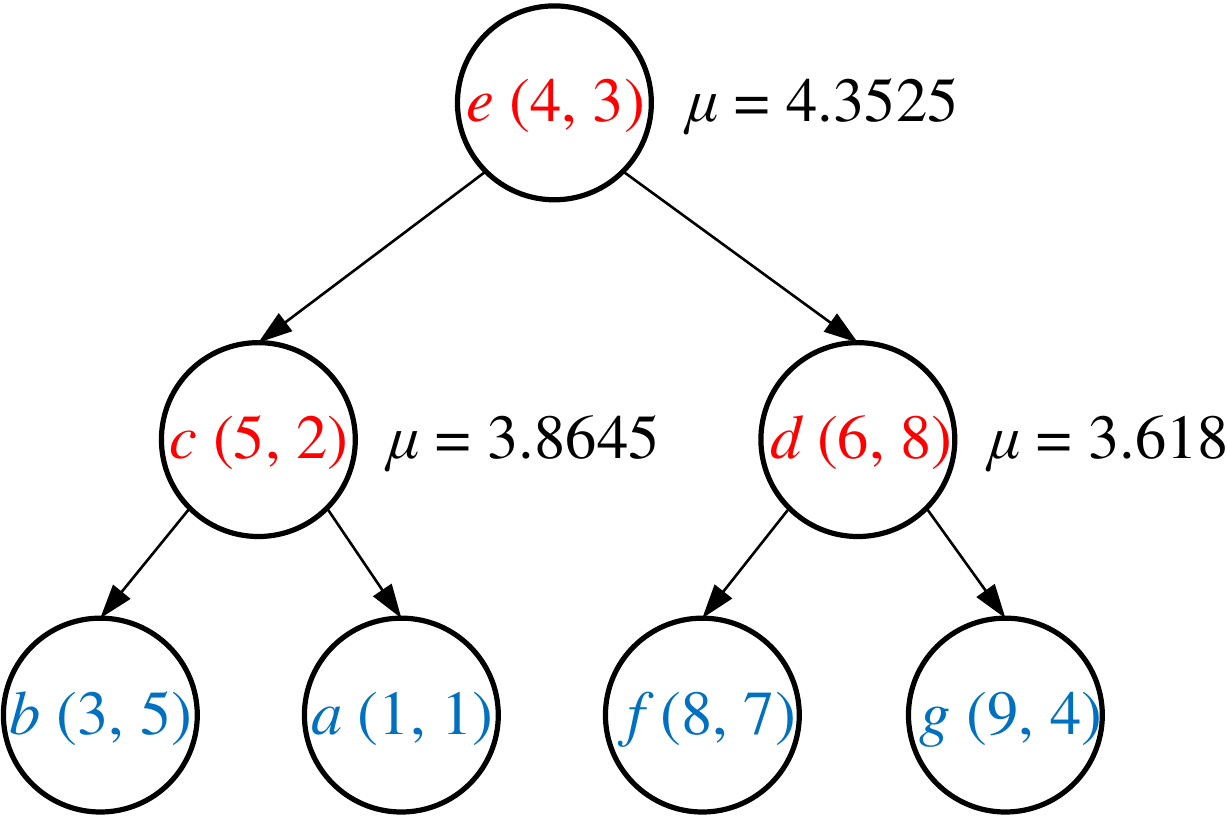}
        \label{FIG:vp-2s}
    }

    \subfigure[$\varepsilon$-ary VP-tree structure]
    {
        \includegraphics[width=0.38\textwidth]{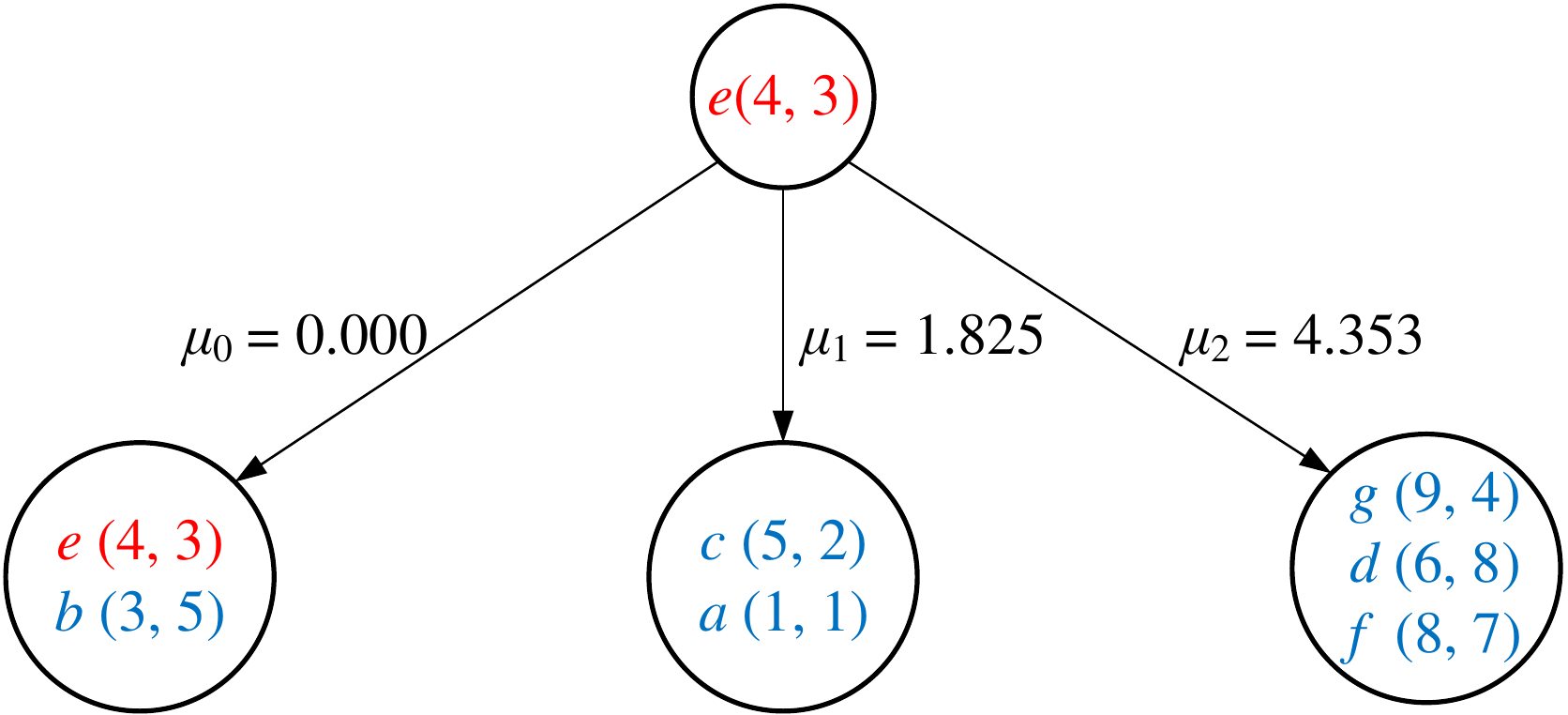}
        \label{FIG:vp-ns}
    }
    \caption{Example of the comparison between binary VP-tree structure and $\varepsilon$-ary VP-tree structure in a 2-dimensional input domain.}
    \label{FIG:vptree structure strategy}
\end{figure}

To illustrate the VP-tree construction process, consider a set of data points $\mathcal{D}=\{a, b, c, d, e, f, g\}$ in a 2-dimensional vector space.
The binary VPP process is illustrated in Figure~\ref{FIG:vp-2-p}:
(1) Point $e$ is randomly selected as the vantage point;
(2) the distances between $e$ and all other points are calculated, and stored in $\mathcal{S}$ $=$ $\big\{dist(p,e)|p \in \mathcal{D}-\{e\}\big\}$ $=$ $\{1.414,2.236, 3.606,5.099,5.385,5.657\}$;
(3) the median value in $\mathcal{S}$ is calculated
---
 $\mu=(3.606+5.099)/2=4.3525$, in this example;
(4) $\mathcal{D}_1$ $=$ $\{c,b,a\}$ and $\mathcal{D}_2$ $=$ $\{g,d,f\}$ are constructed;
(5) the points in $\mathcal{D}_1$ and $\mathcal{D}_2$ are recursively organized, according to the preceding steps, until each subset contains only one data point.
The corresponding binary VP-tree structure is shown in Figure~\ref{FIG:vp-2s}.

The $\varepsilon$-ary case is illustrated in Figure~\ref{FIG:vp-n-p}, with the final $\varepsilon$-ary VP-tree structure shown in Figure~\ref{FIG:vp-ns}, where $\varepsilon = 3$.
Similar to the binary case:
(1) Point $e$ is randomly selected as the vantage point;
(2) the distances between $e$ and all points are calculated, $\mathcal{S}$ $=$ $\big\{dist(p,e)|p \in \mathcal{D}\big\}$ $=$ $ \{0.000,1.414,2.236, 3.606,5.099,5.385,5.657\}$;
(3) three boundary values are defined
---
$$
    \begin{aligned}
        \mu_0&=0.000,\\
        \mu_1 &= \frac{\mathcal{S}\Big(\big\lfloor 7/3 \big\rfloor \times 1-1\Big) + \mathcal{S}\Big(\big\lfloor 7/3 \big\rfloor \times 1\Big)}{2}\\
        &= \frac{\mathcal{S}(1)+\mathcal{S}(2)}{2}
        = \frac{1.414+2.236}{2}
        = 1.825,\\
        \mu_2 &= \frac{\mathcal{S}\Big(\big\lfloor 7/3 \big\rfloor \times 2-1\Big) + \mathcal{S}\Big(\big\lfloor 7/3 \big\rfloor \times 2\Big)}{2}\\
        &= \frac{\mathcal{S}(3)+\mathcal{S}(4)}{2}
        = \frac{3.606+5.099}{2}
        = 4.353,
    \end{aligned}
$$
partitioning $\mathcal{D}$ into three approximately-equally-sized subsets; (4) the steps above are repeated in each subset.

\subsubsection{Nearest Neighbor Search in a VP-tree
\label{SEC:NNS in VPTREES}}

This section describes the algorithm for an \NN\/ search in a VP-tree, which involves identifying the nearest neighboring point of a query point $q$ with the requirement that the maximum distance between $q$ and the point be less than a specific threshold $\sigma$.
This means that if the distance between $q$ and its neighbor is greater than $\sigma$, then this cannot be the \NN\/.
If $dist(q,vp)$ is the distance between the query point $q$ and the vantage point $vp$, then the algorithm focuses on finding the \NN\/ of $q$ within the range $dist(q,vp)\pm\sigma$.
With these requirements, the search for the \NN\/ of $q$ in a binary VP-tree only needs to explore both $\mathcal{D}_1$ and $\mathcal{D}_2$ if $q$ is in the range of $[\mu-\sigma,\mu+\sigma]$ (as shown in Figure~\ref{FIG:vp-2-p1})
---
otherwise, only one subset needs to be searched, which effectively prunes one half of the input space.
The approach is based on the principles of triangular inequality.
Formally, if $dist(q,vp)\leq \mu-\sigma$, for $p \in \mathcal{D}_2$, the distance between $p$ and $q$ is lower-bounded by $\sigma$~\cite{Chiueh1994}:
\begin{equation}
    \begin{aligned}
        dist(p,q)  \geq&~   \Big|\big|dist(p,vp)\big|-\big|dist(q,vp)\big|\Big|\\
                \geq&~   \Big|\big|dist(p,vp)\big|-(\mu-\sigma)\Big|\\
                   >&~   |\mu-\mu+\sigma|\\
                   =&~   \sigma,\\
    \end{aligned}
\end{equation}
therefore, the subset $\mathcal{D}_2$ can be ignored.
Similarly, if $dist(q,vp)> \mu+\sigma$, for $p \in \mathcal{D}_1$, then $\mathcal{D}_1$ can be ignored.
For the $\varepsilon$-ary cases, the system needs to explore $\mathcal{D}_i$ if:
\begin{equation}\label{EQ:NVPSEARCHCONDITION}
    \mu_i-\sigma < d(q,vp)\leq \mu_{i+1}+\sigma,
\end{equation}
for
$i=0,1,\cdots,\varepsilon-2$
(the special case of $i= \varepsilon-1$ will be discussed in Section~\ref{SEC:ANNS in VPTREES}):
It is also based on triangular inequality, and can be derived in a similar way to the binary case.

\section{\VPP\/: ART based on Vantage Point Partitioning
\label{SEC:VPPS-ART}}

In this section, we present our proposed ART approach,  \textit{Vantage Point Partitioning ART} (\VPP\/).
\begin{figure*}[t]
    \centering
    \graphicspath{{VPP-ART-Figures/}}
    \includegraphics{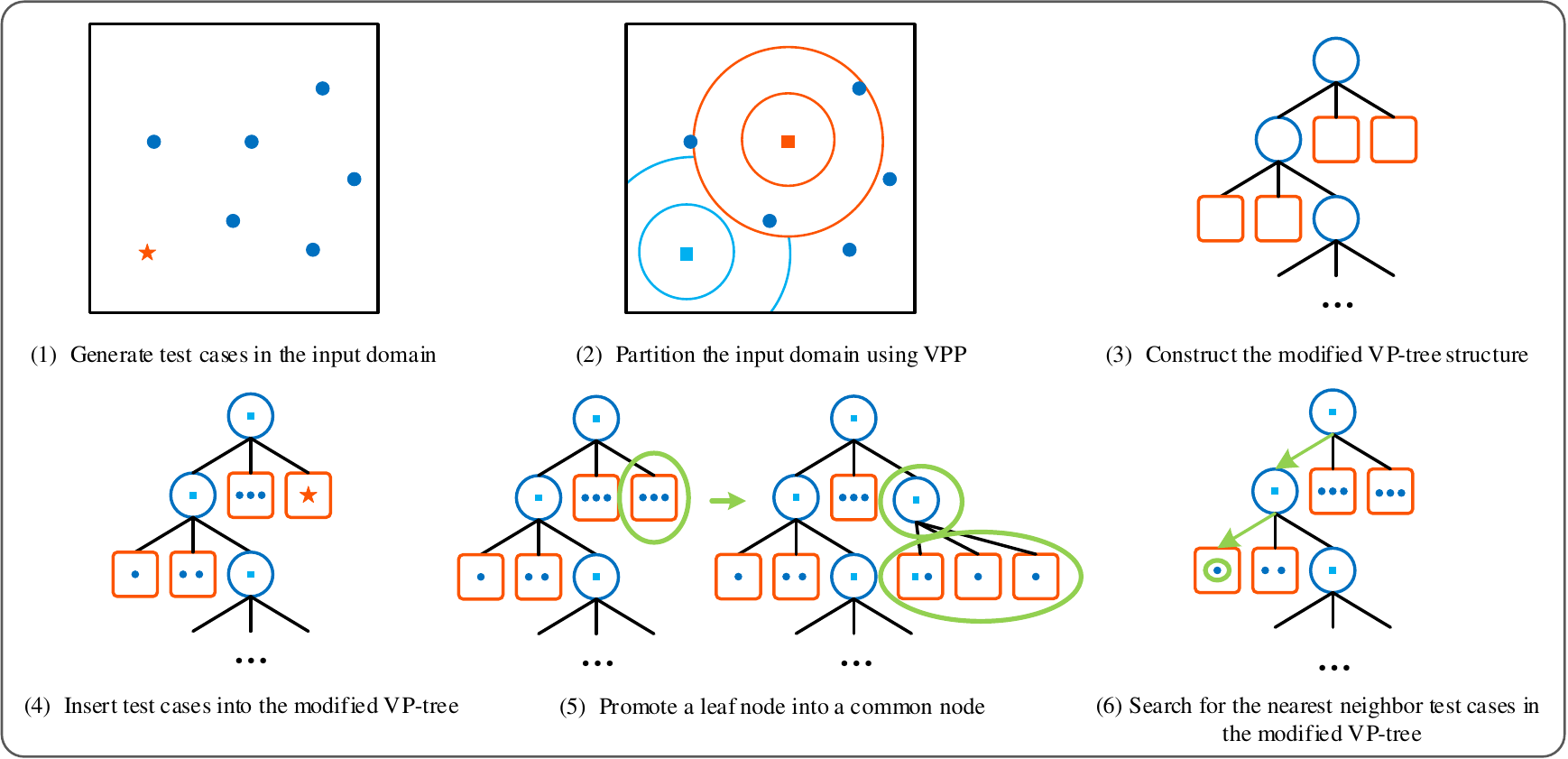}
    \caption{The \VPP\/ framework, with six main stages.}
    \label{FIG:VPPS-ART-framework}
\end{figure*}

\subsection{Framework
\label{SEC:vp framework}}

A main challenge for \FSCS\/ lies in the high time cost in generating test cases.
In this paper, we combine vantage point partitioning with the original \FSCS\/ to improve the efficiency, mainly by organizing the set of executed test cases into a new storage structure.
As noted, ART requires that the data structure used to store the executed test cases be able support dynamic insertion.

The entire \VPP\/ approach is similar to that of \FSCS\/.
(1) Initially, a test case is randomly generated within the input domain.
This test case is used to execute the SUT, and, if no testing termination condition is met, then this test case is added to the modified VP-tree.
(2) Thereafter, $k$ candidate test cases are randomly generated in each round, and (temporarily) stored in the candidate test case set $C$.
(3) The best candidate in $C$ is determined using a search strategy to find all candidates' \NN\/s in the modified VP-tree,
with the candidate whose minimum distance is greatest then selected (Equation~\ref{EQ:2.1}).

In general, typical testing termination conditions include:
(1) at least one failure has been detected in the program under test; and
(2) the number of executed test cases has reached some predetermined threshold value.
When the termination condition is satisfied, the entire testing process ends.


The framework is shown in Figure~\ref{FIG:VPPS-ART-framework}.
In the figure, the small squares represent the vantage points, the dots represent  executed test cases, and the star represents a newly-generated test case.
The framework consists of six main stages:
(1) Generate test cases in the input domain;
(2) Partition the input domain using VPP;
(3) Construct the modified VP-tree structure;
(4) Insert test cases into the modified VP-tree;
(5) Promote a leaf node into a common node; and
(6) Search for the \NN\/ test cases in the modified VP-tree.

\textbf{\textit{Stage 1}}:
The test cases are generated in the same way as in \FSCS\/.
The candidate test case set ($C$) contains $k$ test cases randomly generated in the input domain.
The executed test cases are stored in a modified VP-tree.
Similar to \FSCS\/, the Euclidean distance is used to measure the similarity (distance) between test cases.

\textbf{\textit{Stage 2}}:
As the testing process proceeds, VPP is used to partition the input domain into different concentric hypersphere sub-domains, bounded by different vantage points.
Each sub-domain contains far fewer test cases than the number of executed test cases in the entire input domain.

\textbf{\textit{Stage 3}}:
A modified VP-tree structure that supports dynamic data is used to store executed test cases, and to support the \NN\/s searches.

\textbf{\textit{Stage 4}}:
As \VPP\/ proceeds, the candidate test cases ($C$) are generated randomly within the input domain, and the best candidate is identified and applied to the SUT.
If an SUT failure is not revealed (and no other testing termination criteria are met), then the test case is added to the modified VP-tree.
As the testing continues, the number of executed test cases in the VP-tree increases.

\textbf{\textit{Stage 5}}:
Test cases are only stored in leaf nodes of the modified VP-tree.
During test case insertion, the number of test cases in a leaf node may reach the storage capacity, causing a promotion operation to be performed, which expands the storage capacity of the leaf node.
Each round of test case insertion only needs to be executed, at most, once.
After at most one promotion operation, a suitable leaf node (with spare capacity) will be identified and used to store  the current test case.

\textbf{\textit{Stage 6}}:
The executed NN for each candidate test case is identified using the $\varepsilon$-ary VP-tree, with a series of query thresholds $\sigma$  used to perform the searches.
Starting from the root node of the modified VP-tree, the distances to vantage test cases are compared, layer by layer, until the leaf node containing the NN is identified.

\subsection{Algorithm}

The original VP-tree structure is constructed according to the distance criterion.
The $top$-$down$ partitioning strategy makes the management of VP-tree updates complicated
---
the partitioning of upper-level nodes has an impact on the partitioning of lower-level nodes~\cite{Fu2000}.
In a worst-case scenario, reconstruction of the entire tree structure may be required.
The VP-tree update operation remains a problem requiring further study.
Fu \textit{et al.}~\cite{Fu2000} proposed a dynamic VP-tree structure, but this strategy involving upward backtracking and node-splitting/merging, which may incur significant time overheads.

In this paper, we
(1) introduce a \textit{modified VP-tree structure} in which to store the executed \FSCS\/ test cases; and
(2) propose an \textit{insertion approach} for this modified structure, the pseudo-code for which is in Algorithm~\ref{ALG:VPART}.

\begin{algorithm}[t]
    \footnotesize
    \caption{\VPP\/}
    \label{ALG:VPART}
    \algorithmicrequire~\\
    	The candidate set size, $k$;\\
	The number of dimensions in the SUT input domain, $d$;\\
	The capacity of a leaf node, $\lambda$;\\
	The partitioning parameter, $\varepsilon$;

    \algorithmicensure~ \\The VP-tree with executed test cases, $T$;\\
    \begin{algorithmic}[1]
        \STATE Set $T \leftarrow \{\}$; /* \textit{Initially, VP-tree for storing the executed test\\ cases is empty. }*/
        \STATE Set $C \leftarrow \{\}$; /* \textit{Initially, candidate test case set $C$, for storing $k$\\ candidates in each round, is empty.} */
        \STATE Set $min_{\textit{dist}}[k] \leftarrow \infty$; /* \textit{Initially, the minimum distance between\\ each candidate test case and $T$ is set to infinity.} */
        \STATE Randomly generate a test case $tc$ from the input domain
        \STATE Execute $tc$;
        \WHILE {(\textit{No termination condition is satisfied})}
            \STATE \textbf{InsertTCIntoVPtree}($tc$, $\lambda$, $\varepsilon$, $node$);
            \STATE Randomly generate $k$ candidate test cases $c_1, c_2, \cdots, c_k$\\ from the input domain, then set $C \leftarrow \{c_1, c_2, \cdots, c_k\}$;

                \FOR{(each candidate $c_j \in C$, where $j=1,2,\cdots,k$)}
                   \STATE Set $min_{\textit{dist}}[j] \leftarrow$ \textbf{GetMinDistFromVPtree}($c_j$, $node$);
                \ENDFOR
            \STATE Find $c_{\textit{best}}$ from $c_1, c_2, \cdots, c_k$ having the maximum\\ distance from its nearest test case;
            \STATE Set $tc \leftarrow c_{\textit{best}}$ and execute $tc$;
        \ENDWHILE
        \RETURN $T$;
    \end{algorithmic}
\end{algorithm}

\subsubsection{Modified \VPP\/ VP-tree structure
\label{SEC:algorithm}}

\begin{figure}[!b]
    \centering
    \graphicspath{{VPP-ART-Figures/}}
    \includegraphics[width=0.45\textwidth]{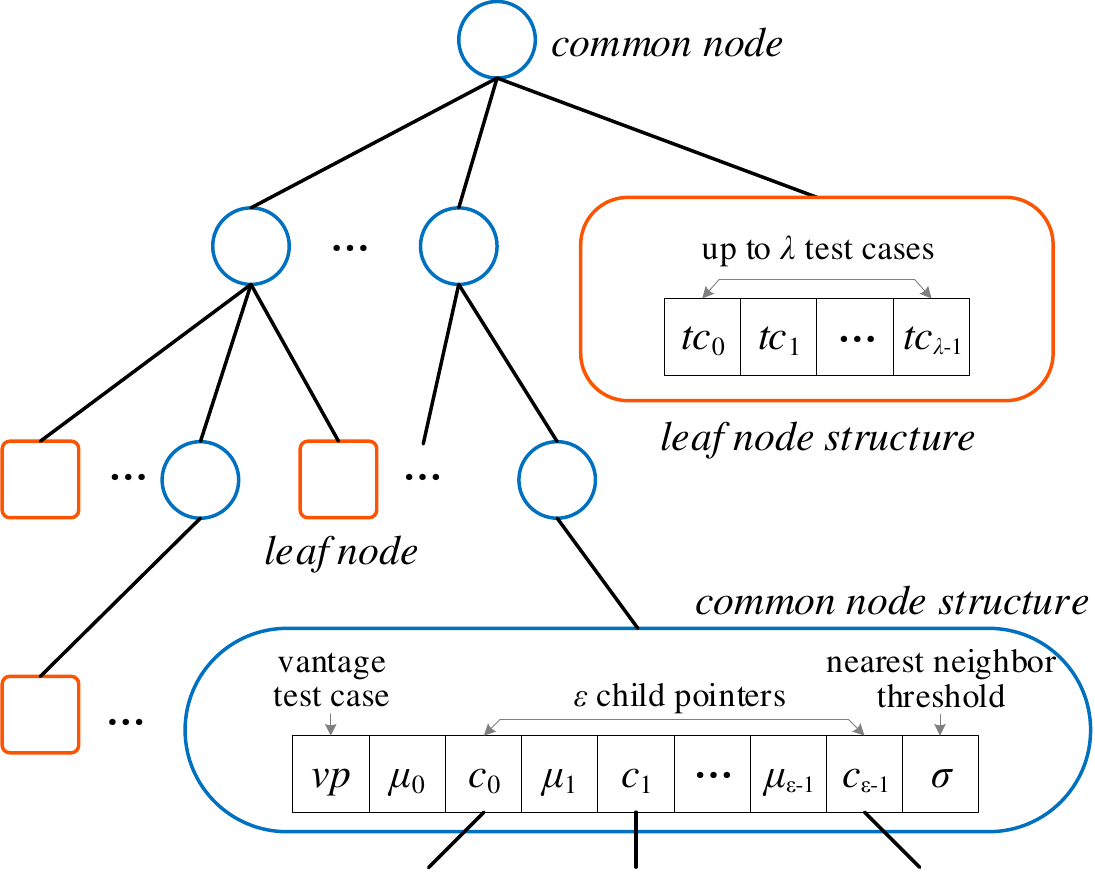}
    \caption{The modified \VPP\/ VP-tree structure.}
    \label{FIG:modified vp tree}
\end{figure}

Figure~\ref{FIG:modified vp tree} shows the modified VP-tree structure, where nodes are divided into \textit{leaf nodes} and \textit{common nodes}, denoted by circles and squares, respectively.
\textit{\textbf{Leaf nodes}} contain only test cases, the maximum number of which that can be stored in a single leaf node is denoted by $\lambda$.
\textit{\textbf{Common nodes}} contain no test cases, but do contain the vantage test case information, several boundary distance values with child pointers, and an \NN\/ threshold.

The \textbf{\textit{partitioning parameter}}, $\varepsilon$, specifies the number of subsets per division.
This determines the number of child pointers in each common node ($c_0,c_1,\cdots,c_{\varepsilon-1}$ in Figure~\ref{FIG:modified vp tree}).
The child pointers point to nodes in the next level of the tree.
The same $\varepsilon$ value is used for each partition, with this value being set by the tester.

\textbf{\textit{Vantage test cases}} are at the core of the structure, and their choice, in each level of the VP-tree, plays an important role in the performance of the algorithm.
An ideal vantage test case should have a uniform distribution of distances between it and other test cases.
This minimizes the number of test cases in concentric regions, thereby reducing the probability that all sub-trees must be explored.
However, finding the ideal vantage test case for a given test case set can require very heavy computational costs.
In practice, therefore, a test case is randomly selected from the $\lambda$ test cases stored in a leaf node, yielding an approximate (instead of optimal) vantage test case.
This selection method has been shown to be effective, experimentally.

\subsubsection{\VPP\/ Test Case Insertion Algorithm}

The VP-tree update operations, especially insertion, have a crucial role in the effectiveness of \VPP\/.
In the following, we focus on the \VPP\/ test case storage process, and propose \textit{insert} and node \textit{promotion} strategies to make this process more dynamic.

\begin{algorithm}[t]
    \footnotesize
    \caption{InsertTCIntoVPtree($tc$, $\lambda$, $\varepsilon$, $node$)}
    \label{ALG:VPARTInsert}
    \algorithmicrequire~ \\The executed test case need to insert into the tree, $tc$;\\ The capacity of a leaf node, $\lambda$;\\ The number of subsets partition, $\varepsilon$;\\A node of the VP-tree, $node$;

    \algorithmicensure~ \\Success flag, FALSE or TRUE;\\
    \begin{algorithmic}[1]
        \IF{(\textit{$node$ is $tc$}-QBN (a leaf node))}
            \IF{($|tc$-QBN$|<\lambda$)}
                \STATE Insert tc into $tc$-QBN;
                \STATE Set $tc$-BN $\leftarrow$ $tc$-QBN;
                \RETURN TRUE;
            \ELSIF{($|tc$-QBN$|=\lambda$) /* \textbf{\textit{Node promotion strategy}} */}
                \STATE Randomly select a test case from $tc$-QBN, as the vantage\\ test case, $vp$;
                \FOR {(each test case $p \in tc$-QBN $\cup \{tc\}$)}
                    \STATE Calculate $dist(p,vp)$;
                    \STATE Sort the test cases in ascending order according to the\\ distance values;
                \ENDFOR
                \STATE Calculate the boundary distance values $\mu_i$ according to\\ Equation~\ref{EQ:insert1};
                \STATE Calculate $\sigma$ of current node, using Equation~\ref{EQ:sigma2};
                \STATE Partition the input domain into $\varepsilon$ sub-domains;\\/* \textit{Each sub-domain contains approximately the same\\ number of test cases} */

                \STATE Reorganize all the test cases to the children nodes of this\\ new common node;
                \IF{(\textit{allocate the new }$tc$-BN)}
                    \RETURN TRUE;
                \ENDIF
            \ENDIF
        \ELSE
            \STATE Calculate $dist(vp,tc)$;
            \IF{($\mu_i < dist(vp,tc)\leq \mu_{i+1}$ where $0 \leq i \leq \varepsilon-2$)}
                \STATE \textbf{InsertTCIntoVPtree}($tc$, $\lambda$, $\varepsilon$, $node.i$-th child node);
            \ENDIF
            \IF{($dist(vp,tc) > \mu_{\varepsilon-1}$)}
                \STATE \textbf{InsertTCIntoVPtree}($tc$, $node.(\varepsilon -1)$-th child node);
            \ENDIF
        \ENDIF
    \end{algorithmic}
\end{algorithm}




\textbf{\textit{Insertion}}:
As testing progresses, it becomes necessary to insert newly-generated test cases into the modified VP-tree.
For an executed test case $e$, if the current node is a leaf node, then the leaf node is said to be the \textit{quasi-belonging-node} for $e$
(denoted $e$-QBN).
If the number of test cases in $e$-QBN is less than the maximum storage capacity ($\alpha < \lambda$), then $e$ is inserted into $e$-QBN, and  $e$-QBN is said to be the \textit{belonging-node} for $e$ (denoted $e$-BN).
If $\alpha=\lambda$, then it is necessary to promote (reconstruct)  $e$-QBN to find $e$-BN.
If the current node is a common node, then $dist(e,vp)$ is calculated (where $vp$ is the vantage test case in this node), and compared with the boundary values $\mu_i$ to determine which child pointer should be followed to find $e$-BN.


\textbf{\textit{Promotion}}:
This step is only performed when $\alpha=\lambda$, which means that it is necessary to transform $e$-QBN from a leaf node into a common node.
A test case $t_i$ $(0 < i \leq \lambda)$ is randomly selected from the test cases stored in $e$-QBN as the vantage test case for this node;
the remaining test cases and $e$ are reorganized into child nodes (new leaf nodes) of this new common node; and
$e$ is assigned to its new $e$-BN.
This promotion process is executed, at most, one time when inserting $e$:
After (at most) one promotion process, \VPP\/ will find an $e$-BN in which to store $e$.
The Insertion/Promotion pseudo-code is listed in Algorithm~\ref{ALG:VPARTInsert}.

\subsubsection{Approximate Nearest Neighbor Search in \VPP\/
\label{SEC:ANNS in VPTREES}}
In this section, the approach to calculate $\sigma$ is discussed, and the \VPP\/ approximate \NN\/ search process is explained.

\textbf{\textit{Nearest neighbor threshold}}:
As discussed in Section~\ref{SEC:NNS in VPTREES}, the threshold value $\sigma$ is the key to the search algorithm.
It can be used to reduce the \NN\/ search effort such that the \NN\/s of a query point are always very close to the query point itself
---
in other words, the smallest possible value of $\sigma$ can be used.
Chiueh \textit{et al.}~\cite{Chiueh1994} proposed a minimum possible value of $\sigma$, which we adopted and modified to match our partitioning strategy.
Upper ($u[i]$) and lower ($l[i]$) bounds exist for distances in the $i$-th subset $\mathcal{D}_i$ of the partitioned input domain:
$u[i]$ is the distance from the farthest test case in the current subset to its corresponding vantage test case, and
$l[i]$ is the distance from the nearest test case.
When traversing the VP-tree, it is not necessary to explore the nearest executed test cases of a candidate $c$ in $\mathcal{D}_i$ if
$dist(c,vp)>u[i]+\sigma$, or
$dist(c,vp)<l[i]-\sigma$.
This use of $\sigma$ guarantees that $dist(c, vp)$ will fall within at least one of the ranges in $\big[l[i] - \sigma, u[i] + \sigma\big]$:
This means that the search operation can be performed (and completed) in one sub-domain of the entire input domain.
For each common node, the upper and lower bounds of $\varepsilon$ partitioned subsets are computed as:
\begin{equation}
\label{EQ:sigma1}
    \sigma_i=\frac{l[i]-u[i-1]}{2},
\end{equation}
and the default value of $\sigma$ (for that node) is chosen to be the maximum value from $(\varepsilon - 1)$ values, as determined by:
\begin{equation}
\label{EQ:sigma2}
    \sigma=\max_{i=1}^{\varepsilon-1}\sigma_i.
\end{equation}

Based on the relationship among $\sigma$, $\mu$, and $dist(c,vp)$, \VPP\/ can start from the root node of the modified VP-tree and recursively search downwards until it finds the leaf node that contains the nearest executed test case of the current candidate.

Combined with the partitioning boundary values in Figure~\ref{FIG:vp-n-p1}, the last sub-domain ($\mathcal{D}_{\varepsilon-1}$) defined in Equation~\ref{EQ:NVPSEARCHCONDITION} is not covered by the search process.
The reason for this is that the possible values of $i$ can only be from $0$ to $\varepsilon-2$.
When $dist(q, vp) > \mu_{\varepsilon-1}$, the \NN\/ search process cannot be executed in $\mathcal{D}_{\varepsilon-1}$.
To solve this, $\mathcal{D}_{\varepsilon-1}$ is treated separately:
If $dist(q, vp) > \mu_{\varepsilon-1}$, then an exhaustive search in $\mathcal{D}_{\varepsilon-1}$ is conducted to obtain the query point $q$, and to get the \NN\/ distance in $\mathcal{D}_{\varepsilon-1}$.
This will mean that the \NN\/ of
$q$
---
which located at the boundary of $\mathcal{D}_{\varepsilon-2}$ and $\mathcal{D}_{\varepsilon-1}$
---
may be in the $\mathcal{D}_{\varepsilon-2}$, while the \VPP\/ only searches in the $\mathcal{D}_{\varepsilon-1}$, and finds an approximate \NN\/, rather than an exact \NN\/.

\begin{algorithm}[t]
    \footnotesize
    \caption{GetMinDistFromVPtree($c_j$, $node$)}
    \label{ALG:VPARTSearch}
    \algorithmicrequire~ \\A candidate test case, $c_j$;\\A node of the VP-tree, $node$;

    \algorithmicensure~ \\The minimum distance between $c_j$ and $E$, $min_{\textit{dist}}[j]$;\\
    \begin{algorithmic}[1]
        \STATE $min_{\textit{dist}}[j] \leftarrow \infty$;
        \IF{(\textit{$node$ is leaf node})}
            \FOR {(each test case $p$ in $node$)}
                \IF{($dist(c_j,p)<min_{\textit{dist}}[j]$)}
                    \STATE $min_{\textit{dist}}[j] \leftarrow dist(c_j,p)$;
                \ENDIF
            \ENDFOR
            \RETURN $min_{\textit{dist}}[j]$;
        \ELSE
            \STATE Calculate $dist(c_j,vp)$;
            \FOR{($0 \leq i \leq \varepsilon-2)$)}
                \IF{($\mu_i-\sigma < dist(vp,c_j)\leq \mu_{i+1}+\sigma$)}
                \STATE Get $\sigma$ of current common node;
                    \STATE \textbf{GetMinDistFromVPtree}($c_j$, $node.i$-th child node);
                \ENDIF
            \ENDFOR
            \IF{($dist(c_j,vp) > \mu_{\varepsilon-1}$)}
                \STATE \textbf{GetMinDistFromVPtree}($c_j$, $node.(\varepsilon -1)$-th child node);
            \ENDIF
        \ENDIF
    \end{algorithmic}
\end{algorithm}

\begin{figure*}[!t]
\graphicspath{{VPP-ART-Figures/}}
  \centering
  \subfigure[Insertion]
  {
    \includegraphics[width=0.235\textwidth]{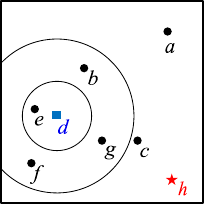}
    \label{insertion-domain}
  }
  \hspace{10mm}
  \subfigure[Promotion]
  {
    \includegraphics[width=0.235\textwidth]{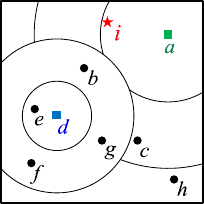}
    \label{promotion-domain}
  }
  \hspace{10mm}
  \subfigure[Search]
  {
    \includegraphics[width=0.235\textwidth]{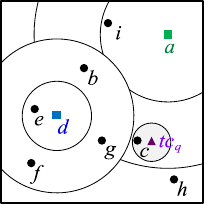}
    \label{search-domain}
  }
  \caption{Illustration of \VPP\/ insertion, promotion, and search strategies in a 2-dimensional input domain.}
  \label{FIG:ips-vpps-art-domain}
\end{figure*}

\begin{figure*}[!b]
\graphicspath{{VPP-ART-Figures/}}
  \centering
  \subfigure[Insert into a leaf node]
  {
    \includegraphics[width=0.22\textwidth]{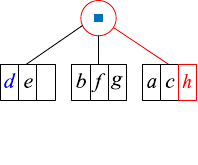}
    \label{insertion-tree}
  }
  \hspace{3mm}
  \subfigure[Promote a leaf node]
  {
    \includegraphics[width=0.29\textwidth]{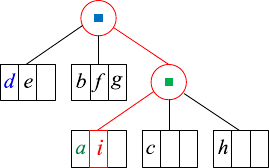}
    \label{promotion-tree}
  }
  \hspace{3mm}
  \subfigure[Search for the \NN\/ in the tree]
  {
    \includegraphics[width=0.29\textwidth]{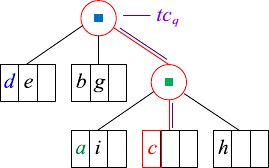}
    \label{search-tree}
  }
  \caption{Illustration of insertion, promotion, and searching strategies in the modified VP-tree (in a 2-dimensional input domain).
  These modified VP-tree structures correspond to Figure~\ref{FIG:ips-vpps-art-domain}.}
  \label{FIG:ips-vpps-art-tree}
\end{figure*}
\textbf{\textit{Approximate Nearest Neighbor Search}}:
The \NN\/ obtained by \VPP\/ could be an \textit{approximate \NN\/}, because this search process is likely to find an inaccurate neighbor in the leaf node that does not contain an \textit{exact \NN\/}, which can be explained as follows:
(1) Some specific sub-domains need special consideration, and the entire search may be carried out in these sub-domains, which may not contain the exact \NN\/s.
Such as the last sub-domains ($\mathcal{D}_{\varepsilon-1}$) after each round of partitioning.
(2) Because the boundary distance $\mu$ and the \NN\/ threshold $\sigma$ of each node are calculated according to the fixed number of test cases, these two values of each node will be affected when a new point is inserted.
To a certain extent, \VPP\/ deliberately ignores the impact of the insertion process on these two values.
Specifically, when a new point is inserted, \VPP\/ ignores the change of the two values of the QBN's upper-level nodes
---
it does not update their $\mu$ and $\sigma$ values.
This may cause the \NN\/ of the query point to no longer be in the original sub-domain, which reduces the accuracy of the search algorithm.

Although \VPP\/ adopts an approximate \NN\/ search, it still has  advantages in some cases:
(1) For an exact \NN\/, an exhaustive search will be executed
---
the distances between the executed test cases and the candidate will all be calculated, and the closest executed test case will be identified as the \NN\/.
When the number of test cases is very large, or the dimensionality is relatively high, the search efficiency decreases sharply.
However, \VPP\/ can obtain a better efficiency by using an approximate search:
\VPP\/ can identify an acceptable \NN\/ using fewer distance calculations when searching for an approximate \NN\/.
(2) An approximate \NN\/ search process, by not being limited to identifying the exact \NN\/, can improve the search efficiency at an acceptable cost in the accuracy.
When searching some sub-domains, the accuracy of some \NN\/s may be lost, but \VPP\/ may achieve similar, or even better, results than other exact \NN\/-based ART algorithms.
The difference between approximate \NN\/ and exact \NN\/ can be very small, enabling \VPP\/ to have comparable failure-detection effectiveness.
(3) Using the characteristic that test cases closer to the vantage point are more likely to be divided into the same sub-domain.
When the number of test cases increases, test cases with greater similarity will aggregate together.
Using the VP-tree structure,
\VPP\/, according to the distance relationship between candidate test cases and vantages points,
will search the possible sub-domains (leaf nodes in the VP-tree) of executed test cases with greater similarity to candidate test cases.

The pseudo-code for the search process is listed in Algorithm~\ref{ALG:VPARTSearch}.

\subsubsection{Examples of \VPP\/ operations}

Figure~\ref{FIG:ips-vpps-art-domain} shows examples of insertion, promotion, and searching with \VPP\/, in a $2$-dimensional input domain.
The corresponding VP-tree structures are shown in Figure~\ref{FIG:ips-vpps-art-tree},
where $\lambda=3$, $\varepsilon=3$.

\textbf{ \textit{Insertion}}:
Test case $h$ is a newly-generated test case for which no software failure is found and needs to be inserted into the VP-tree (Figure~\ref{insertion-domain}).
At this point, the third leaf node of the tree defined by the vantage test case $d$ still has space for the test case, so it is inserted directly (Figure~\ref{insertion-tree}).

\textbf{\textit{Promotion}}:
Test case $i$ is the newly-generated test case that has revealed no software failure.
It needs to be inserted into the VP-tree (Figure~\ref{promotion-domain}).
The third leaf node of the VP-tree defined by the vantage test case $d$ has no space to accommodate $i$, so the promotion operation is executed.
During promotion, $a$ is randomly selected as the vantage test case from $a$, $c$, and $h$).
The third sub-domain defined by the vantage test case $d$ is redistributed, and the four test cases are reorganized to three leaf nodes according to the distance values in $\mathcal{S}=\big\{dist(i,a), dist(c,a),dist(h,a)\big\}$.
Finally, test case $i$ is then inserted into the first node of $a$ (Figure~\ref{promotion-tree}).

\textbf{\textit{Search}}:
For queries with nearest executed test cases, a candidate test case $tc_q$ needs to find the \NN\/ in the input domain (Figure~\ref{search-domain}).
Starting from the root of the tree, the leaf nodes containing the \NN\/s are searched, layer by layer, based on the partitioning radius $\mu$, the query threshold $\sigma$, and the distance relations defined in Section~\ref{SEC:NNS in VPTREES}.
In this example, if the second leaf node defined by the vantage point $a$ contains the \NN\/, and this leaf node contains only $c$, then $c$ is considered the \NN\/ for $tc_q$ (Figure~\ref{search-tree}).

\section{Experimental Studies
\label{SEC:experimental studies}}

This section introduces the design and settings of the simulations and experiments that we conducted to evaluate \VPP\/.

\subsection{Research Questions
\label{SEC:RQ}}

The proposed \VPP\/ algorithm aims to reduce the time overheads of the original \FSCS\/ algorithm, thus, measurement and examination
of the test case generation time is necessary.
\VPP\/ is also expected to maintain the \FSCS\/ failure-detection effectiveness, which requires evaluation and verification in various scenarios.
The experimental studies were guided by the following research questions:

\begin{itemize}
  \item [\textbf{RQ1}]
  	\textit{How well does \VPP\/ perform, in terms of software failure-detection, compared with other ART algorithms?}
  \item [\textbf{RQ2}]
  	\textit{Compared with \FSCS\/ and \KDFC\/, to what extent can \VPP\/ reduce computational overheads?}
\end{itemize}

\subsection{Variables and Evaluation Metrics}

This section describes the independent and dependent variables in our research.
The evaluation metrics used to examine the effectiveness and efficiency of the different ART algorithms are also introduced.

\subsubsection{Independent Variable
\label{independent_v}}

The independent variable in the experimental study are the different ART algorithms used to generate test cases.
\VPP\/, the new algorithm proposed in this paper, is compared with \FSCS\/~\cite{Chen2004a} and \KDFC\/~\cite{Mao2019}.

\VPP\/ is an enhanced version of \FSCS\/, and we want to know the effects of using VPP on \FSCS\/.
The KD-tree is an efficient spatial indexing mechanism.
Mao \textit{et al.}~\cite{Mao2019} introduced KD-trees into ART, and proposed three \KDFC\/ algorithms:
\Naive\/; \SemiBal\/; and \LimBal\/.
\Naive\/ and \SemiBal\/ search for the exact \NN\/ of candidate test cases, and thus their generated test cases are the same as those generated by  \FSCS\/.
\LimBal\/ uses a limited backtracking method, identifying an {\em approximate} \NN\/, similar to \VPP\/.
The simulations and experiments sought to examine two things:
(1) the impact of the difference between exact and approximate \NN\/ searching on ART effectiveness and efficiency; and
(2) the differences in effectiveness and efficiency between the two approximate \NN\/-search-based ART algorithms, \VPP\/ and \LimBal\/.

\subsubsection{Dependent Variables}

The dependent variables in our studies are the evaluation metrics, for both effectiveness and efficiency.

\textbf{Effectiveness Metrics}
The \textit{F-measure}~\cite{Chen2004c} gives the number of test case executions before detecting the first failure in the SUT, and has been widely used in ART studies \cite{Huang2019}.
We also used the \textit{F-measure} as an evaluation metric in our study, with $F_{\textit{RT}}$ and $F_{\textit{ART}}$ denoting the \textit{F-measure} when conducting RT and ART, respectively.
Theoretically, $F_{\textit{RT}}$ equals to $1/\theta$ (where $\theta$ is the SUT failure rate).
The \textit{\fratio\/}~\cite{Chen2004c} denotes the ratio of $F_{\textit{ART}}$ to $F_{\textit{RT}}$, showing the improvement of ART over RT:
A lower \textit{\fratio\/} indicates better ART performance.

\textbf{Efficiency Metrics}
There are several measurements commonly taken when examining the efficiency of a testing methodology, including:
\textit{generation time};
\textit{execution time}; and
\textit{F-time}~\cite{Huang2019}~\cite{Chen2006}.
The generation time refers to the cost of generating $\textbf{\textit{k}}$ test cases;
the execution time refers to the time cost of executing the SUT with $\textbf{\textit{k}}$ test cases; and
the \textit{F-time}~\cite{Chen2006} is defined as the entire time cost for finding the first failure in the SUT.
Generally speaking, the test case generation time has a great impact on the entire test cost.
For the simulation studies, we recorded the average generation time to generate a certain number of test cases.

\subsection{Experimental Environment}

The simulations and experiments were conducted using a 16-GB RAM laptop PC with an i7 CPU, running at 2.20 GHz, running under the 64-bit Windows 10 operating system.
All the algorithms under study were implemented in Java with JDK 1.8.
The IDEs used were Eclipse (Version 4.15.0) and Microsoft Visual Studio 2019.

\subsection{Data Collection and Statistical Analysis}

The \textit{F-measure} and \textit{F-time} values were calculated by running  each of the ART algorithms until a failure was detected.
In the simulations, a failure was considered to be detected whenever a test case was generated from within a simulated failure region.
In the experiments, the actual output was compared with the expected output (the test oracle~\cite{Weyuker1982},~\cite{BarrHMSY2015},~\cite{Jahangirova2017}):
A difference indicated a failure being detected.
To minimize the error caused by randomness, and to provide confidence in the comparison, each experiment was run 3000 times, with the average being recorded.

When analyzing the experimental data, the \textit{p-value} (probability value) and \textit{\es\/} for the different ART algorithms were calculated \cite{leech2002call},~\cite{grissom2005effect},~\cite{Arcuri2014}.
These can describe any significant differences or improvements between the two compared methods \cite{Menzies2009}.
The unpaired two-tailed Mann-Whitney-Wilcoxon test~\cite{Arcuri2014} was used to verify whether or not there was a significant differences among the investigated ART algorithms.
A \textit{p-value} less than $0.05$ indicates a significant difference between the two algorithms~\cite{cowles1982origins}.
The \textit{\es\/}~\cite{Arcuri2014} shows the possibility that one method is better than another:
We used the non-parametric Vargha and Delaney \es\/~\cite{Vargha2000}.
For two methods, \textit{A} and \textit{B},
an \textit{\es\/} between \textit{A} and \textit{B} of $0.50$ means that \textit{A} and \textit{B} are equivalent;
if the value is greater than $0.50$, \textit{A} is better than method \textit{B}; and
if the value is less than $0.50$, \textit{B} is better than method \textit{A}.

\begin{table*}[!t]
\scriptsize
\centering
\setlength{\tabcolsep}{0.45mm}
    \caption{22 subject programs used in experiments}
    \label{Table:22SUT}
    \begin{tabular}{@{}cccccccc@{}}
    \toprule
        \multirow{2}{*}{\begin{tabular}[c]{@{}c@{}}\\[-3pt]\textbf{No.}\end{tabular}}
        & \multirow{2}{*}{\begin{tabular}[c]{@{}c@{}}\\[-3pt]\textbf{Program}\end{tabular}}& \multirow{2}{*}{\begin{tabular}[c]{@{}c@{}}\\[-5pt]\textbf{Dimension}\\ $(d)$\end{tabular}} & \multicolumn{2}{c}{\textbf{Input domain}}  & \multirow{2}{*}{\begin{tabular}[c]{@{}c@{}}\\[-5pt]\textbf{Size}\\ \textbf{(LOC)}\end{tabular}} &
        \multirow{2}{*}{\begin{tabular}[c]{@{}c@{}}\\[-3pt]\textbf{Mutant Operators}\end{tabular}} &
        \multirow{2}{*}{\begin{tabular}[c]{@{}c@{}}\\[-5pt]\textbf{Total}\\\textbf{Faults}\end{tabular}} \\ \cmidrule{4-5}
         &  &  & \textbf{From} & \textbf{To} &  &  & \\
    \midrule

       1 & \texttt{airy} & 1 & -5000 & 5000 & 43 & CR & 1\\
        2 & \texttt{bessj0} & 1 & -300000 & 300000 & 28 & AOR, ROR, SVR, CR & 5\\
        3 & \texttt{erfcc} & 1 & -30000 & 30000 & 14 & AOR, ROR, SVR, CR & 4\\
        4 & \texttt{probks} & 1 & -50000 & 50000 & 22 & AOR, ROR, SVR, CR & 4\\
        5 & \texttt{tanh} & 1 & -500 & 500 & 18 & AOR, ROR, SVR, CR & 4\\
        6 & \texttt{bessj} & 2 & (2, -1000) & (300, 15000) & 99 & AOR, ROR, CR & 4\\
        7 & \texttt{gammq} & 2 & (0, 0) & (1700, 40) & 106 & ROR, CR & 4\\
        8 & \texttt{sncndn} & 2 & (-5000, -5000) & (5000, 5000) & 64 & ROR, CR & 5\\
        9 & \texttt{golden} & 3 & (-100, -100, -100) & (60, 60, 60) & 80 & ROR, SVR, CR & 5\\
        10 & \texttt{plgndr} & 3 & (10, 0, 0) & (500, 11, 1) & 36 & AOR, ROR, CR & 5\\
        11 & \texttt{cel} & 4 & (0.001, 0.001, 0.001, 0.001) & (1, 300, 10000, 1000) & 49 & AOR, ROR, CR & 3\\
        12 & \texttt{el2} & 4 & (0, 0, 0, 0) & (250, 250, 250, 250) & 78 & AOR, ROR, SVR, CR & 9\\
        13 & \texttt{calDay} & 5 & (1, 1, 1, 1, 1800) & (12, 31, 12, 31, 2200) & 37 & SDL & 1\\
        14 & \texttt{complex} & 6 & (-20, -20, -20, -20, -20, -20) & (20, 20, 20, 20, 20, 20) & 68 & SVR & 1\\
        15 & \texttt{pntLinePos} & 6 & (-25, -25, -25, -25, -25, -25) & (25, 25, 25, 25, 25, 25) & 23 & CR & 1\\
        16 & \texttt{triangle} & 6 & (-25, -25, -25, -25, -25, -25) & (25, 25, 25, 25, 25, 25) & 21 & CR & 1\\
        17 & \texttt{line} & 8 & (-10, -10, -10, -10, -10, -10, -10, -10) & (10, 10, 10, 10, 10, 10, 10, 10) & 86 & ROR & 1\\
        18 & \texttt{pntTrianglePos} & 8 & (-10, -10, -10, -10, -10, -10, -10, -10) & (10, 10, 10, 10, 10, 10, 10, 10) & 68 & CR & 1\\
        19 & \texttt{twoLinesPos} & 8 & (-15, -15, -15, -15, -15, -15, -15, -15) & (15, 15, 15, 15, 15, 15, 15, 15) & 28 & CR & 1\\
        20 & \texttt{nearestDistance} & 10 & (1, 1, 1, 1, 1, 1, 1, 1, 1, 1) & (15, 15, 15, 15, 15, 15, 15, 15, 15, 15) & 26 & CR & 1\\
        21 & \texttt{calGCD} & 10 & (1, 1, 1, 1, 1, 1, 1, 1, 1, 1) & (1000, 1000, 1000, 1000, 1000, 1000, 1000, 1000, 1000, 1000) & 24 & AOR & 1\\
        22 & \texttt{select} & 11 & (1, 1, 1, 1, 1, 1, 1, 1, 1, 1, 1) & (10, 100, 100, 100, 100, 100, 100, 100, 100, 100, 100) & 117 & RSR, CR & 2\\
     \bottomrule

    \end{tabular}
\end{table*}

\subsection{Simulations and Experiments}

The original \FSCS\/ and \KDFC\/ algorithms were compared with our proposed algorithm, \VPP\/, through a series of simulations and experiments~\cite{Huang2019}.
The simulations involved simulated software faults, while the experiments used real-life subject programs altered through mutation analysis techniques~\cite{Jia2010}.

\subsubsection{Simulations Design}

The simulations used a $d$-dimensional hypercube as the program input domain ($\mathcal{D}$).
$\mathcal{D}$ was set as $\big\{(x_1,x_2,\cdots,x_d)|0.0$ $\leq$ $x_1,x_2,\cdots,x_d <$ $1.0\big\}$, with the dimensionality, $d$,  set as $1$, $2$, $3$, $4$, $5$, $8$ and $10$.

To address RQ1, the simulated failure regions were randomly placed in the input domain $\mathcal{D}$.
Once a program has been written, the failure regions are fixed, but their locations are unknown to developers and testers (before testing).
Failure regions have geometric shape (described by the \textit{failure patterns}) and size (described by the \textit{failure rate}), and distribution~\cite{Huang2019}.
In the simulations, the failure rate and pattern were set in advance, allowing the failure regions to be located randomly in $\mathcal{D}$.
As described in Section~\ref{failure rate and failure patterns}, failure patterns have often been categorized into three main types:
\textit{strip}; \textit{block}; and \textit{point}.
The simulations included all three failure patterns types.
Block patterns used a randomly-located, single-solid shape with equal lengths of side
---
a square in $2$-dimensions, cube in $3$-dimensions, etc.
The strip patterns were each constructed using two points on adjacent boundaries that were connected with a width/volume to yield the desired size.
According to the predetermined failure rate $\theta$, the point patterns used $25$ randomly-located regions.
The simulations used seven different $\theta$ settings:
$0.01$, $0.005$, $0.002$, $0.001$, $0.0005$, $0.0002$ and $0.0001$.

To address RQ2, the simulations recorded the test case generation times for both \FSCS\/ and \VPP\/.
A total of $20$,$000$ test cases were generated by each algorithm in the $d$-dimensional input domain, and the generation times recorded at intervals of $500$.

\subsubsection{Experiments Design}

Although simulations can simulate the performance of the algorithms in different scenarios, the failure types may not be representative of real-life situations for example, in reality, the failure types can be categorized into \textit{regular} and \textit{irregular} types~\cite{Chen2006b}~\cite{chen2006c}).
In addition to the simulations, we conducted experiments using 22 real-life programs with faults seeded in using mutation operators \cite{Jia2010}.
Table~\ref{Table:22SUT} presents the detailed information of these programs.
12 of the programs come from \textit{Numerical Recipes}~\cite{Press1986} and ACM's \textit{Collected Algorithms}~\cite{ACM1980}, and have been widely studied in ART research \cite{Chen2004a,Chen2007,Chan2006a,Huang2015}.
Programs \texttt{calDay}, \texttt{complex}, and \texttt{line} are from Ferrer \textit{et al.}~\cite{Ferrer2012}.
The \texttt{pntLinePos}, \texttt{pntTrianglePos} and \texttt{twoLinesPos} programs describe the positional relationships between a point and a line, a point and a triangle, and between two lines, respectively~\cite{Liang2017}.
The \texttt{triangle} program classifies a triangle into one of three types (acute, right and obtuse)~\cite{Liang2017}.
The \texttt{nearestDistance} program uses five points to realize the nearest point pair function.
The \texttt{calGCD} program calculates the greatest common divisor of $10$ integers, and \texttt{select} returns the $i$-th largest element from an unordered array~\cite{May2007}.

All subject programs were implemented in Java or C++, and had previously been used in the \KDFC\/ experiments \cite{Mao2019}.
Six mutation operators were used to generate mutants of the original subject programs~\cite{Jia2010}:
(1) arithmetic operator replacement (AOR);
(2) relational operator replacement (ROR);
(3) scalar variable replacement (SVR);
(4) constant replacement (CR);
(5) statement deletion (SDL); and
(6) return statement replacement (RSR).
Five algorithms were applied in the experiments:
the original \FSCS\/;
the three \KDFC\/ algorithms; and
our proposed \VPP\/.

\begin{table*}
\scriptsize
\centering
\setlength{\tabcolsep}{3.2mm}
 \caption{\fratio\/ values of \VPP\/ with different $\langle \varepsilon,\lambda \rangle$ parameter pair values}
 \label{TAB:pre}
    \begin{tabular}{@{}cccccccccccc@{}}
   \toprule
    \multirow{3}{*}{\begin{tabular}[c]{@{}c@{}}\\[-3pt]\textbf{Partitioning}\\ \textbf{Number} \\ $(\varepsilon)$\end{tabular}} &  \multirow{3}{*}{\begin{tabular}[c]{@{}c@{}}\\[1pt]\textbf{Dimension}\\ $(d)$\end{tabular}} & \multicolumn{10}{c}{\textbf{\fratio\/}} \\\cmidrule{3-12}
    &  & \multirow{2}{*}{\begin{tabular}[c]{@{}c@{}}\\[-3pt]{\textit{\FSCS\/}}\end{tabular}} & \multicolumn{9}{c}{\textit{\VPP\/}}\\ \cmidrule{4-12}
    &  & & \textbf{$\lambda = 10$} & $\lambda = 15$ & $\lambda = 20$ & $\lambda = 25$ & $\lambda = 30$ & $\lambda = 35$ & $\lambda = 40$ & $\lambda = 45$ & $\lambda = 50$ \\ \midrule
        \multirow{7}{*}{\textbf{$\varepsilon=2$}}
        &$d=1$&0.5564&0.5706&0.5652&0.5523&0.5611&0.5675&0.5747&0.5631&0.5634&0.5702\\
        &$d=2$&0.6391&0.6790&0.6584&0.6752&0.6712&0.6645&0.6547&0.6641&0.6516&0.6440\\
        &$d=3$&0.7549&0.8056&0.7890&0.7809&0.7787&0.7964&0.8053&0.7957&0.8096&0.7970\\
        &$d=4$&0.9033&0.9611&0.9329&0.9243&0.9391&0.9506&0.9539&0.9280&0.9368&0.9462\\
        &$d=5$&1.0462&1.0824&1.0945&1.0984&1.0852&1.0978&1.1147&1.1260&1.1200&1.1069\\
        &$d=8$&1.8607&1.7050&1.6886&1.7668&1.7673&1.8200&1.7956&1.7711&1.8460&1.8289\\
        &$d=10$&2.6138&2.4504&2.4896&2.6366&2.6495&2.6355&2.6268&2.7046&2.7357&2.7719\\\midrule
        \multirow{7}{*}{\textbf{$\varepsilon=3$}}
        &$d=1$&0.5564&0.5555&0.5605&0.5686&0.5584&0.5557&0.5573&0.5599&0.5514&0.5582\\
        &$d=2$&0.6391&0.6510&0.6601&0.6447&0.6559&0.6377&0.6661&0.6459&0.6725&0.6523\\
        &$d=3$&0.7549&0.7553&0.7534&0.8083&0.7832&0.7753&0.7663&0.8063&0.8037&0.7940\\
        &$d=4$&0.9033&0.9390&0.9448&0.9510&0.9385&0.9385&0.9345&0.9355&0.9484&0.9421\\
        &$d=5$&1.0462&1.0605&1.0890&1.0709&1.0698&1.1085&1.0876&1.0997&1.1178&1.1053\\
        &$d=8$&1.8607&1.6631&1.8007&1.7354&1.7543&1.8558&1.8406&1.8833&1.8870&1.8493\\
        &$d=10$&2.6138&2.2916&2.4821&2.4765&2.5670&2.6265&2.6319&2.7906&2.7058&2.7080\\\midrule
        \multirow{7}{*}{\textbf{$\varepsilon=4$}}
        &$d=1$&0.5564&0.5577&0.5636&0.5621&0.5613&0.5529&0.5726&0.5567&0.5607&0.5588\\
        &$d=2$&0.6391&0.6570&0.6587&0.6590&0.6739&0.6534&0.6296&0.6340&0.6448&0.6529\\
        &$d=3$&0.7549&0.7617&0.7680&0.7571&0.7604&0.7659&0.7519&0.7900&0.7709&0.7754\\
        &$d=4$&0.9033&0.9144&0.9159&0.9272&0.9086&0.9418&0.9572&0.9427&0.9418&0.9393\\
        &$d=5$&1.0462&1.0896&1.0891&1.0734&1.0727&1.1486&1.0854&1.0964&1.0948&1.1133\\
        &$d=8$&1.8607&1.6900&1.6744&1.8339&1.8043&1.8048&1.8067&1.8578&1.9174&1.9717\\
        &$d=10$&2.6138&2.3144&2.3562&2.6492&2.6325&2.8174&2.6989&2.8207&2.8412&2.9358\\\midrule
        \multirow{7}{*}{\textbf{$\varepsilon=5$}}
        &$d=1$&0.5564&0.5573&0.5542&0.5507&0.5822&0.5566&0.5639&0.5558&0.5585&0.5496\\
        &$d=2$&0.6391&0.6467&0.6594&0.6416&0.6360&0.6536&0.6427&0.6412&0.6433&0.6447\\
        &$d=3$&0.7549&0.7912&0.7747&0.7547&0.7745&0.7674&0.7518&0.7610&0.7671&0.7726\\
        &$d=4$&0.9033&0.9276&0.9167&0.9024&0.9194&0.9582&0.9239&0.9559&0.9592&0.9391\\
        &$d=5$&1.0462&1.0932&1.0849&1.0923&1.1254&1.1063&1.1072&1.1258&1.1370&1.1280\\
        &$d=8$&1.8607&1.7753&1.7982&1.8214&1.8652&1.8581&2.0066&1.9264&1.9821&1.9547\\
        &$d=10$&2.6138&2.4743&2.5811&2.6273&2.7442&2.7095&2.8409&2.9523&2.8931&2.9514\\
    \bottomrule
    \end{tabular}
\end{table*}

\section{Experimental Results
\label{SEC:results}}

This section presents the results from the simulations and experiments.
The differences in effectiveness and efficiency among \FSCS\/, \KDFC\/, and \VPP\/ are discussed, and answers are provided to the two research questions from Section~\ref{SEC:RQ}.
In the tables in this section, a \textit{blue bold} number indicates the minimum value of \textit{\fratio}, \textit{F-measure} or \textit{F-time} across the several ART algorithms; and
a \textit{red bold} number indicates that the \textit{p-value} of the comparison between\VPP\/ and the corresponding ART algorithm is less than $0.05$, indicating significance.

\subsection{\VPP\/ Parameter Settings
\label{Determining-Parameters-of-VPPS-ART}}

As explained, the \VPP\/ performance is strongly impacted by two parameters:
the partitioning parameter, $\varepsilon$; and
the maximum test case capacity of a leaf node, $\lambda$.
These two values play important roles in the partitioning of the input domain, and also limit the amount of executed test cases in each sub-domain, which will affect the accuracy of the \NN\/ returned by \VPP\/.
When \VPP\/ performs the \NN\/ search in one sub-domain, the approximate \NN\/ is identified, as explained in Section~\ref{SEC:ANNS in VPTREES}.
When there are more executed test cases in this sub-domain,
---
which is directly influenced by the static values of $\varepsilon$ and $\lambda$,
---
then \VPP\/ will have a greater probability of finding a more accurate \NN\/ (of course, it may also be an approximate \NN\/ due to the construction of the modified VP-tree).
If the approximate \NN\/ returned by \VPP\/ is similar to the exact \NN\/, the failure-detection effectiveness of \VPP\/ and \FSCS\/ will be comparable.
Therefore, this section focuses on the influence of different $\langle \varepsilon,\lambda \rangle$ parameter pair values on \VPP\/.
The specific parameter settings for the simulations were as follows:
\begin{itemize}
    \item Dimension: $d=1,2,3,4,5,8,10$;
    \item Failure rate: $\theta=0.0005$;
    \item Partitioning parameter, $\varepsilon=2,3,4,5$;
    \item Maximum test case capacity of a leaf node, $\lambda=10,15,20,25,30,35,40,45,50$.
\end{itemize}

Table~\ref{TAB:pre} presents the \textit{\fratio\/} simulation results of \VPP\/ for the different parameter values.
Based on the data in the table, some observations can be summarized as follows:

(1) As the maximum test case capacity of leaf nodes ($\lambda$) increases,
the \VPP\/ \textit{\fratio\/} differences (when $1 \leq d \leq 5$) are not significant,
but are, on the whole, slightly higher than \FSCS\/;
For $d = 8,10$, the \textit{\fratio\/} values increase gradually, and are lower than \FSCS\/ when $\lambda$ is small.
This shows that changes in $\lambda$ have little impact on the failure-detection effectiveness of \VPP\/ in low dimensional input domains, but can have a great impact in high dimensions.
Therefore, a small $\lambda$ value can effectively ensure \VPP\/ failure-detection effectiveness in low dimensions, and improve \FSCS\/ performance in high dimensions.

(2) When the partition parameter ($\varepsilon$) increases above $2$, the \VPP\/ \textit{\fratio\/} values  are not significantly different when $1\leq d \leq 5$.
When $d = 8,10$, the \textit{\fratio\/} values show a trend of first decreasing, and then increasing;
an inflection point appears when $\varepsilon = 3$.
Similar to $\lambda$, changes in $\varepsilon$ have little impact on \VPP\/ failure-detection effectiveness in low dimensions, but show a certain change trend in the high dimensions.
Therefore, a smaller $\varepsilon$ value can enhance the \VPP\/ performance.

Based on the above, the parameter pair $\langle \varepsilon,\lambda \rangle$ were assigned $\langle 3,10 \rangle$ for the simulations and experiments.

\subsection{Comparisons of Failure-Detection Effectiveness}

This section reports on the effectiveness comparisons between \VPP\/ and \FSCS\/, and between \VPP\/ and the three kinds of \KDFC\/.
The results and main findings address RQ1, as follows.

\subsubsection{Answer to RQ1 - Part 1: Results of Simulations}

\begin{table*}[t]
\scriptsize
\centering
\setlength{\tabcolsep}{1.8mm}
 \caption{\fratio\/ values  and statistical analysis comparisons among \VPP\/ and other ART algorithms for \textit{block patterns}}
 \label{TAB:FR-block}
    \begin{tabular}{@{}ccccccccccccccc@{}}
    \toprule
        \multirow{3}{*}{\begin{tabular}[c]{@{}c@{}}\\[1pt]\textbf{Dimension}\\ $(d)$\end{tabular}}
        & \multirow{3}{*}{\begin{tabular}[c]{@{}c@{}}\\[1pt]\textbf{Failure Rate}\\ $(\theta)$\end{tabular}} & \multicolumn{5}{c}{\textbf{\fratio\/}}  & \multicolumn{8}{c}{\textbf{Statistical Analysis}}\\\cmidrule{3-15}
         &  & \multirow{2}{*}{\begin{tabular}[r]{@{}r@{}}\\[-5pt]\textit{VPP-}\\\textit{ART}\end{tabular}} & \multirow{2}{*}{\begin{tabular}[r]{@{}r@{}}\\[-5pt]\textit{FSCS-}\\\textit{ART}\end{tabular}} & \multirow{2}{*}{\begin{tabular}[r]{@{}r@{}}\\[-5pt]\textit{Naive-}\\\textit{KDFC}\end{tabular}} & \multirow{2}{*}{\begin{tabular}[r]{@{}r@{}}\\[-5pt]\textit{SemiBal-}\\\textit{KDFC}\end{tabular}} & \multirow{2}{*}{\begin{tabular}[r]{@{}r@{}}\\[-5pt]\textit{LimBal-}\\\textit{KDFC}\end{tabular}} & \multicolumn{2}{r}{\textit{vs.} \textit{\FSCS\/}} & \multicolumn{2}{r}{\textit{vs.} \textit{\Naive\/}} & \multicolumn{2}{r}{\textit{vs.} \textit{\SemiBal\/}} & \multicolumn{2}{r}{\textit{vs.} \textit{\LimBal\/}}\\\cmidrule{8-15}
         &  &  &  &  &  &  & \textit{p-value} & \textit{\es\/} & \textit{p-value} & \textit{\es\/} & \textit{p-value} & \textit{\es\/} & \textit{p-value} & \textit{\es\/}\\
         \midrule
        \multirow{7}{*}{$d=1$}
        & 0.01 & \textcolor{blue}{\textbf{0.5634}} & 0.5729 & 0.5664 & 0.5714 & 0.5658 & 0.3679 & 0.5067 & 0.7899 & 0.5020 & 0.4101 & 0.5061 & 0.5294 & 0.5047\\
        & 0.005 & 0.5666 & 0.5633 & 0.5670 & 0.5696 & \textcolor{blue}{\textbf{0.5619}} & 0.8994 & 0.4991 & 0.7826 & 0.5021 & 0.4768 & 0.5053 & 0.8362 & 0.4985\\
        & 0.002 & 0.5639 & 0.5683 & 0.5665 & 0.5723 & \textcolor{blue}{\textbf{0.5605}} & 0.5704 & 0.5042 & 0.9981 & 0.5000 & 0.5034 & 0.5050 & 0.7817 & 0.4979\\
        & 0.001 & 0.5634 & 0.5720 & \textcolor{blue}{\textbf{0.5549}} & 0.5690 & 0.5617 & 0.6041 & 0.5039 & 0.1890 & 0.4902 & 0.5659 & 0.5043 & 0.7915 & 0.4980\\
        & 0.0005 & \textcolor{blue}{\textbf{0.5555}} & 0.5564 & 0.5629 & 0.5556 & 0.5619 & 0.5919 & 0.5040 & 0.4510 & 0.5056 & 0.7454 & 0.5024 & 0.4847 & 0.5052\\
        & 0.0002 & \textcolor{blue}{\textbf{0.5520}} & 0.5700 & 0.5658 & 0.5662 & 0.5614 & 0.0602 & 0.5140 & 0.1353 & 0.5111 & 0.2506 & 0.5086 & 0.2506 & 0.5086\\
        & 0.0001 & 0.5766 & 0.5765 & \textcolor{blue}{\textbf{0.5527}} & 0.5545 & 0.5569 & 0.8939 & 0.4990 & \textcolor{red}{\textbf{0.0327}} & 0.4841 & 0.0559 & 0.4857 & 0.1115 & 0.4881\\\midrule
        \multirow{7}{*}{$d=2$}
        & 0.01 & \textcolor{blue}{\textbf{0.6820}} & 0.6911 & 0.6822 & 0.6904 & 0.6953 & 0.1867 & 0.5098 & 0.4907 & 0.5051 & 0.1303 & 0.5113 & 0.0805 & 0.5130\\
        & 0.005 & 0.6750 & 0.6613 & \textcolor{blue}{\textbf{0.6561}} & 0.6635 & 0.6671 & 0.7798 & 0.4979 & 0.3619 & 0.4932 & 0.6590 & 0.4967 & 0.8597 & 0.4987\\
        & 0.002 & 0.6712 & \textcolor{blue}{\textbf{0.6536}} & 0.6574 & 0.6633 & 0.6561 & 0.5933 & 0.4960 & 0.7498 & 0.4976 & 0.5793 & 0.5041 & 0.5865 & 0.4959\\
        & 0.001 & 0.6742 & 0.6573 & \textcolor{blue}{\textbf{0.6449}} & 0.6557 & 0.6595 & 0.5260 & 0.4953 & 0.1795 & 0.4900 & 0.6460 & 0.4966 & 0.6193 & 0.4963\\
        & 0.0005 & 0.6510 & \textcolor{blue}{\textbf{0.6391}} & 0.6525 & 0.6484 & 0.6492 & 0.9636 & 0.4997 & 0.2177 & 0.5092 & 0.4341 & 0.5058 & 0.3357 & 0.5072\\
        & 0.0002 & 0.6489 & \textcolor{blue}{\textbf{0.6268}} & 0.6409 & 0.6414 & 0.6388 & 0.2310 & 0.4911 & 0.9428 & 0.4995 & 0.8521 & 0.4986 & 0.5000 & 0.4950\\
        & 0.0001 & \textcolor{blue}{\textbf{0.6244}} & 0.6248 & 0.6531 & 0.6389 & 0.6313 & 0.5895 & 0.5040 & \textcolor{red}{\textbf{0.0030}} & 0.5222 & 0.0847 & 0.5129 & 0.4206 & 0.5060\\\midrule
        \multirow{7}{*}{$d=3$}
        & 0.01 & 0.8840 & 0.8641 & 0.8431 & 0.8504 & \textcolor{blue}{\textbf{0.8391}} & 0.6879 & 0.5030 & 0.2535 & 0.4915 & 0.7570 & 0.4977 & 0.2964 & 0.4922\\
        & 0.005 & 0.8391 & 0.8314 & \textcolor{blue}{\textbf{0.8176}} & 0.8195 & 0.8177 & 0.3398 & 0.5071 & 0.9302 & 0.5007 & 0.6204 & 0.5037 & 0.5875 & 0.4960\\
        & 0.002 & 0.8214 & 0.7847 & \textcolor{blue}{\textbf{0.7778}} & 0.7948 & 0.8052 & 0.2653 & 0.4917 & 0.1122 & 0.4882 & 0.5126 & 0.4951 & 0.3310 & 0.4928\\
        & 0.001 & 0.8189 & 0.7735 & \textcolor{blue}{\textbf{0.7720}} & 0.7735 & 0.7772 & 0.1464 & 0.4892 & 0.0981 & 0.4877 & 0.1562 & 0.4894 & 0.0703 & 0.4865\\
        & 0.0005 & 0.7553 & 0.7549 & \textcolor{blue}{\textbf{0.7504}} & 0.7618 & 0.7615 & 0.2115 & 0.5093 & 0.3457 & 0.5070 & 0.2778 & 0.5081 & 0.2614 & 0.5084\\
        & 0.0002 & 0.7881 & 0.7499 & 0.7603 & \textcolor{blue}{\textbf{0.7441}} & 0.7464 & 0.5469 & 0.4955 & 0.5871 & 0.4960 & 0.1881 & 0.4902 & 0.2464 & 0.4914\\
        & 0.0001 & 0.7688 & 0.7358 & \textcolor{blue}{\textbf{0.7252}} & 0.7518 & 0.7387 & 0.6934 & 0.4971 & 0.1312 & 0.4887 & 0.7021 & 0.4971 & 0.2108 & 0.4907\\\midrule
        \multirow{7}{*}{$d=4$}
        & 0.01 & 1.0886 & 1.0786 & 1.0739 & 1.0711 & \textcolor{blue}{\textbf{1.0666}} & 0.9147 & 0.5008 & 0.6676 & 0.5032 & 0.9517 & 0.4995 & 0.9651 & 0.5003\\
        & 0.005 & 1.0523 & 1.0272 & 1.0350 & \textcolor{blue}{\textbf{1.0200}} & 1.0202 & 0.5394 & 0.4954 & 0.8604 & 0.5013 & 0.4641 & 0.4945 & 0.5352 & 0.4954\\
        & 0.002 & 0.9948 & 0.9606 & \textcolor{blue}{\textbf{0.9497}} & 0.9711 & 0.9754 & 0.8243 & 0.4983 & 0.5681 & 0.4957 & 0.9008 & 0.5009 & 0.9820 & 0.4998\\
        & 0.001 & 0.9398 & 0.9155 & \textcolor{blue}{\textbf{0.9122}} & 0.9190 & 0.9366 & 0.7930 & 0.5020 & 0.4660 & 0.4946 & 0.8987 & 0.4991 & 0.8987 & 0.5009\\
        & 0.0005 & 0.9390 & 0.9033 & 0.8908 & \textcolor{blue}{\textbf{0.8904}} & 0.9067 & 0.1379 & 0.4889 & 0.1067 & 0.4880 & \textcolor{red}{\textbf{0.0464}} & 0.4852 & 0.2141 & 0.4907\\
        & 0.0002 & 0.8965 & 0.8522 & \textcolor{blue}{\textbf{0.8494}} & 0.8651 & 0.8708 & 0.0963 & 0.4876 & 0.2810 & 0.4920 & 0.1513 & 0.4893 & 0.6204 & 0.4963\\
        & 0.0001 & 0.8857 & 0.8357 & \textcolor{blue}{\textbf{0.8234}} & 0.8687 & 0.8491 & 0.3925 & 0.4936 & 0.3509 & 0.4930 & 0.4141 & 0.5061 & 0.8148 & 0.5017\\\midrule
        \multirow{7}{*}{$d=5$}
        & 0.01 & 1.3417 & 1.3346 & 1.3416 & 1.3268 & \textcolor{blue}{\textbf{1.3209}} & 0.6070 & 0.5038 & 0.9107 & 0.5008 & 0.8170 & 0.4983 & 0.5754 & 0.5042\\
        & 0.005 & 1.2809 & 1.2694 & 1.2638 & 1.2632 & \textcolor{blue}{\textbf{1.2550}} & 0.7747 & 0.4979 & 0.8919 & 0.4990 & 0.6168 & 0.4963 & 0.7876 & 0.5020\\
        & 0.002 & 1.1671 & 1.1661 & 1.1932 & 1.1685 & \textcolor{blue}{\textbf{1.1550}} & 0.3047 & 0.5077 & 0.1059 & 0.5121 & 0.5585 & 0.5044 & 0.5848 & 0.5041\\
        & 0.001 & 1.1130 & 1.1097 & 1.1185 & 1.1317 & \textcolor{blue}{\textbf{1.0850}} & 0.9833 & 0.5002 & 0.6000 & 0.5039 & 0.2411 & 0.5087 & 0.6566 & 0.4967\\
        & 0.0005 & 1.0605 & 1.0462 & 1.0498 & 1.0584 & \textcolor{blue}{\textbf{1.0217}} & 0.7815 & 0.5021 & 0.7631 & 0.5022 & 0.8088 & 0.5018 & 0.4692 & 0.4946\\
        & 0.0002 & 1.0404 & 1.0156 & \textcolor{blue}{\textbf{0.9930}} & 1.0215 & 1.0054 & 0.8649 & 0.4987 & 0.5564 & 0.4956 & 0.5235 & 0.5048 & 0.5712 & 0.4958\\
        & 0.0001 & 1.0223 & 0.9935 & 0.9833 & \textcolor{blue}{\textbf{0.9810}} & 0.9867 & 0.8739 & 0.5012 & 0.7783 & 0.4979 & 0.3574 & 0.4931 & 0.4456 & 0.4943\\\midrule
        \multirow{7}{*}{$d=8$}
        & 0.01 & \textcolor{blue}{\textbf{2.4832}} & 2.6802 & 2.6413 & 2.6390 & 2.5701 & \textcolor{red}{\textbf{0.0000}} & 0.5319 & \textcolor{red}{\textbf{0.0030}} & 0.5221 & \textcolor{red}{\textbf{0.0046}} & 0.5211 & \textcolor{red}{\textbf{0.0026}} & 0.5225\\
        & 0.005 & \textcolor{blue}{\textbf{2.2558}} & 2.4032 & 2.4134 & 2.3672 & 2.2685 &  \textcolor{red}{\textbf{0.0081}} & 0.5197 & \textcolor{red}{\textbf{0.0010}} & 0.5246 & \textcolor{red}{\textbf{0.0079}} & 0.5198 & 0.1401 & 0.5110\\
        & 0.002 & \textcolor{blue}{\textbf{1.9843}} & 2.1176 & 2.1177 & 2.0986 & 2.0333 & 0.0672 & 0.5136 & 0.0511 & 0.5145 & 0.1144 & 0.5118 & 0.1073 & 0.5120\\
        & 0.001 & \textcolor{blue}{\textbf{1.7889}} & 1.9526 & 1.9312 & 1.9525 & 1.8306 &  \textcolor{red}{\textbf{0.0004}} & 0.5263 & \textcolor{red}{\textbf{0.0027}} & 0.5224 & \textcolor{red}{\textbf{0.0068}} & 0.5202 & 0.1988 & 0.5096\\
        & 0.0005 & \textcolor{blue}{\textbf{1.6631}} & 1.8607 & 1.8474 & 1.8163 & 1.7219 &  \textcolor{red}{\textbf{0.0000}} & 0.5352 & \textcolor{red}{\textbf{0.0000}} & 0.5303 & \textcolor{red}{\textbf{0.0015}} & 0.5237 & 0.1321 & 0.5112\\
        & 0.0002 & \textcolor{blue}{\textbf{1.5212}} & 1.7096 & 1.7099 & 1.6956 & 1.5956 &  \textcolor{red}{\textbf{0.0000}} & 0.5340 & \textcolor{red}{\textbf{0.0000}} & 0.5314 & \textcolor{red}{\textbf{0.0000}} & 0.5318 & 0.1752 & 0.5101\\
        & 0.0001 & \textcolor{blue}{\textbf{1.3741}} & 1.6325 & 1.5772 & 1.6110 & 1.5027 &  \textcolor{red}{\textbf{0.0000}} & 0.5510 & \textcolor{red}{\textbf{0.0000}} & 0.5452 & \textcolor{red}{\textbf{0.0000}} & 0.5502 & \textcolor{red}{\textbf{0.0003}} & 0.5267\\\midrule
        \multirow{7}{*}{$d=10$}
        & 0.01 & \textcolor{blue}{\textbf{3.3475}} & 3.9718 & 3.9114 & 3.9454 & 3.8735 &  \textcolor{red}{\textbf{0.0000}} & 0.5539 & \textcolor{red}{\textbf{0.0000}} & 0.5552 & \textcolor{red}{\textbf{0.0000}} & 0.5567 & \textcolor{red}{\textbf{0.0000}} & 0.5583\\
        & 0.005 & \textcolor{blue}{\textbf{3.1357}} & 3.5995 & 3.4993 & 3.5591 & 3.4597 &  \textcolor{red}{\textbf{0.0000}} & 0.5456 & \textcolor{red}{\textbf{0.0000}} & 0.5342 & \textcolor{red}{\textbf{0.0000}} & 0.5447 & \textcolor{red}{\textbf{0.0000}} & 0.5434\\
        & 0.002 & \textcolor{blue}{\textbf{2.8812}} & 3.1565 & 3.1775 & 3.1184 & 2.9093 &  \textcolor{red}{\textbf{0.0000}} & 0.5372 & \textcolor{red}{\textbf{0.0000}} & 0.5367 & \textcolor{red}{\textbf{0.0000}} & 0.5306 & \textcolor{red}{\textbf{0.0239}} & 0.5168\\
        & 0.001 & \textcolor{blue}{\textbf{2.6297}} & 2.8741 & 2.8712 & 2.9142 & 2.6868 &  \textcolor{red}{\textbf{0.0004}} & 0.5263 & \textcolor{red}{\textbf{0.0011}} & 0.5243 & \textcolor{red}{\textbf{0.0005}} & 0.5259 & 0.1794 & 0.5100\\
        & 0.0005 & \textcolor{blue}{\textbf{2.2916}} & 2.7049 & 2.7083 & 2.7002 & 2.4310 &  \textcolor{red}{\textbf{0.0000}} & 0.5432 & \textcolor{red}{\textbf{0.0000}} & 0.5466 & \textcolor{red}{\textbf{0.0000}} & 0.5443 & 0.0728 & 0.5134\\
        & 0.0002 & \textcolor{blue}{\textbf{1.9882}} & 2.4118 & 2.3827 & 2.4727 & 2.1770 &  \textcolor{red}{\textbf{0.0000}} & 0.5536 & \textcolor{red}{\textbf{0.0000}} & 0.5528 & \textcolor{red}{\textbf{0.0000}} & 0.5614 & \textcolor{red}{\textbf{0.0007}} & 0.5252\\
        & 0.0001 & \textcolor{blue}{\textbf{1.8902}} & 2.2568 & 2.2235 & 2.3156 & 2.0289 &  \textcolor{red}{\textbf{0.0000}} & 0.5481 & \textcolor{red}{\textbf{0.0000}} & 0.5466 & \textcolor{red}{\textbf{0.0000}} & 0.5549 & \textcolor{red}{\textbf{0.0150}} & 0.5181\\
        \bottomrule
    \end{tabular}
\end{table*}

\begin{table*}[]
\scriptsize
\centering
\setlength{\tabcolsep}{1.8mm}
 \caption{\fratio\/ values  and statistical analysis comparisons among \VPP\/ and other ART algorithms for \textit{strip patterns}}
 \label{TAB:FR-strip}
    \begin{tabular}{@{}ccccccccccccccc@{}}
    \toprule
        \multirow{3}{*}{\begin{tabular}[c]{@{}c@{}}\\[1pt]\textbf{Dimension}\\ $(d)$\end{tabular}}
        & \multirow{3}{*}{\begin{tabular}[c]{@{}c@{}}\\[1pt]\textbf{Failure Rate}\\ $(\theta)$\end{tabular}} & \multicolumn{5}{c}{\textbf{\fratio\/}}  & \multicolumn{8}{c}{\textbf{Statistical Analysis}}\\\cmidrule{3-15}
         &  & \multirow{2}{*}{\begin{tabular}[r]{@{}r@{}}\\[-5pt]\textit{VPP-}\\\textit{ART}\end{tabular}} & \multirow{2}{*}{\begin{tabular}[r]{@{}r@{}}\\[-5pt]\textit{FSCS-}\\\textit{ART}\end{tabular}} & \multirow{2}{*}{\begin{tabular}[r]{@{}r@{}}\\[-5pt]\textit{Naive-}\\\textit{KDFC}\end{tabular}} & \multirow{2}{*}{\begin{tabular}[r]{@{}r@{}}\\[-5pt]\textit{SemiBal-}\\\textit{KDFC}\end{tabular}} & \multirow{2}{*}{\begin{tabular}[r]{@{}r@{}}\\[-5pt]\textit{LimBal-}\\\textit{KDFC}\end{tabular}} & \multicolumn{2}{r}{\textit{vs.} \textit{\FSCS\/}} & \multicolumn{2}{r}{\textit{vs.} \textit{\Naive\/}} & \multicolumn{2}{r}{\textit{vs.} \textit{\SemiBal\/}} & \multicolumn{2}{r}{\textit{vs.} \textit{\LimBal\/}}\\\cmidrule{8-15}
         &  &  &  &  &  &  & \textit{p-value} & \textit{\es\/} & \textit{p-value} & \textit{\es\/} & \textit{p-value} & \textit{\es\/} & \textit{p-value} & \textit{\es\/}\\
         \midrule
        \multirow{7}{*}{$d=1$}
        & 0.01 & \textcolor{blue}{\textbf{0.5634}} & 0.5729 & 0.5664 & 0.5714 & 0.5658 & 0.3679 & 0.5067 & 0.7899 & 0.5020 & 0.4101 & 0.5061 & 0.5294 & 0.5047\\
        & 0.005 & 0.5666 & 0.5633 & 0.5670 & 0.5696 & \textcolor{blue}{\textbf{0.5619}} & 0.8994 & 0.4991 & 0.7826 & 0.5021 & 0.4768 & 0.5053 & 0.8362 & 0.4985\\
        & 0.002 & 0.5639 & 0.5683 & 0.5665 & 0.5723 & \textcolor{blue}{\textbf{0.5605}} & 0.5704 & 0.5042 & 0.9981 & 0.5000 & 0.5034 & 0.5050 & 0.7817 & 0.4979\\
        & 0.001 & 0.5634 & 0.5720 & \textcolor{blue}{\textbf{0.5549}} & 0.5690 & 0.5617 & 0.6041 & 0.5039 & 0.1890 & 0.4902 & 0.5659 & 0.5043 & 0.7915 & 0.4980\\
        & 0.0005 & \textcolor{blue}{\textbf{0.5555}} & 0.5564 & 0.5629 & 0.5556 & 0.5619 & 0.5919 & 0.5040 & 0.4510 & 0.5056 & 0.7454 & 0.5024 & 0.4847 & 0.5052\\
        & 0.0002 & \textcolor{blue}{\textbf{0.5520}} & 0.5700 & 0.5658 & 0.5662 & 0.5614 & 0.0602 & 0.5140 & 0.1353 & 0.5111 & 0.2506 & 0.5086 & 0.2506 & 0.5086\\
        & 0.0001 & 0.5766 & 0.5765 & \textcolor{blue}{\textbf{0.5527}} & 0.5545 & 0.5569 & 0.8939 & 0.4990 & \textcolor{red}{\textbf{0.0327}} & 0.4841 & 0.0559 & 0.4857 & 0.1115 & 0.4881\\\midrule
        \multirow{7}{*}{$d=2$}
        & 0.01&0.9302&0.9816&0.9365&0.9490& \textcolor{blue}{\textbf{0.9276}} &\textcolor{red}{\textbf{0.0415}}&0.5152&0.3121&0.5075&0.2007&0.5095&0.4732&0.5053\\
        & 0.005&0.9434&0.9716&0.9521& \textcolor{blue}{\textbf{0.9279}} &0.9456&0.2303&0.5089&0.7102&0.5028&0.9005&0.5009&0.4696&0.5054\\
        & 0.002& \textcolor{blue}{\textbf{0.9457}} &0.9961&0.9749&0.9611&0.9859&0.0644&0.5138&0.8804&0.5011&0.3378&0.5071&0.3354&0.5072\\
        & 0.001&0.9852&0.9561&0.9783&0.9775& \textcolor{blue}{\textbf{0.9547}} &0.2154&0.4908&0.9536&0.5004&0.9391&0.4994&0.9948&0.5000\\
        & 0.0005&0.9978&0.9784&0.9769& \textcolor{blue}{\textbf{0.9641}} &0.9808&0.4873&0.4948&0.3322&0.4928&0.0716&0.4866&0.7090&0.4972\\
        & 0.0002&0.9871&0.9827& \textcolor{blue}{\textbf{0.9574}} &0.9915&0.9811&0.2026&0.4905&0.1222&0.4885&0.5323&0.4953&0.3852&0.4935\\
        & 0.0001&0.9678&1.0130&0.9726& \textcolor{blue}{\textbf{0.9534}} &0.9760&0.2234&0.5091&0.4883&0.5052&0.4522&0.4944&0.5044&0.5050\\\midrule
        \multirow{7}{*}{$d=3$}
        & 0.01&0.9515&0.9639&0.9606&0.9850& \textcolor{blue}{\textbf{0.9491}} &0.6975&0.5029&0.8514&0.5014&0.2837&0.5080&0.6560&0.4967\\
        & 0.005&0.9844& \textcolor{blue}{\textbf{0.9404}} &0.9817&0.9803&0.9809&0.5083&0.4951&0.6035&0.5039&0.4249&0.5059&0.3965&0.5063\\
        & 0.002&0.9946&0.9853& \textcolor{blue}{\textbf{0.9569}} &0.9918&0.9653&0.3922&0.4936&\textcolor{red}{\textbf{0.0275}}&0.4836&0.9863&0.4999&0.1312&0.4887\\
        & 0.001&0.9861& \textcolor{blue}{\textbf{0.9514}} &0.9852&0.9757&1.0010&0.1059&0.4879&0.5522&0.5044&0.8288&0.4984&0.7370&0.5025\\
        & 0.0005&1.0068&0.9978&0.9859& \textcolor{blue}{\textbf{0.9510}} &0.9832&0.7362&0.4975&0.4889&0.4948&0.1556&0.4894&0.2067&0.4906\\
        & 0.0002&0.9862&0.9734&0.9834& \textcolor{blue}{\textbf{0.9730}} &0.9974&0.6592&0.4967&0.7665&0.4978&0.5442&0.4955&0.7425&0.5024\\
        & 0.0001&0.9996& \textcolor{blue}{\textbf{0.9945}} &1.0162&1.0572&1.0066&0.9807&0.4998&0.4034&0.5062&\textcolor{red}{\textbf{0.0476}}&0.5148&0.6856&0.4970\\\midrule
        \multirow{7}{*}{$d=4$}
        & 0.01&0.9984&0.9733&1.0022&0.9895& \textcolor{blue}{\textbf{0.9723}} &0.7517&0.4976&0.5908&0.5040&0.9209&0.5007&0.4972&0.5051\\
        & 0.005&0.9913&0.9830&0.9971&0.9604& \textcolor{blue}{\textbf{0.9602}} &0.9922&0.4999&0.7150&0.5027&0.6021&0.4961&0.3329&0.4928\\
        & 0.002&0.9949&1.0274&1.0084& \textcolor{blue}{\textbf{0.9749}} &0.9919&0.2591&0.5084&0.6052&0.5039&0.8414&0.4985&0.7604&0.5023\\
        & 0.001&0.9839&0.9982&0.9874& \textcolor{blue}{\textbf{0.9767}} &0.9807&0.9649&0.5003&0.9953&0.5000&0.4106&0.4939&0.9186&0.4992\\
        & 0.0005&0.9997&1.0038&1.0264&0.9968& \textcolor{blue}{\textbf{0.9792}} &0.8139&0.5018&0.2896&0.5079&0.8391&0.4985&0.4397&0.4942\\
        & 0.0002& \textcolor{blue}{\textbf{0.9832}} &1.0081&1.0013&1.0117&1.0206&0.1872&0.5098&0.5527&0.5044&0.0689&0.5136&0.2540&0.5085\\
        & 0.0001& \textcolor{blue}{\textbf{0.9693}} &1.0268&0.9943&1.0038&0.9911&\textcolor{red}{\textbf{0.0456}}&0.5149&0.5598&0.5043&0.5370&0.5046&0.3980&0.5063\\\midrule
        \multirow{7}{*}{$d=5$}
        & 0.01& \textcolor{blue}{\textbf{0.9714}} &1.0162&0.9736&0.9806&1.0228&0.1548&0.5106&0.3827&0.5065&0.4749&0.5053&\textcolor{red}{\textbf{0.0056}}&0.5206\\
        & 0.005&0.9787&1.0210&1.0097&1.0002& \textcolor{blue}{\textbf{0.9613}} &0.3629&0.5068&0.1643&0.5104&0.5924&0.5040&0.6791&0.4969\\
        & 0.002&1.0196&1.0108&0.9807& \textcolor{blue}{\textbf{0.9871}} &1.0363&0.6384&0.5035&0.4244&0.4940&0.5290&0.4953&0.3940&0.5064\\
        & 0.001&1.0095& \textcolor{blue}{\textbf{0.9791}} &1.0039&1.0275&1.0236&0.4323&0.4941&0.8983&0.4990&0.5204&0.5048&0.5039&0.5050\\
        & 0.0005&0.9939&1.0236&1.0298& \textcolor{blue}{\textbf{0.9708}} &1.0223&0.3717&0.5067&0.2420&0.5087&0.7043&0.4972&0.4610&0.5055\\
        & 0.0002&0.9987& \textcolor{blue}{\textbf{0.9751}} &0.9881&1.0208&0.9881&0.5492&0.4955&0.9281&0.5007&0.3255&0.5073&0.6977&0.5029\\
        & 0.0001&0.9954&1.0039& \textcolor{blue}{\textbf{0.9648}} &0.9953&0.9832&0.7696&0.4978&0.4521&0.4944&0.6092&0.4962&0.4384&0.4942\\\midrule
        \multirow{7}{*}{$d=8$}
        & 0.01& \textcolor{blue}{\textbf{0.9446}} &0.9907&0.9836&0.9847&1.0045&0.0642&0.5138&0.3624&0.5068&0.2449&0.5087&0.2238&0.5091\\
        & 0.005&1.0329&1.0145& \textcolor{blue}{\textbf{0.9723}} &1.0094&0.9781&0.3021&0.4923&\textcolor{red}{\textbf{0.0229}}&0.4830&0.6840&0.4970&0.3628&0.4932\\
        & 0.002& \textcolor{blue}{\textbf{0.9842}} &0.9905&1.0159&1.0024&1.0316&0.3255&0.5073&0.1617&0.5104&0.6831&0.5030&0.3427&0.5071\\
        & 0.001&1.0189&1.0107& \textcolor{blue}{\textbf{0.9999}} &1.0069&1.0411&0.4169&0.4939&0.7538&0.4977&0.6408&0.4965&0.9866&0.5001\\
        & 0.0005&1.0247&1.0123&1.0064& \textcolor{blue}{\textbf{0.9830}} &0.9866&0.1190&0.4884&0.1358&0.4889&\textcolor{red}{\textbf{0.0036}}&0.4783&\textcolor{red}{\textbf{0.0290}}&0.4837\\
        & 0.0002&0.9792&1.0166&0.9967& \textcolor{blue}{\textbf{0.9596}} &0.9942&0.5020&0.5050&0.8309&0.5016&0.5052&0.4950&0.6009&0.5039\\
        & 0.0001& \textcolor{blue}{\textbf{0.9932}} &1.0207&1.0074&0.9979&0.9935&0.3650&0.5068&0.3201&0.5074&0.4655&0.5054&0.5424&0.5045\\\midrule
        \multirow{7}{*}{$d=10$}
        & 0.01& \textcolor{blue}{\textbf{0.9760}} &1.0068&1.0196&0.9771&0.9967&0.8753&0.5012&0.1357&0.5111&0.3455&0.4930&0.9348&0.4994\\
        & 0.005&1.0035&1.0265& \textcolor{blue}{\textbf{0.9950}} &1.0067&1.0089&0.7691&0.5022&0.5492&0.5045&0.8914&0.5010&0.5979&0.4961\\
        & 0.002&0.9856&0.9933& \textcolor{blue}{\textbf{0.9827}} &0.9882&0.9974&0.5222&0.5048&0.8152&0.4983&0.8856&0.5011&0.7253&0.5026\\
        & 0.001&1.0031&1.0083&1.0074&0.9946& \textcolor{blue}{\textbf{0.9834}} &0.6407&0.5035&0.7539&0.5023&0.9126&0.5008&0.7832&0.4979\\
        & 0.0005&1.0161&1.0054& \textcolor{blue}{\textbf{1.0052}} &1.0388&1.0265&0.6510&0.4966&0.4760&0.4947&0.2862&0.5079&0.4809&0.5053\\
        & 0.0002&1.0008&1.0073&1.0096& \textcolor{blue}{\textbf{0.9823}} &1.0246&0.5084&0.5049&0.9860&0.5001&0.6003&0.4961&0.8356&0.5015\\
        & 0.0001& \textcolor{blue}{\textbf{0.9743}} &0.9945&1.0097&0.9832&1.0149&0.2287&0.5090&0.1881&0.5098&0.9243&0.4993&0.3204&0.5074\\
        \bottomrule
    \end{tabular}
\end{table*}

\begin{table*}[]
\scriptsize
\centering
\setlength{\tabcolsep}{1.8mm}
\caption{\fratio\/ values and statistical analysis comparisons among \VPP\/ and other ART algorithms for \textit{point patterns}}
 \label{TAB:FR-point}
    \begin{tabular}{@{}ccccccccccccccc@{}}
    \toprule
        \multirow{3}{*}{\begin{tabular}[c]{@{}c@{}}\\[1pt]\textbf{Dimension}\\ $(d)$\end{tabular}}
        & \multirow{3}{*}{\begin{tabular}[c]{@{}c@{}}\\[1pt]\textbf{Failure Rate}\\ $(\theta)$\end{tabular}} & \multicolumn{5}{c}{\textbf{\fratio\/}}  & \multicolumn{8}{c}{\textbf{Statistical Analysis}}\\\cmidrule{3-15}
         &  & \multirow{2}{*}{\begin{tabular}[r]{@{}r@{}}\\[-5pt]\textit{VPP-}\\\textit{ART}\end{tabular}} & \multirow{2}{*}{\begin{tabular}[r]{@{}r@{}}\\[-5pt]\textit{FSCS-}\\\textit{ART}\end{tabular}} & \multirow{2}{*}{\begin{tabular}[r]{@{}r@{}}\\[-5pt]\textit{Naive-}\\\textit{KDFC}\end{tabular}} & \multirow{2}{*}{\begin{tabular}[r]{@{}r@{}}\\[-5pt]\textit{SemiBal-}\\\textit{KDFC}\end{tabular}} & \multirow{2}{*}{\begin{tabular}[r]{@{}r@{}}\\[-5pt]\textit{LimBal-}\\\textit{KDFC}\end{tabular}} & \multicolumn{2}{r}{\textit{vs.} \textit{\FSCS\/}} & \multicolumn{2}{r}{\textit{vs.} \textit{\Naive\/}} & \multicolumn{2}{r}{\textit{vs.} \textit{\SemiBal\/}} & \multicolumn{2}{r}{\textit{vs.} \textit{\LimBal\/}}\\\cmidrule{8-15}
         &  &  &  &  &  &  & \textit{p-value} & \textit{\es\/} & \textit{p-value} & \textit{\es\/} & \textit{p-value} & \textit{\es\/} & \textit{p-value} & \textit{\es\/}\\
         \midrule
       \multirow{7}{*}{$d=1$}
        &0.01& \textcolor{blue}{\textbf{0.9592}} &0.9607&0.9755&0.9621&0.9763&0.6231&0.5037&0.4279&0.5059&0.3374&0.5072&0.0750&0.5133\\
        &0.005& \textcolor{blue}{\textbf{0.9262}} &0.9543&0.9563&0.9576&0.9320&0.0654&0.5137&0.0993&0.5123&0.1321&0.5112&0.6065&0.5038\\
        &0.002& \textcolor{blue}{\textbf{0.9355}} &0.9568&0.9627&0.9825&0.9788&0.7519&0.4976&0.8121&0.5018&0.6477&0.5034&0.1144&0.5118\\
        &0.001&1.0026& \textcolor{blue}{\textbf{0.9346}} &0.9771&0.9623&0.9651&\textcolor{red}{\textbf{0.0445}}&0.4850&0.8614&0.4987&0.1676&0.4897&0.2653&0.4917\\
        &0.0005&0.9779& \textcolor{blue}{\textbf{0.9380}} &0.9815&0.9446&0.9422&0.3095&0.4924&0.5943&0.5040&0.7309&0.4974&0.2594&0.4916\\
        &0.0002&0.9708&0.9693&0.9798& \textcolor{blue}{\textbf{0.9282}} &0.9655&0.9067&0.4991&0.4938&0.5051&0.1021&0.4878&0.9205&0.5007\\
        &0.0001&0.9750&0.9721&0.9807&0.9610& \textcolor{blue}{\textbf{0.9530}} &0.8204&0.5017&0.9610&0.4996&0.9196&0.5008&0.6786&0.4969\\\midrule
        \multirow{7}{*}{$d=2$}
        &0.01&0.9979&0.9988&0.9918& \textcolor{blue}{\textbf{0.9894}} &1.0207&0.2747&0.5081&0.2338&0.5089&0.8688&0.5012&0.3148&0.5075\\
        &0.005& \textcolor{blue}{\textbf{0.9662}} &0.9762&1.0042&0.9825&0.9917&0.8852&0.5011&0.1460&0.5108&0.5882&0.5040&0.2836&0.5080\\
        &0.002&0.9918&0.9675&0.9718& \textcolor{blue}{\textbf{0.9557}} &0.9877&0.7730&0.4979&0.4953&0.4949&0.1776&0.4900&0.6302&0.5036\\
        &0.001&0.9688&0.9995& \textcolor{blue}{\textbf{0.9550}} &0.9672&0.9817&0.0655&0.5137&0.7319&0.4974&0.5542&0.4956&0.9392&0.4994\\
        &0.0005& \textcolor{blue}{\textbf{0.9427}} &0.9663&0.9650&0.9806&0.9777&0.0864&0.5128&0.1931&0.5097&\textcolor{red}{\textbf{0.0339}}&0.5158&\textcolor{red}{\textbf{0.0245}}&0.5168\\
        &0.0002&0.9681&1.0034&0.9522& \textcolor{blue}{\textbf{0.9392}} &0.9428&0.1068&0.5120&0.7450&0.4976&0.1525&0.4893&0.2421&0.4913\\
        &0.0001&0.9758&0.9792& \textcolor{blue}{\textbf{0.9511}} &0.9673&0.9556&0.9347&0.5006&0.4807&0.4947&0.8478&0.5014&0.7290&0.4974\\\midrule
        \multirow{7}{*}{$d=3$}
        &0.01& \textcolor{blue}{\textbf{1.0376}}&1.1231&1.0930&1.1084&1.0795&\textcolor{red}{\textbf{0.0005}}&0.5260&\textcolor{red}{\textbf{0.0149}}&0.5182&\textcolor{red}{\textbf{0.0038}}&0.5216&0.1723&0.5102\\
        &0.005& \textcolor{blue}{\textbf{1.0609}} &1.0744&1.0973&1.0665&1.1051&0.5259&0.5047&0.0861&0.5128&0.9106&0.4992&\textcolor{red}{\textbf{0.0198}}&0.5174\\
        &0.002&1.0269& \textcolor{blue}{\textbf{1.0235}} &1.0297&1.0746&1.0499&0.7553&0.5023&0.5115&0.5049&\textcolor{red}{\textbf{0.0372}}&0.5155&0.1239&0.5115\\
        &0.001&1.0221&1.0343& \textcolor{blue}{\textbf{1.0151}} &1.0548&1.0551&0.6355&0.5035&0.6821&0.5031&0.1631&0.5104&0.1841&0.5099\\
        &0.0005& \textcolor{blue}{\textbf{0.9988}} &1.0017&1.0121&1.0113&1.0077&0.8080&0.5018&0.2859&0.5080&0.3116&0.5075&0.8003&0.5019\\
        &0.0002&1.0183&1.0093& \textcolor{blue}{\textbf{1.0036}} &1.0122&1.0074&0.8180&0.4983&0.5799&0.4959&0.7378&0.4975&0.7097&0.4972\\
        &0.0001&1.0023&1.0072& \textcolor{blue}{\textbf{0.9795}} &0.9905&0.9824&0.9596&0.4996&0.2742&0.4918&0.9740&0.5002&0.5906&0.4960\\\midrule
        \multirow{7}{*}{$d=4$}
        &0.01& \textcolor{blue}{\textbf{1.2336}} &1.3211&1.2789&1.3035&1.3037&\textcolor{red}{\textbf{0.0023}}&0.5227&0.1548&0.5106&\textcolor{red}{\textbf{0.0181}}&0.5176&\textcolor{red}{\textbf{0.0014}}&0.5238\\
        &0.005& \textcolor{blue}{\textbf{1.2100}} &1.2614&1.2517&1.2633&1.2192&\textcolor{red}{\textbf{0.0265}}&0.5165&0.2072&0.5094&\textcolor{red}{\textbf{0.0424}}&0.5151&0.2278&0.5090\\
        &0.002&1.1283&1.1809&1.1735&1.1524& \textcolor{blue}{\textbf{1.1212}} &0.1155&0.5117&\textcolor{red}{\textbf{0.0355}}&0.5157&0.3352&0.5072&0.8672&0.4988\\
        &0.001& \textcolor{blue}{\textbf{1.0915}} &1.1137&1.1287&1.1401&1.1370&0.0935&0.5125&0.1586&0.5105&\textcolor{red}{\textbf{0.0229}}&0.5170&\textcolor{red}{\textbf{0.0257}}&0.5166\\
        &0.0005& \textcolor{blue}{\textbf{1.0788}} &1.1117&1.1062&1.0980&1.1065&0.1022&0.5122&0.3322&0.5072&0.2515&0.5085&0.0720&0.5134\\
        &0.0002& \textcolor{blue}{\textbf{1.0444}} &1.0521&1.1007&1.0487&1.0510&0.5517&0.5044&\textcolor{red}{\textbf{0.0073}}&0.5200&0.4581&0.5055&0.5910&0.5040\\
        &0.0001&1.0506&1.0837& \textcolor{blue}{\textbf{1.0500}} &1.0589&1.0509&0.2403&0.5088&0.9512&0.5005&0.6854&0.5030&0.6532&0.5033\\\midrule
        \multirow{7}{*}{$d=5$}
        &0.01& \textcolor{blue}{\textbf{1.4384}} &1.5695&1.5413&1.5603&1.5243&\textcolor{red}{\textbf{0.0003}}&0.5273&\textcolor{red}{\textbf{0.0012}}&0.5241&\textcolor{red}{\textbf{0.0007}}&0.5253 &\textcolor{red}{\textbf{0.0206}}&0.5173\\
        &0.005& \textcolor{blue}{\textbf{1.3364}} &1.4785&1.4519&1.4385&1.4456&\textcolor{red}{\textbf{0.0002}}&0.5278&\textcolor{red}{\textbf{0.0032}}&0.5219&\textcolor{red}{\textbf{0.0154}}&0.5181&\textcolor{red}{\textbf{0.0030}}&0.5222\\
        &0.002& \textcolor{blue}{\textbf{1.2637}} &1.3549&1.3642&1.3691&1.3510&\textcolor{red}{\textbf{0.0000}}&0.5332&\textcolor{red}{\textbf{0.0001}}&0.5288&\textcolor{red}{\textbf{0.0000}}&0.5324&\textcolor{red}{\textbf{0.0000}}&0.5365\\
        &0.001& \textcolor{blue}{\textbf{1.1862}} &1.2964&1.2948&1.3005&1.2110&\textcolor{red}{\textbf{0.0002}}&0.5282&\textcolor{red}{\textbf{0.0015}}&0.5237&\textcolor{red}{\textbf{0.0000}}&0.5312&0.0569&0.5142\\
        &0.0005& \textcolor{blue}{\textbf{1.1718}} &1.2559&1.2236&1.2361&1.1938&\textcolor{red}{\textbf{0.0006}}&0.5256&\textcolor{red}{\textbf{0.0107}}&0.5190&\textcolor{red}{\textbf{0.0092}}&0.5194&0.1488&0.5108\\
        &0.0002&1.1638&1.1746&1.1636& \textcolor{blue}{\textbf{1.1562}} &1.1708&0.4006&0.5063&0.3808&0.5065&0.9149&0.4992&0.4725&0.5054\\
        &0.0001&1.1276&1.1257&1.1474&1.1553& \textcolor{blue}{\textbf{1.1074}} &0.6017&0.5039&0.2233&0.5091&0.1501&0.5107&0.5762&0.4958\\\midrule
        \multirow{7}{*}{$d=8$}
        &0.01& \textcolor{blue}{\textbf{2.0945}} &2.4049&2.3487&2.4215&2.3374&\textcolor{red}{\textbf{0.0000}}&0.5377&\textcolor{red}{\textbf{0.0000}}&0.5366&\textcolor{red}{\textbf{0.0000}}&0.5400&\textcolor{red}{\textbf{0.0000}}&0.5339\\
        &0.005& \textcolor{blue}{\textbf{2.0220}} &2.3711&2.3543&2.3313&2.2386&\textcolor{red}{\textbf{0.0000}}&0.5443&\textcolor{red}{\textbf{0.0000}}&0.5396&\textcolor{red}{\textbf{0.0000}}&0.5387&\textcolor{red}{\textbf{0.0000}}&0.5333\\
        &0.002& \textcolor{blue}{\textbf{1.8539}} &2.1827&2.2076&2.1710&2.0722&\textcolor{red}{\textbf{0.0000}}&0.5515&\textcolor{red}{\textbf{0.0000}}&0.5491&\textcolor{red}{\textbf{0.0000}}&0.5432&\textcolor{red}{\textbf{0.0000}}&0.5380\\
        &0.001& \textcolor{blue}{\textbf{1.7151}} &2.0761&2.1198&2.0804&1.9098&\textcolor{red}{\textbf{0.0000}}&0.5534&\textcolor{red}{\textbf{0.0000}}&0.5613&\textcolor{red}{\textbf{0.0000}}&0.5519&\textcolor{red}{\textbf{0.0000}}&0.5372\\
        &0.0005& \textcolor{blue}{\textbf{1.6117}} &1.9976&1.9948&1.9881&1.8038&\textcolor{red}{\textbf{0.0000}}&0.5626&\textcolor{red}{\textbf{0.0000}}&0.5643&\textcolor{red}{\textbf{0.0000}}&0.5633&\textcolor{red}{\textbf{0.0000}}&0.5368\\
        &0.0002& \textcolor{blue}{\textbf{1.4995}} &1.7979&1.7829&1.8695&1.6424&\textcolor{red}{\textbf{0.0000}}&0.5528&\textcolor{red}{\textbf{0.0000}}&0.5546&\textcolor{red}{\textbf{0.0000}}&0.5619&\textcolor{red}{\textbf{0.0000}}&0.5303\\
        &0.0001& \textcolor{blue}{\textbf{1.4719}} &1.7575&1.7223&1.7567&1.5938&\textcolor{red}{\textbf{0.0000}}&0.5539&\textcolor{red}{\textbf{0.0000}}&0.5450&\textcolor{red}{\textbf{0.0000}}&0.5512&\textcolor{red}{\textbf{0.0037}}&0.5216\\\midrule
        \multirow{7}{*}{$d=10$}
        &0.01& \textcolor{blue}{\textbf{2.3399}} &2.5080&2.5382&2.5738&2.5994&\textcolor{red}{\textbf{0.0271}}&0.5165&\textcolor{red}{\textbf{0.0069}}&0.5201&\textcolor{red}{\textbf{0.0002}}&0.5277&\textcolor{red}{\textbf{0.0000}}&0.5320\\
        &0.005& \textcolor{blue}{\textbf{2.4368}} &2.8216&2.6878&2.7286&2.7150&\textcolor{red}{\textbf{0.0000}}&0.5437&\textcolor{red}{\textbf{0.0000}}&0.5313&\textcolor{red}{\textbf{0.0000}}&0.5428&\textcolor{red}{\textbf{0.0000}}&0.5379\\
        &0.002& \textcolor{blue}{\textbf{2.3325}} &2.8479&3.0133&2.8899&2.6569&\textcolor{red}{\textbf{0.0000}}&0.5561&\textcolor{red}{\textbf{0.0000}}&0.5735&\textcolor{red}{\textbf{0.0000}}&0.5577&\textcolor{red}{\textbf{0.0000}}&0.5387\\
        &0.001& \textcolor{blue}{\textbf{2.2444}} &2.8835&2.8983&2.8322&2.6249&\textcolor{red}{\textbf{0.0000}}&0.5657&\textcolor{red}{\textbf{0.0000}}&0.5743&\textcolor{red}{\textbf{0.0000}}&0.5672&\textcolor{red}{\textbf{0.0000}}&0.5439\\
        &0.0005& \textcolor{blue}{\textbf{2.0856}} &2.7033&2.7247&2.6135&2.4393&\textcolor{red}{\textbf{0.0000}}&0.5710&\textcolor{red}{\textbf{0.0000}}&0.5743&\textcolor{red}{\textbf{0.0000}}&0.5602&\textcolor{red}{\textbf{0.0000}}&0.5408\\
        &0.0002& \textcolor{blue}{\textbf{1.8928}} &2.4606&2.4871&2.5065&2.2301&\textcolor{red}{\textbf{0.0000}}&0.5711&\textcolor{red}{\textbf{0.0000}}&0.5723&\textcolor{red}{\textbf{0.0000}}&0.5738&\textcolor{red}{\textbf{0.0000}}&0.5446\\
        &0.0001& \textcolor{blue}{\textbf{1.7778}} &2.0912&2.3241&2.3366&2.1341&\textcolor{red}{\textbf{0.0000}}&0.5445&\textcolor{red}{\textbf{0.0000}}&0.5708&\textcolor{red}{\textbf{0.0000}}&0.5744&\textcolor{red}{\textbf{0.0000}}&0.5553\\
        \bottomrule

    \end{tabular}
\end{table*}

Tables \ref{TAB:FR-block} to \ref{TAB:FR-point} present the simulation \textit{\fratio\/} value comparisons
among \FSCS\/, \KDFC\/, and \VPP\/,
for \textit{block}, \textit{strip}, and \textit{point patterns},
respectively.
In the tables, the \textit{\es\/} values represent the probability that \VPP\/ outperforms the compared ART algorithm; and
the \textit{p-value} indicates the statistical significance of the \textit{\fratio\/} differences.
Based on the data in these three tables, we have the following observations:

In general, for all five ART algorithms, the \textit{\fratio\/} values decrease as $\theta$ decreases, for both block and point patterns:
This indicates that ART has better failure-detection effectiveness when the failure rate is small.
As the dimensionality, $d$, increases, the \textit{\fratio\/} values generally increase,
showing that $d$ has an important negative impact on the effectiveness of ART algorithms.

\noindent\textit{\textbf{A. \VPP\/ versus \FSCS\/}}

(1) Block-pattern simulation findings:
When considering a fixed failure rate $\theta$, the \VPP\/ \textit{\fratio\/} values increase with increases in the dimension $d$:
The higher dimension is, the worse the failure-detection effectiveness of \VPP\/ is.
When $1 \leq d \leq 4$, the failure-detection effectiveness of \VPP\/ is significantly better than RT, in most cases.
Except in some cases ($d = 4, \theta = 0.01,0.005$), the \textit{\fratio\/} values are less than $1.0$, but when $d$ increases, then the \VPP\/ failure-detection effectiveness becomes weaker than RT.
When $1 \leq d \leq 5$, the failure-detection effectiveness of \VPP\/ and \FSCS\/ are similar.
When $d = 8,10$, the failure-detection effectiveness of \VPP\/ is obviously better than that of \FSCS\/, which is shown by statistical analysis of the \textit{\fratio\/} value comparisons between \VPP\/ and \FSCS\/.
When $1 \leq d \leq 5$, all \textit{p-values} are greater than $0.05$; however,
when $d = 8,10$,
the \textit{\es\/} values are all greater than $0.5$, and
the \textit{p-values} are all significantly less than $0.05$ (except for $d = 8,\theta = 0.002$).

\begin{table*}[t]
\scriptsize
\centering
 \caption{F-measures and statistical analysis comparisons among \VPP\/ and other different ART algorithms with the $22$ subject programs}
 \label{effectiveness_real}
\setlength{\tabcolsep}{1.13mm}
    \begin{tabular}{@{}cccccccccccccccc@{}}
    \toprule
        \multirow{3}{*}{\begin{tabular}[c]{@{}c@{}}\\[1pt]\textbf{No.}\end{tabular}} & \multirow{3}{*}{\begin{tabular}[c]{@{}c@{}}\\[1pt]\textbf{Program}\end{tabular}} &
        \multirow{3}{*}{\begin{tabular}[c]{@{}c@{}}\\[1pt]\textbf{Dimension}\\ $(d)$\end{tabular}}&
        \multicolumn{5}{c}{\textbf{F-measure}}  & \multicolumn{8}{c}{\textbf{Statistical Analysis}}\\\cmidrule{4-16}
         & & & \multirow{2}{*}{\begin{tabular}[r]{@{}r@{}}\\[-5pt]\textit{VPP-}\\\textit{ART}\end{tabular}} & \multirow{2}{*}{\begin{tabular}[r]{@{}r@{}}\\[-5pt]\textit{FSCS-}\\\textit{ART}\end{tabular}} & \multirow{2}{*}{\begin{tabular}[r]{@{}r@{}}\\[-5pt]\textit{Naive-}\\\textit{KDFC}\end{tabular}} & \multirow{2}{*}{\begin{tabular}[r]{@{}r@{}}\\[-5pt]\textit{SemiBal-}\\\textit{KDFC}\end{tabular}} & \multirow{2}{*}{\begin{tabular}[r]{@{}r@{}}\\[-5pt]\textit{LimBal-}\\\textit{KDFC}\end{tabular}} & \multicolumn{2}{r}{\textit{vs.} \textit{\FSCS\/}} & \multicolumn{2}{r}{\textit{vs.} \textit{\Naive\/}} & \multicolumn{2}{r}{\textit{vs.} \textit{\SemiBal\/}} & \multicolumn{2}{r}{\textit{vs.} \textit{\LimBal\/}}\\\cmidrule{9-16}
         &  & & &  &  &  &  & \textit{p-value} & \textit{\es\/} & \textit{p-value} & \textit{\es\/} & \textit{p-value} & \textit{\es\/} & \textit{p-value} & \textit{\es\/}\\
         \midrule
        1 &\texttt{airy}&1&\textcolor{blue}{\textbf{797.28}}&816.03&806.91&803.29&809.33&0.1793 &0.5100 &0.3791 &0.5066 &0.4055 &0.5062 &0.3163 &0.5075 \\
        2 &\texttt{bessj0}&1&450.05&448.20&443.44&\textcolor{blue}{\textbf{440.31}}&449.13&0.8286 &0.5016 &0.9175 &0.5008 &0.4290 &0.4941 &0.5561 &0.5044 \\
        3 &\texttt{erfcc}&1&\textcolor{blue}{\textbf{1019.32}}&1054.65&1040.58&1045.86&1033.00&\textcolor{red}{\textbf{0.0301}}&0.5162 &0.3323 &0.5072 &0.1319 &0.5112 &0.2828 &0.5080 \\
        4 &\texttt{probks}&1&1452.91&1469.21&\textcolor{blue}{\textbf{1450.82}}&1452.57&1475.86&0.4970 &0.5051 &0.9523 &0.5004 &0.9427 &0.5005 &0.4633 &0.5055 \\
        5 &\texttt{tanh}&1&312.42&319.82&\textcolor{blue}{\textbf{306.91}}&309.85&309.36&0.3084 &0.5076 &0.3982 &0.4937 &0.7942 &0.4981 &0.4720 &0.4946 \\
        6 &\texttt{bessj}&2&462.96&\textcolor{blue}{\textbf{452.49}}&457.52&457.60&461.69&0.4585 &0.4945 &0.9216 &0.5007 &0.6456 &0.4966 &0.5489 &0.4955 \\
        7 &\texttt{gammq}&2&\textcolor{blue}{\textbf{1063.88}}&1087.52&1066.34&1100.74&1172.50&0.2777 &0.5081 &0.8889 &0.5010 &0.0757 &0.5132 &\textcolor{red}{\textbf{0.0011}} &0.5243 \\
        8 &\texttt{sncndn}&2&640.22&643.40&\textcolor{blue}{\textbf{629.63}}&649.75&655.74&0.5157 &0.5048 &0.9029 &0.4991 &0.3750 &0.5066 &0.1915 &0.5097 \\
        9 &\texttt{golden}&3&1824.80&1831.29&\textcolor{blue}{\textbf{1806.09}}&1816.82&1902.52&0.4552 &0.5056 &0.9594 &0.4996 &0.5315 &0.5047 &\textcolor{red}{\textbf{0.0038}} &0.5216 \\
        10 &\texttt{plgndr}&3&1733.83&\textcolor{blue}{\textbf{1572.82}}&1648.26&1618.40&1665.89&\textcolor{red}{\textbf{0.0076}}&0.4801 &0.1738 &0.4899 &0.0951 &0.4876 &0.6289 &0.4964 \\
        11 &\texttt{cel}&4&1628.50&\textcolor{blue}{\textbf{1572.56}}&1577.88&1593.71&1586.13&0.5594 &0.4956 &0.7333 &0.4975 &0.7584 &0.4977 &0.7521 &0.4976 \\
        12 &\texttt{el2}&4&796.70&714.58&\textcolor{blue}{\textbf{710.58}}&714.01&749.98&\textcolor{red}{\textbf{0.0003}}&0.4730 &\textcolor{red}{\textbf{0.0009}} &0.4753 &\textcolor{red}{\textbf{0.0001}} &0.4702 &\textcolor{red}{\textbf{0.0429}} &0.4849 \\
        13 &\texttt{calDay}&5&\textcolor{blue}{\textbf{1101.30}}&1312.37&1295.35&1262.40&1226.75&\textcolor{red}{\textbf{0.0000}} &0.5380 &\textcolor{red}{\textbf{0.0000}} &0.5380 &\textcolor{red}{\textbf{0.0000}} &0.5354 &\textcolor{red}{\textbf{0.0006}} &0.5256 \\
        14 &\texttt{complex}&6&\textcolor{blue}{\textbf{1134.81}}&1283.25&1155.82&1150.68&1142.01&\textcolor{red}{\textbf{0.0002}} &0.5278 &0.9074 &0.5009 &0.8793 &0.5011 &0.7912 &0.5020 \\
        15 &\texttt{pntLinePos}&6&\textcolor{blue}{\textbf{1397.57}}&1589.92&1444.04&1490.97&1458.47&\textcolor{red}{\textbf{0.0000}} &0.5431 &0.5160 &0.5048 &\textcolor{red}{\textbf{0.0299}} &0.5162 &0.0980 &0.5123 \\
        16 &\texttt{triangle}&6&1411.86&1396.55&1415.63&1389.04&\textcolor{blue}{\textbf{1324.95}}&0.4483 &0.4943 &0.7890 &0.4980 &0.1772 &0.4899 &\textcolor{red}{\textbf{0.0175}} &0.4823 \\
        17 &\texttt{line}&8&3322.27&3435.07&3343.48&\textcolor{blue}{\textbf{3269.93}}&3370.50&0.1796 &0.5100 &0.7428 &0.5024 &0.3340 &0.4928 &0.2840 &0.5080 \\
        18 &\texttt{pntTrianglePos}&8&\textcolor{blue}{\textbf{4584.95}}&5067.49&5046.12&4955.54&4659.86&0.1498 &0.5107 &0.2122 &0.5093 &0.3528 &0.5069 &0.3202 &0.4926 \\
        19 &\texttt{TwoLinesPos}&8&\textcolor{blue}{\textbf{7415.10}}&9814.67&8297.09&8430.90&8909.71&\textcolor{red}{\textbf{0.0000}} &0.5922 &\textcolor{red}{\textbf{0.0000}} &0.5304 &\textcolor{red}{\textbf{0.0000}} &0.5337 &\textcolor{red}{\textbf{0.0000}} &0.5522 \\
        20 &\texttt{nearestDistance}&10&\textcolor{blue}{\textbf{2145.31}}&2259.57&2277.88&2161.20&2188.53&\textcolor{red}{\textbf{0.0000}} &0.5489 &\textcolor{red}{\textbf{0.0176}} &0.5177 &0.4080 &0.5062 &0.1908 &0.5098 \\
        21 &\texttt{calGCD}&10&\textcolor{blue}{\textbf{1004.15}}&1016.98&1016.14&1017.74&1056.02&0.5202 &0.5048 &0.6552 &0.5033 &0.3601 &0.5068 &\textcolor{red}{\textbf{0.0242}}&0.5168 \\
        22 &\texttt{select}&11&\textcolor{blue}{\textbf{5490.70}}&5599.86&5907.50&5634.03&5808.85&0.6502 &0.5034 &\textcolor{red}{\textbf{0.0368}} &0.5156 &0.5461 &0.5045 &0.0885 &0.5127 \\
        \bottomrule
    \end{tabular}
\end{table*}

(2) Strip-pattern simulation findings:
When $d=1$, both \VPP\/ and \FSCS\/ have better failure-detection effectiveness than RT,  with \textit{\fratio\/} values of about $0.55$.
In other dimensions, regardless of dimension $d$ and failure rate $\theta$, the \textit{\fratio\/} values of \VPP\/ are similar to those of \FSCS\/.
Nearly all \textit{p-values} are greater than $0.05$ (except $d=2,\theta=0.01$ and $d=4,\theta=0.0001$),
indicating that the failure-detection performance of \VPP\/ is not significantly different from \FSCS\/.

(3) Point-pattern simulation findings:
Similar to with the block pattern,
for a give failure rate $\theta$, the failure-detection effectiveness of both \VPP\/ and \FSCS\/ decrease as the dimension $d$ increases
---
as seen from the increasing \textit{\fratio\/} values for increasing dimensionality.
When $d=1,2$, both \VPP\/ and \FSCS\/ outperform RT, but when $d\geq3$, \VPP\/ has similar, or better, performance than \FSCS\/
---
reflected in \VPP\/ having lower \textit{\fratio} values than \FSCS\/, in most cases.
When $1\leq d \leq 4$, the \textit{p-values} are usually greater than $0.05$, and all \textit{\es\/} are about $0.50$.
When $d \geq 5$, the \textit{p-values} are less than $0.05$ (except $d=5,\theta=0.0002,0.0001$), and all \textit{\es\/} values are greater than $0.50$.
In summary,
when $1 \leq d \leq 4$, there is no significant difference in the failure-detection effectiveness of \VPP\/ and \FSCS\/;
but when $d\geq5$, \VPP\/ has a much better performance than \FSCS\/.

\noindent\textit{\textbf{B. \VPP\/ versus \KDFC\/}}

(1) Block-pattern simulation findings:
For a given failure rate $\theta$, the failure-detection effectiveness of \FSCS\/, \VPP\/, and all three \KDFC\/ versions appear similar, as the dimension $d$ increases.
When $1\leq d \leq 5$, the three \KDFC\/ versions have lower \textit{\fratio\/} values than \VPP\/, indicating that \KDFC\/ has similar, or better, failure-detection in these dimensions.
However, when $d=8,10$, \VPP\/ has the lower \textit{\fratio\/} values, showing a much better failure-detection effectiveness.
When $1\leq d \leq 5$, nearly all \KDFC\/ \textit{p-values} are greater than $0.05$, and all \KDFC\/ \textit{\es\/} values are around $0.50$.
When $d=8,10$, the \Naive\/ and \SemiBal\/ \textit{p-values} are less than $0.05$ (except $d=8,\theta=0.002$), and all \textit{\es\/} values are greater than $0.50$.
For \LimBal\/, when $d =8$, all \textit{\es\/} values are greater than $0.50$, and \textit{p-values} are greater than $0.05$ (except $d=8,\theta=0.01,0.0001$).
When $d=10$, all \LimBal\/ \textit{\es\/} values are greater than $0.50$, and all \textit{p-values} are less than for $d=8$ (except $\theta=0.0001$); furthermore,
except for $\theta=0.001,0.0005$, all \textit{p-values} are less than $0.05$.
Overall, \VPP\/ has similar failure-detection effectiveness to \KDFC\/ in lower-dimensional input domains, and better performance than \Naive\/ and \SemiBal\/ in higher dimensions.
\VPP\/ also performs similar or better than \LimBal\/ in some high-dimensional cases.

(2) Strip-pattern simulation findings:
Similar to other ART algorithms, there is no significant difference in \textit{\fratio\/} values among \VPP\/ and \KDFC\/.
When $d=1$, \VPP\/ and  \KDFC\/ have lower \textit{F-measures} than RT.
In other dimensions, the \textit{\fratio\/} values are about $1.0$, indicating similar failure-detection effectiveness to RT.
The statistical analyses also indicate no significant difference among the four ART algorithms.

(3) Point-pattern simulation findings:
For a given failure rate $\theta$, as the dimension $d$ increases, the failure-detection effectiveness of both \VPP\/ and \KDFC\/  declines.
When $d=1,2$, the effectiveness of \VPP\/ and \KDFC\/ is better than RT, with their \textit{\fratio} values being less than $1.0$;
when $d \geq 3$, the  \VPP\/ \textit{\fratio} values are similar or better than those of \KDFC\/.
When $1 \leq d \leq 4$, the \textit{p-values} are usually greater than $0.05$, and the \textit{\es\/} values are around $0.50$;
When $d=5$, in the cases of the lower failure rates ($\theta=0.0002,0.0001$ for \Naive\/ and \SemiBal\/, and $\theta= 0.0005,0.0002,0.0001$ for \LimBal\/), the \textit{p-values} are  greater than $0.05$, indicating no statistically-significant difference.
When $d=8,10$, all \textit{p-values} are less than $0.05$, and all \textit{\es\/} values are greater than $0.50$.
In summary, when the dimensionality is low, \VPP\/ and \KDFC\/ have comparable failure-detection effectiveness; and
in high dimensions, \VPP\/ performs better.

\textbf{\textit{Discussion of effectiveness simulation results}:}
We examined the failure-detection effectiveness of \VPP\/, \FSCS\/, and three versions of \KDFC\/ through simulations.
\Naive\/ and \SemiBal\/ search for the exact \NN\/ of a candidate test case (the same as \FSCS\/); but
\VPP\/ and \LimBal\/ use approximate \NN\/ searches.
Through the simulations, it was found that \VPP\/ has a similar failure-detection effectiveness to \FSCS\/ and other ART algorithms that use the exact \NN\/ search approach, when the input domain dimensionality is low,
and has better effectiveness in high dimensions.
There was no significant difference between \VPP\/ and \LimBal\/, in low dimensions, in terms of their failure-detection effectiveness.
However, in high dimensions, \VPP\/ can perform comparably or better than \LimBal\/.

\subsubsection{Answer to RQ1 - Part 2: Results of Experiments}
Table~\ref{effectiveness_real} presents the \textit{\fratio\/} data and statistical analyses for \VPP\/, \FSCS\/, and \KDFC\/ with the $22$ subject programs.
Based on the data in this table, we have the following observations:

\noindent\textit{\textbf{A. \VPP\/ versus \FSCS\/}}


\VPP\/ has lower \textit{F-measures} than \FSCS\/ in $16$ of the $22$ real-life programs ($73\%$).
The \textit{\es} values for comparisons between \VPP\/ and \FSCS\/ range from $0.47$ to about $0.60$, indicating that \VPP\/ can effectively ensure or improve on the failure-detection effectiveness of \FSCS\/ in real-life programs.
Six of the eight programs
 (\texttt{erfcc}, \texttt{calDay}, \texttt{complex}, \texttt{pntLinePos}, \texttt{twoLinesPos}, and \texttt{nearestDistance}) whose \textit{p-values} are less than $0.05$ have \textit{\es} values  greater than $0.5$, indicating that \VPP\/ outperforms \FSCS\ in the tests of these programs;
\FSCS\/ performs better for the programs \texttt{plgndr} and \texttt{el2}.
Generally speaking, \VPP\/ performs similarly or slightly worse than \FSCS\/ in low-dimensional programs, but has comparable or better failure-detection effectiveness in high-dimensional programs
---
from the perspective of the \textit{F-measure}, \VPP\/ may require fewer test cases to find the first program failure in high-dimensional programs.


\noindent\textit{\textbf{B. \VPP\/ versus \KDFC\/}}

In low-dimensional input domain programs, \KDFC\/ can usually achieve lower \textit{F-measure} values,
but this changes for high dimensional input domains, when \VPP\/ performs best.



\textbf{\textit{Discussion of effectiveness experiments using subject programs}:}
In the experiments using real-life programs, the \VPP\/ failure-detection effectiveness was found to be comparable to that of other ART algorithms.
\VPP\/ had comparable failure-detection effectiveness in low-dimensional programs, and often had better performance than other ART algorithms with the high-dimensional programs
---
\VPP\/ will use fewer test cases to detect the first failure in the high-dimensional programs.


\textbf{\textit{Summary and Discussions of answering to RQ1}:}
Both the simulations and experimental data suggest that \VPP\/ can achieve comparable, or better, failure-detection effectiveness, especially in higher dimensions.

\subsection{Comparisons of Efficiency}

This section reports on the efficiency comparisons between \VPP\/ and the other ART algorithms.
The results and main findings address RQ2, as follows.

\begin{figure}[!t]
\graphicspath{{VPP-ART-Figures/}}
  \centering
  \subfigure[$d=1$]
  {
    \includegraphics[width=0.47\textwidth]{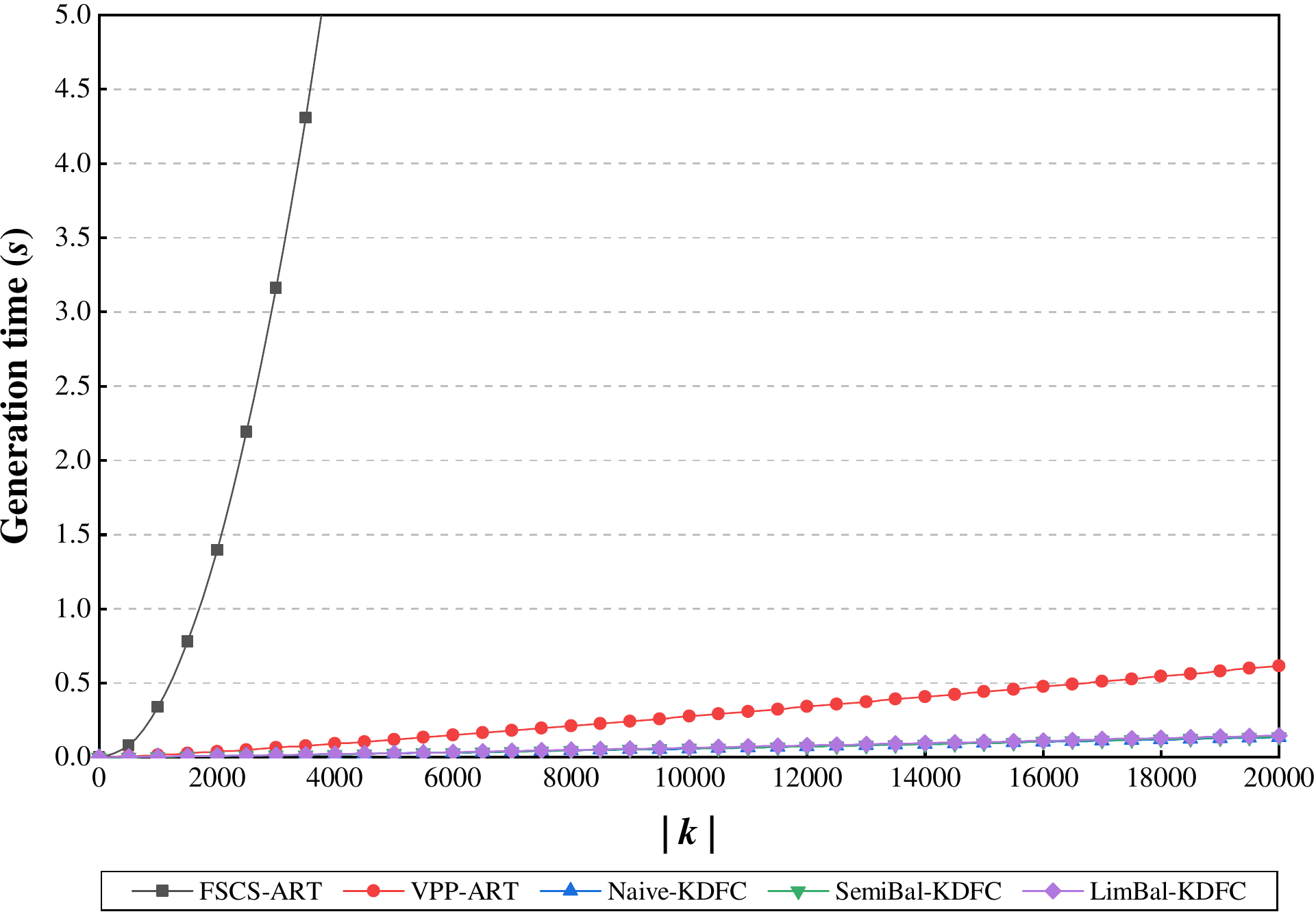}
    \label{1dtime}
  }
  \hspace{5mm}
  \subfigure[$d=2$]
  {
    \includegraphics[width=0.47\textwidth]{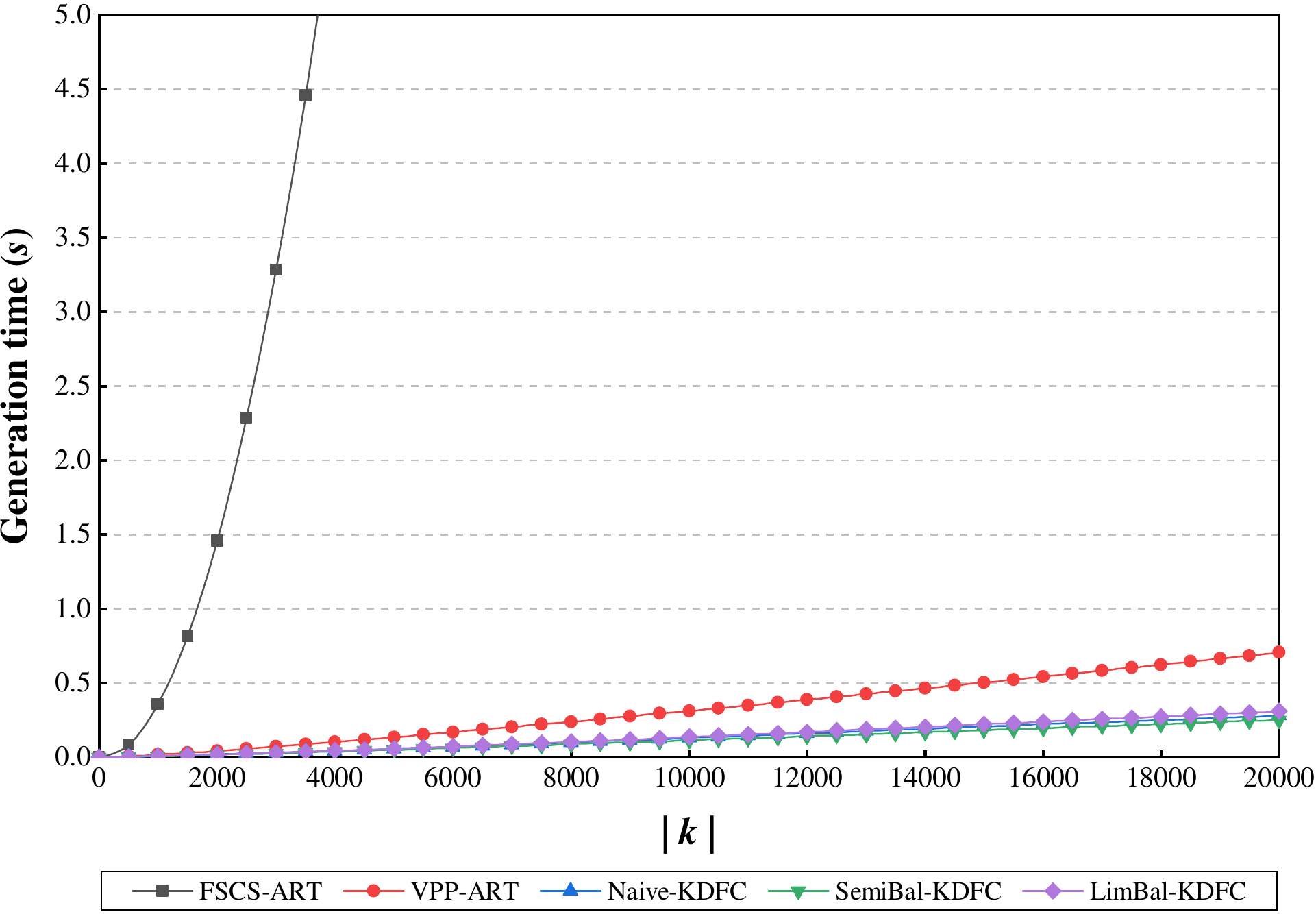}
    \label{2dtime}
  }
  \subfigure[$d=3$]
  {
    \includegraphics[width=0.47\textwidth]{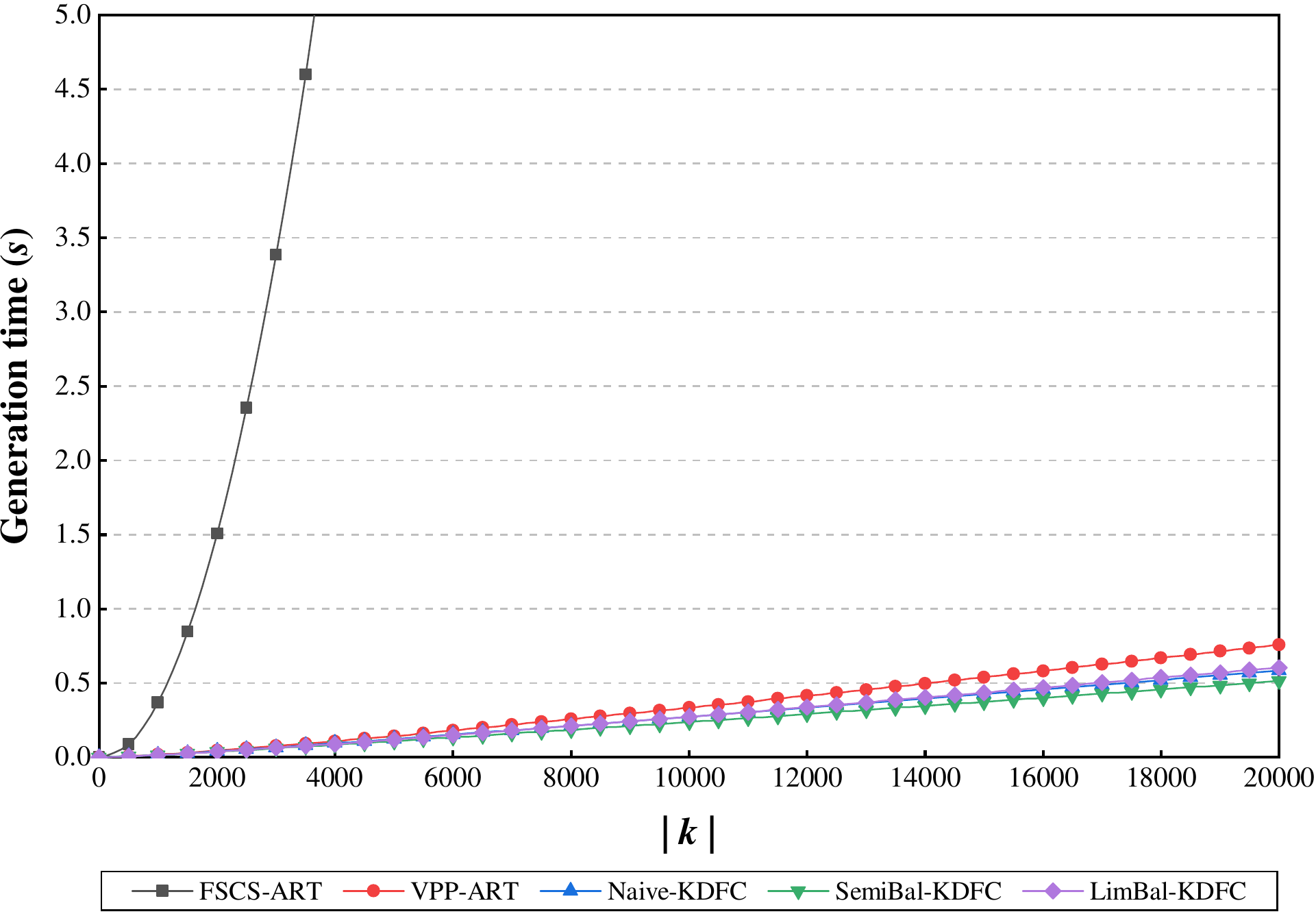}
    \label{3dtime}
  }
  \hspace{5mm}
  \caption{Generation times for various test suite sizes, for $d=1$, $2$, and $3$.}
  \label{FIG:Generation-TIME-1/2/3}
\end{figure}

\subsubsection{Answer to RQ2 - Part 1: Results of Simulations}
\begin{figure*}[!t]
\graphicspath{{VPP-ART-Figures/}}
  \centering
  \subfigure[$d=4$]
  {
    \includegraphics[width=0.47\textwidth]{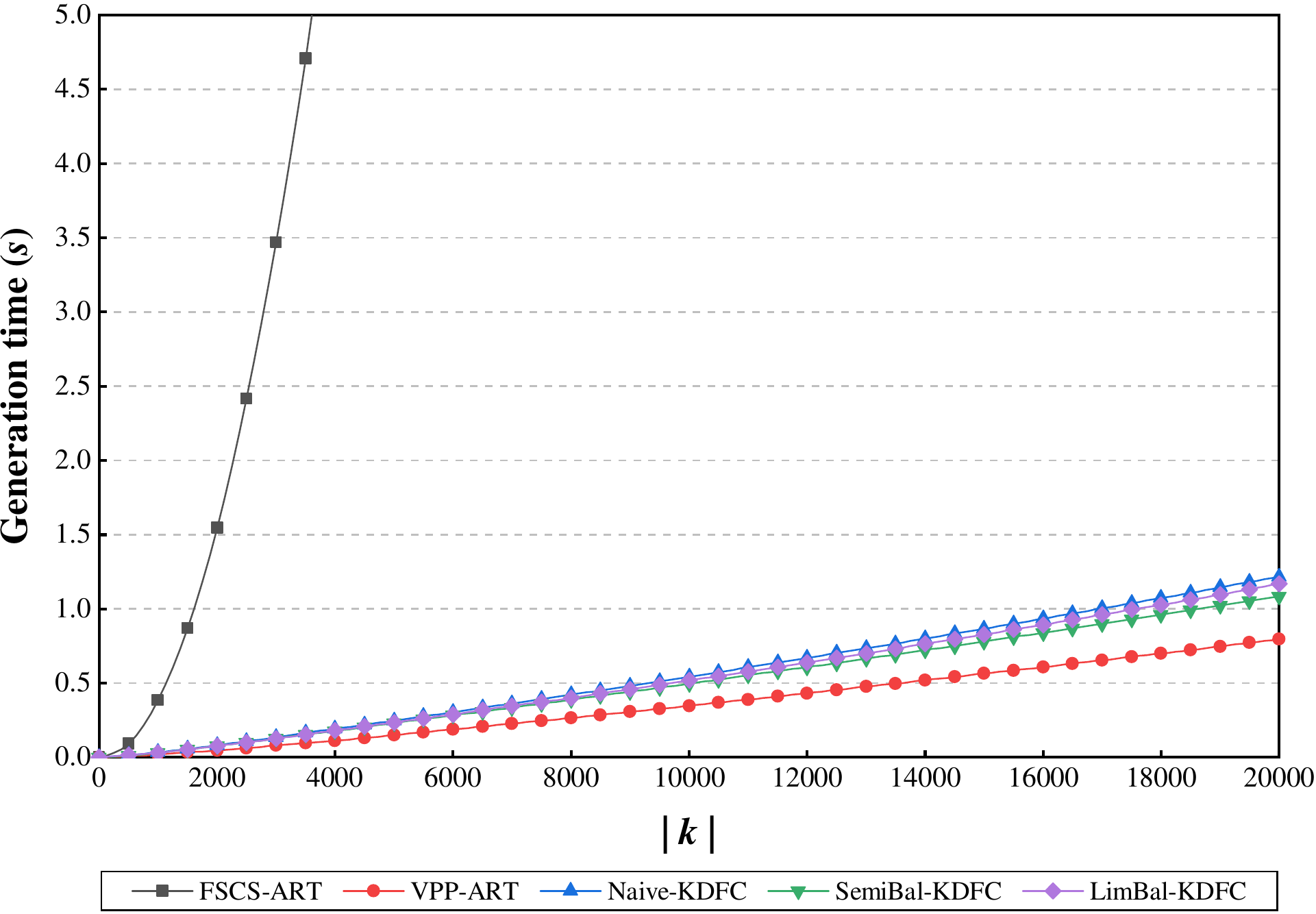}
    \label{4dtime}
  }
  \subfigure[$d=5$]
  {
    \includegraphics[width=0.47\textwidth]{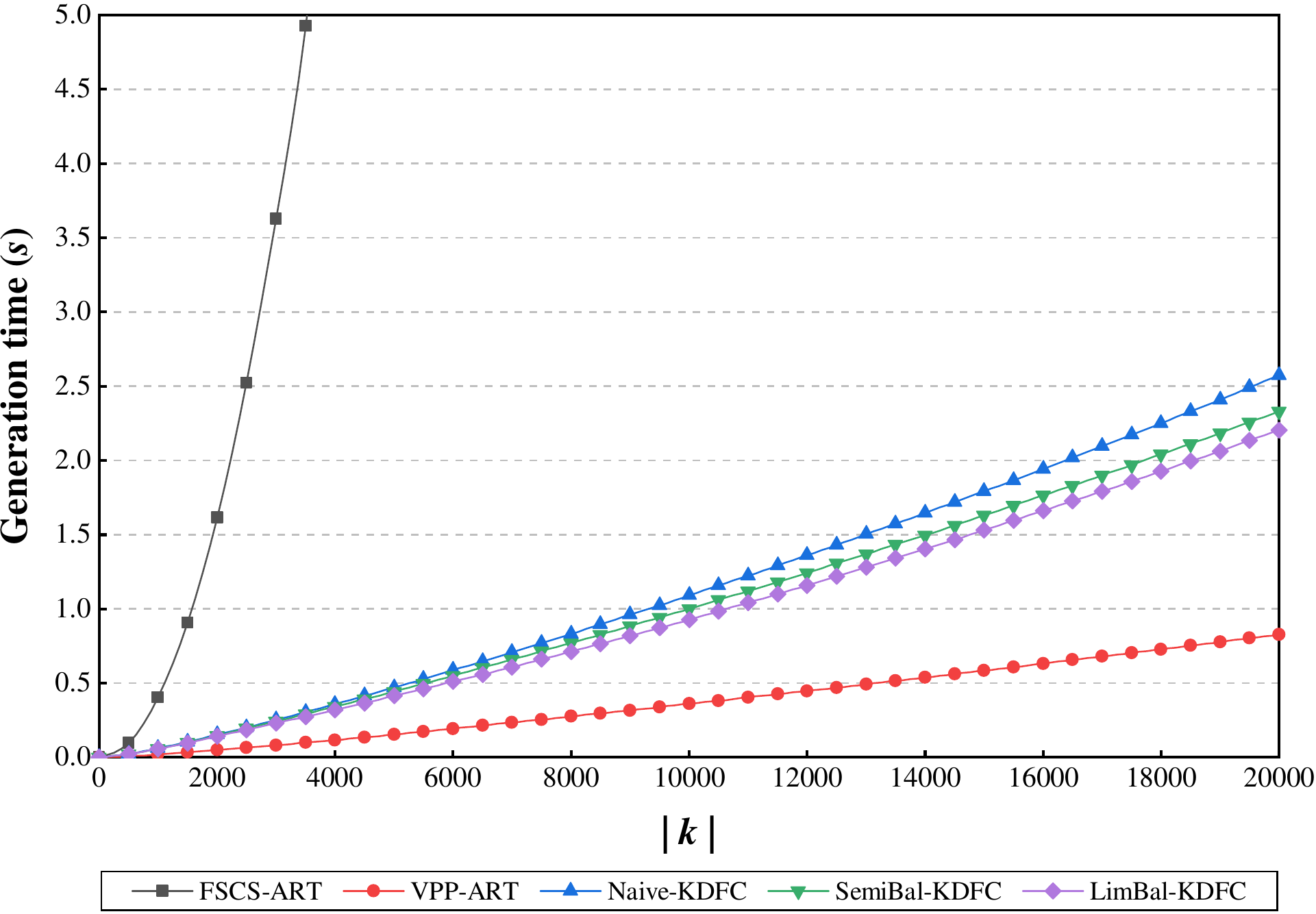}
    \label{5dtime}
  }
  \subfigure[$d=8$]
  {
    \includegraphics[width=0.47\textwidth]{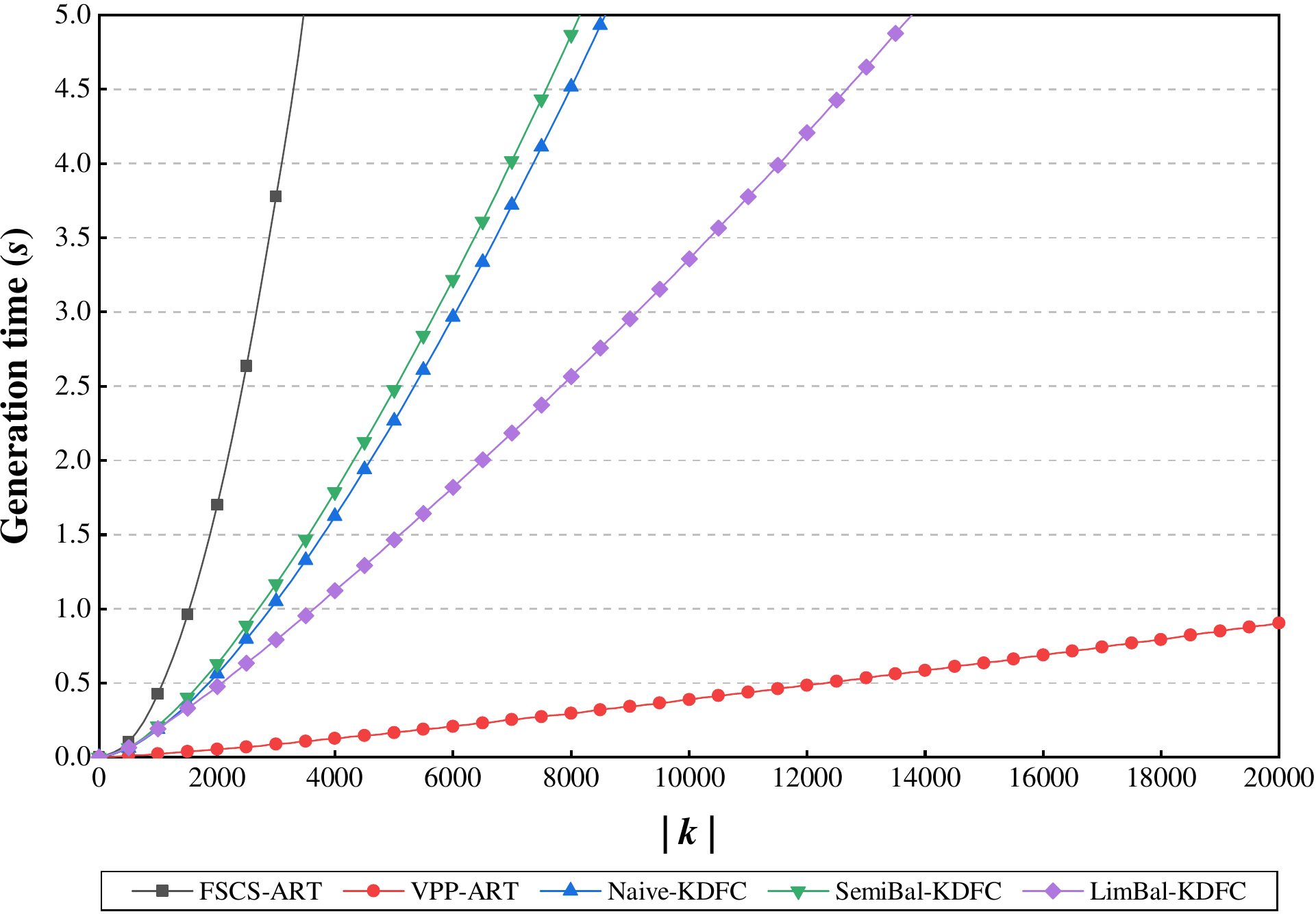}
    \label{8dtime}
  }
  \subfigure[$d=10$]
  {
    \includegraphics[width=0.47\textwidth]{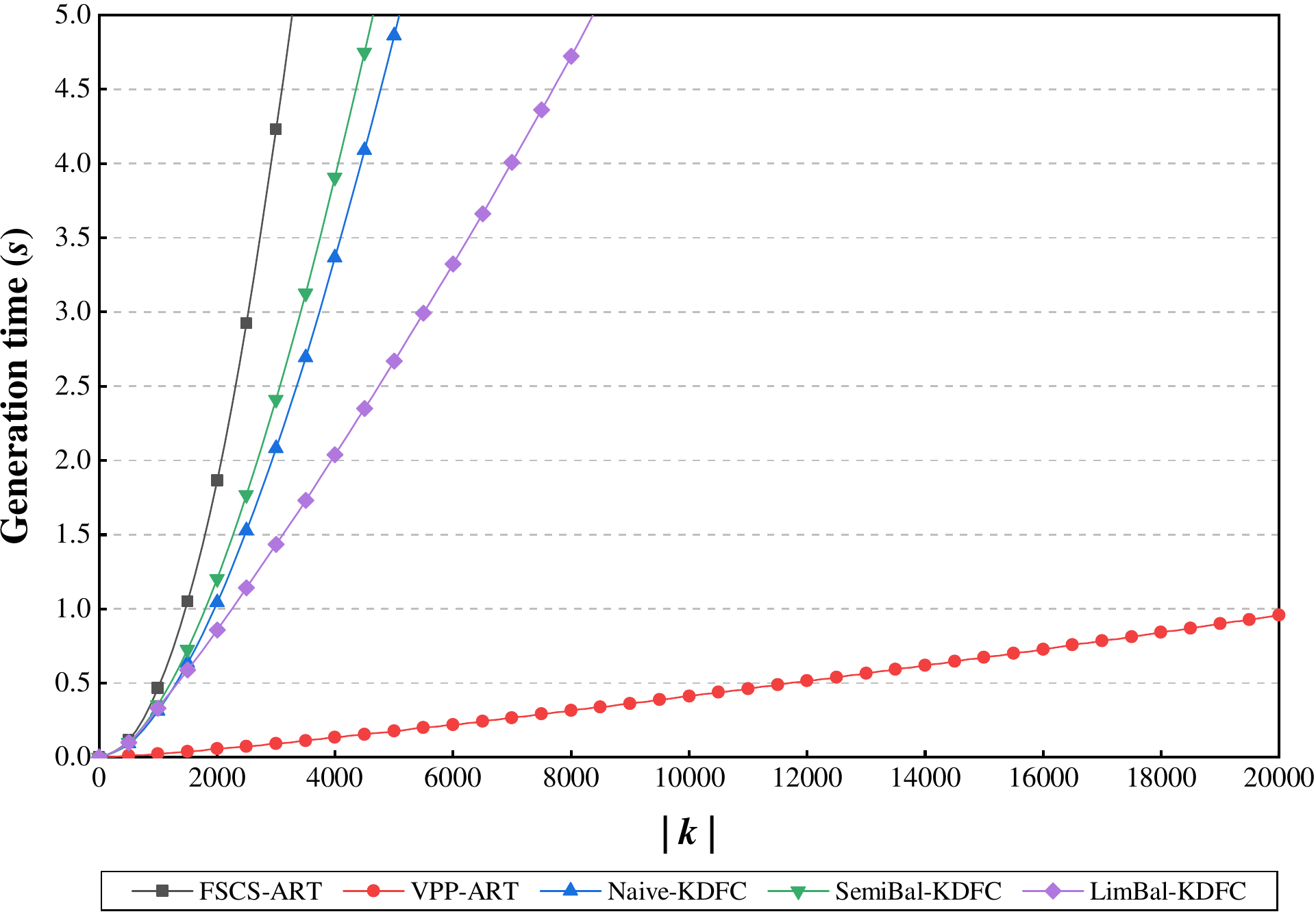}
    \label{10dtime}
  }
  \caption{Generation times for various test suite sizes, for $d=4$, $5$, $8$, and $10$.}
  \label{FIG:Generation-TIME-4/5/8/10}
\end{figure*}

Figure~\ref{FIG:Generation-TIME-1/2/3} shows the test case generation times of \VPP\/, \FSCS\/, and the three \KDFC\/ algorithms, for various test suite sizes, in $1$, $2$, and $3-$dimensional input domains.
Figure~\ref{FIG:Generation-TIME-4/5/8/10} shows the data for $4$, $5$, $8$, and $10-$dimensional input domains.
In the figures, the $x$-axis shows the size of the test suite ($\textbf{\textit{k}}$), and the $y$-axis shows the time taken to generate the $\textbf{\textit{k}}$ test cases.
Based on the simulation results, we have the following observations:

\noindent\textit{\textbf{A. \VPP\/ versus \FSCS\/}}

\VPP\/ has a significant advantage over \FSCS\/ in terms of test case generation time, across different input domain dimensionalities.
Compared with FSCS-ART, \VPP\/ uses vantage point partitioning and approximate \NN\/ strategies to increase the efficiency.
The simulation results indicate that the \VPP\/ test case generation time is not obviously impacted by increases in dimensionality:
\VPP\/ still maintains a strong ability to reduce the time overheads in high dimensions.

\noindent\textit{\textbf{B. \VPP\/ versus \KDFC\/}}

Figure~\ref{FIG:Generation-TIME-1/2/3} shows that when $d = 1$, $2$, or $3$, there is little difference in test case generation time for the three \KDFC\/ versions, all of which have better performance than \VPP\/.
Figure~\ref{FIG:Generation-TIME-4/5/8/10} shows that when the input domain dimension increases to $d = 4$, the \KDFC\/ performance becomes worse than \VPP\/.
As the dimensionality increases ($d \geq 5$), the performance differences among the three \KDFC\/ versions gradually increases, with \LimBal\/ emerging as the best.
However, the performance of all three \KDFC\/ versions is worse than that of \VPP\/, with the difference increasing as the dimensionality increases.

\textbf{\textit{Discussion of efficiency simulation results}:}
The simulation results show that \VPP\/ requires much less time than \FSCS\/ to generate an equal number of test cases.
\VPP\/ performs slightly worse than \KDFC\/ when the dimensionality is low, but increasingly becomes the better performer as the dimensions increase.
The two ART algorithms based on approximate \NN\/ (\LimBal\/ and \VPP\/) have better test case generation time.
As the number of test cases get larger, the time cost of the approximate \NN\/ search is lower than that of the exact \NN\/ search.

\subsubsection{Answer to RQ2 - Part 2: Results of Experiments}
This section reports on the investigation into the efficiency of the five ART algorithms (\VPP\/, \FSCS\/, \Naive\/, \SemiBal\/, and \LimBal\/) on the $22$ subject programs.
Table~\ref{f-time-empirical-studies} reports the average time taken to detect the first failure (\textit{F-time}) in each program, by each algorithm.
Based on the data in the table, the main findings are as follows.

\noindent\textit{\textbf{A. \VPP\/ versus \FSCS\/}}

The experiment data show the \VPP\/ \textit{F-time} results to be much lower than those of \FSCS\/:
The \textit{\es\/} scores show \VPP\/ to be better than \FSCS\/ $78\%$ to $93\%$ of the time.
The \textit{p-values} for both algorithms are $0.0000$, which shows a significant difference between \VPP\/ and \FSCS\/.
In summary, \VPP\/ greatly improves on the testing efficiency of \FSCS\/ for the real-life subject programs.

\begin{table*}[!t]
\scriptsize
\centering
 \caption{F-times and statistical analysis comparisons among \VPP\/ and other different ART algorithms with the $22$ subject programs}
 \label{f-time-empirical-studies}
\setlength{\tabcolsep}{1.1mm}
    \begin{tabular}{@{}cccccccccccccccc@{}}
    \toprule
        \multirow{3}{*}{\begin{tabular}[c]{@{}c@{}}\\[1pt]\textbf{No.}\end{tabular}} & \multirow{3}{*}{\begin{tabular}[c]{@{}c@{}}\\[1pt]\textbf{Program}\end{tabular}} &
        \multirow{3}{*}{\begin{tabular}[c]{@{}c@{}}\\[1pt]\textbf{Dimension}\\$(d)$\end{tabular}} &
        \multicolumn{5}{c}{\textbf{F-time}}  & \multicolumn{8}{c}{\textbf{Statistical Analysis}}\\\cmidrule{4-16}
         &&  & \multirow{2}{*}{\begin{tabular}[r]{@{}r@{}}\\[-5pt]\textit{VPP-}\\\textit{ART}\end{tabular}} & \multirow{2}{*}{\begin{tabular}[r]{@{}r@{}}\\[-5pt]\textit{FSCS-}\\\textit{ART}\end{tabular}} & \multirow{2}{*}{\begin{tabular}[r]{@{}r@{}}\\[-5pt]\textit{Naive-}\\\textit{KDFC}\end{tabular}} & \multirow{2}{*}{\begin{tabular}[r]{@{}r@{}}\\[-5pt]\textit{SemiBal-}\\\textit{KDFC}\end{tabular}} & \multirow{2}{*}{\begin{tabular}[r]{@{}r@{}}\\[-5pt]\textit{LimBal-}\\\textit{KDFC}\end{tabular}} & \multicolumn{2}{r}{\textit{vs.} \textit{\FSCS\/}} & \multicolumn{2}{r}{\textit{vs.} \textit{\Naive\/}} & \multicolumn{2}{r}{\textit{vs.} \textit{\SemiBal\/}} & \multicolumn{2}{r}{\textit{vs.} \textit{\LimBal\/}}\\\cmidrule{9-16}
         &  &&  &  &  &  &  & \textit{p-value} & \textit{\es\/} & \textit{p-value} & \textit{\es\/} & \textit{p-value} & \textit{\es\/} & \textit{p-value} & \textit{\es\/}\\
         \midrule
        1&\texttt{airy}&1&12.74&329.08&\textcolor{blue}{\textbf{3.31}}&3.33&3.71&\textcolor{red}{\textbf{0.0000}}&0.8740&\textcolor{red}{\textbf{0.0000}}&0.2000&\textcolor{red}{\textbf{0.0000}}&0.2008&\textcolor{red}{\textbf{0.0000}}&0.2182\\
        2&\texttt{bessj0}&1&6.38&98.95&\textcolor{blue}{\textbf{1.66}}&1.66&1.76&\textcolor{red}{\textbf{0.0000}}&0.8382&\textcolor{red}{\textbf{0.0000}}&0.2062&\textcolor{red}{\textbf{0.0000}}&0.2061&\textcolor{red}{\textbf{0.0000}}&0.2151\\
        3&\texttt{erfcc}&1&17.33&529.69&\textcolor{blue}{\textbf{4.46}}&4.60&4.80&\textcolor{red}{\textbf{0.0000}}&0.8904&\textcolor{red}{\textbf{0.0000}}&0.1919&\textcolor{red}{\textbf{0.0000}}&0.1971&\textcolor{red}{\textbf{0.0000}}&0.2039\\
        4&\texttt{probks}&1&37.18&1081.17&18.11&\textcolor{blue}{\textbf{17.95}}&18.57&\textcolor{red}{\textbf{0.0000}}&0.8845&\textcolor{red}{\textbf{0.0000}}&0.3000&\textcolor{red}{\textbf{0.0000}}&0.2973&\textcolor{red}{\textbf{0.0000}}&0.3057\\
        5&\texttt{tanh}&1&3.98&50.72&\textcolor{blue}{\textbf{1.08}}&1.17&1.21&\textcolor{red}{\textbf{0.0000}}&0.8273&\textcolor{red}{\textbf{0.0000}}&0.2044&\textcolor{red}{\textbf{0.0000}}&0.2173&\textcolor{red}{\textbf{0.0000}}&0.2218\\
        6&\texttt{bessj}&2&7.55&132.59&8.43&\textcolor{blue}{\textbf{3.08}}&3.97&\textcolor{red}{\textbf{0.0000}}&0.7889&\textcolor{red}{\textbf{0.0000}}&0.5326&\textcolor{red}{\textbf{0.0000}}&0.3214&\textcolor{red}{\textbf{0.0000}}&0.3647\\
        7&\texttt{gammq}&2&20.60&775.74&16.10&\textcolor{blue}{\textbf{13.08}}&14.28&\textcolor{red}{\textbf{0.0000}}&0.8319&\textcolor{red}{\textbf{0.0000}}&0.4520&\textcolor{red}{\textbf{0.0000}}&0.4153&\textcolor{red}{\textbf{0.0000}}&0.4330\\
        8&\texttt{sncndn}&2&11.18&287.05&\textcolor{blue}{\textbf{5.53}}&9.03&9.06&\textcolor{red}{\textbf{0.0000}}&0.7967&\textcolor{red}{\textbf{0.0000}}&0.3773&\textcolor{red}{\textbf{0.0000}}&0.4399&\textcolor{red}{\textbf{0.0000}}&0.4528\\
        9&\texttt{golden}&3&45.73&2417.30&\textcolor{blue}{\textbf{45.16}}&72.10&66.76&\textcolor{red}{\textbf{0.0000}}&0.8518&0.1062&0.5120&\textcolor{red}{\textbf{0.0000}}&0.5623&\textcolor{red}{\textbf{0.0000}}&0.5760\\
        10&\texttt{plgndr}&3&42.52&1714.59&94.44&\textcolor{blue}{\textbf{19.43}}&21.65&\textcolor{red}{\textbf{0.0000}}&0.8155&\textcolor{red}{\textbf{0.0000}}&0.6282&\textcolor{red}{\textbf{0.0000}}&0.3541&\textcolor{red}{\textbf{0.0000}}&0.3745\\
        11&\texttt{cel}&4&36.14&1858.20&155.37&\textcolor{blue}{\textbf{23.46}}&28.03&\textcolor{red}{\textbf{0.0000}}&0.8601&\textcolor{red}{\textbf{0.0000}}&0.7358&\textcolor{red}{\textbf{0.0000}}&0.4126&\textcolor{red}{\textbf{0.0000}}&0.4457\\
        12&\texttt{el2}&4&\textcolor{blue}{\textbf{15.42}}&379.12&25.32&29.34&28.23&\textcolor{red}{\textbf{0.0000}}&0.7910&\textcolor{red}{\textbf{0.0000}}&0.5933&\textcolor{red}{\textbf{0.0000}}&0.5869&\textcolor{red}{\textbf{0.0000}}&0.5968\\
        13&\texttt{calDay}&5&\textcolor{blue}{\textbf{27.57}}&968.30&169.88&36.91&36.18&\textcolor{red}{\textbf{0.0000}}&0.8371&\textcolor{red}{\textbf{0.0000}}&0.7733&\textcolor{red}{\textbf{0.0004}}&0.5264&\textcolor{red}{\textbf{0.0088}}&0.5195\\
        14&\texttt{complex}&6&\textcolor{blue}{\textbf{27.62}}&1317.92&136.16&136.54&113.95&\textcolor{red}{\textbf{0.0000}}&0.8536&\textcolor{red}{\textbf{0.0000}}&0.7358&\textcolor{red}{\textbf{0.0000}}&0.7446&\textcolor{red}{\textbf{0.0000}}&0.7431\\
        15&\texttt{pntLinePos}&6&\textcolor{blue}{\textbf{34.46}}&1895.50&182.67&193.30&155.95&\textcolor{red}{\textbf{0.0000}}&0.8563&\textcolor{red}{\textbf{0.0000}}&0.7496&\textcolor{red}{\textbf{0.0000}}&0.7640&\textcolor{red}{\textbf{0.0000}}&0.7651\\
        16&\texttt{triangle}&6&\textcolor{blue}{\textbf{35.23}}&1562.98&177.93&180.97&142.86&\textcolor{red}{\textbf{0.0000}}&0.8339&\textcolor{red}{\textbf{0.0000}}&0.7444&\textcolor{red}{\textbf{0.0000}}&0.7453&\textcolor{red}{\textbf{0.0000}}&0.7372\\
        17&\texttt{line}&8&\textcolor{blue}{\textbf{102.64}}&10274.08&1542.17&1659.12&936.30&\textcolor{red}{\textbf{0.0000}}&0.8931&\textcolor{red}{\textbf{0.0000}}&0.8484&\textcolor{red}{\textbf{0.0000}}&0.8432&\textcolor{red}{\textbf{0.0000}}&0.8511\\
        18&\texttt{pntTrianglePos}&8&\textcolor{blue}{\textbf{153.93}}&21846.93&2864.84&3077.41&1450.36&\textcolor{red}{\textbf{0.0000}}&0.8972&\textcolor{red}{\textbf{0.0000}}&0.8594&\textcolor{red}{\textbf{0.0000}}&0.8676&\textcolor{red}{\textbf{0.0000}}&0.8547\\
        19&\texttt{TwoLinesPos}&8&\textcolor{blue}{\textbf{272.39}}&70499.32&6015.84&6919.79&3177.67&\textcolor{red}{\textbf{0.0000}}&0.9347&\textcolor{red}{\textbf{0.0000}}&0.8831&\textcolor{red}{\textbf{0.0000}}&0.8946&\textcolor{red}{\textbf{0.0000}}&0.8870\\
        20&\texttt{nearestDistance}&10&\textcolor{blue}{\textbf{69.81}}&3493.54&3113.37&3467.71&1067.29&\textcolor{red}{\textbf{0.0000}}&0.8596&\textcolor{red}{\textbf{0.0000}}&0.8503&\textcolor{red}{\textbf{0.0000}}&0.8540&\textcolor{red}{\textbf{0.0000}}&0.8608\\
        21&\texttt{calGCD}&10&\textcolor{blue}{\textbf{27.96}}&916.10&483.69&554.96&364.48&\textcolor{red}{\textbf{0.0000}}&0.8139&\textcolor{red}{\textbf{0.0000}}&0.8033&\textcolor{red}{\textbf{0.0000}}&0.8119&\textcolor{red}{\textbf{0.0000}}&0.8055\\
        22&\texttt{select}&11&\textcolor{blue}{\textbf{216.93}}&26894.43&24325.93&23561.25&4338.21&\textcolor{red}{\textbf{0.0000}}&0.9035&\textcolor{red}{\textbf{0.0000}}&0.9088&\textcolor{red}{\textbf{0.0000}}&0.9086&\textcolor{red}{\textbf{0.0000}}&0.9039\\
        \bottomrule

\end{tabular}
\end{table*}

\noindent\textit{\textbf{B. \VPP\/ versus \KDFC\/}}

The results show that when the input domain dimensionality is low, all three \KDFC\/ versions require less time than \VPP\/ to find the first failure in the subject programs.
When $d = 1, 2$, \Naive\/ has better performance than \VPP\/, except for the \texttt{bessj} program, but for higher dimensions, according to the
\textit{\es\/} scores, \VPP\/ outperforms \Naive\/ approximately $51\%$ to $91\%$ of the time.
When  $1\leq d \leq 4$,  except for the \texttt{golden} and \texttt{el2} programs, \SemiBal\/ and \LimBal\/ outperform \VPP\/.
For the other programs/dimensions, \VPP\/ outperforms \SemiBal\/ and \LimBal\/ in (approximately) $50\%$ to $91\%$, and (approximately) $51\%$ to $91\%$, of the time, respectively.
All \textit{p-values} are less than $0.05$, except for \Naive\/'s performance with \texttt{golden}, which shows that the differences in \VPP\/ and \KDFC\/ performances are significant.


\textbf{\textit{Discussion of efficiency experiments using subject programs}:}
In general, for all subject programs, \VPP\/ outperforms \FSCS\/, in terms of the time taken to find the first failure.
In low dimensions, \KDFC\/ shows better efficiency than \VPP\/, but as the dimensionality increases, \VPP\/ outperforms \KDFC\/.

\textbf{\textit{Summary and Discussions of Answers to RQ2}:}
\VPP\/ is more efficient than \FSCS\/, according to both the simulation and experimental results.
\KDFC\/ is more efficient in low-dimensional input domains, but \VPP\/ outperforms \KDFC\/ when the dimensionality is higher.

\textbf{\textit{Conclusion}:}
Compared with \FSCS\/, the proposed \VPP\/ approach not only shows comparable or better failure-detection effectiveness in most cases, but also significantly improves on the \FSCS\/ efficiency in most cases.
\VPP\/ has similar failure-detection effectiveness to \KDFC\/, outperforming \KDFC\/ in higher dimensions.
\VPP\/, due to the use of vantage point partitioning and the approximate \NN\/ search strategy, has the following advantages:
(1) \VPP\/ can generate an equivalent number of test cases to \FSCS\/ in very short time;
(2) \VPP\/ can find software failures more efficiently than \FSCS\/; and
(3) \VPP\/ can relieve the influence of dimensionality on the efficiency of \KDFC\/, with \VPP\/ not showing performance degradation as the input domain dimensionality increases.

In conclusion, \VPP\/ is more cost-effective than \FSCS\/; and in high dimensions, \VPP\/ is more cost-effective than \KDFC\/.

\section{Threats to Validity
\label{SEC:Threat-to-Validity}}

This section examines some potential threats to the validity of our study.

\subsection{Construct Validity}

\textit{Construct validity} refers to how well a study examines and measures its assertions.
In this paper, we used the \textit{F-measure} (and related \textit{\fratio\/})~\cite{Chen2004c} to measure the failure-detection effectiveness of RT and different ART algorithms.
In addition to the \textit{F-measure}, two other failure-detection effectiveness metrics that are commonly used are the \textit{E-measure} and the \textit{P-measure}~\cite{Chen2006a}.
The \textit{E-measure} refers to the expected number of failures to be identified by a set of test cases; and
the \textit{P-measure} is the probability of a test set identifying at least one program failure.
The \textit{E-measure} and \textit{P-measure} are most suitable for the evaluation of automated testing systems when the number of test cases is fixed~\cite{Shahbazi2013}.
The \textit{F-measure}, in contrast, is more suitable when testing continues until a failure is detected.
Previous studies have shown that ART outperforms RT in terms of the \textit{P-measure}~\cite{Chen2007a}.
A criticism of the \textit{E-measure} is that multiple failures may be associated with a single fault~\cite{Shahbazi2013}.
In this paper, we examined \FSCS\/ and the enhanced RT algorithms \VPP\/ and \KDFC\/ in simulations and experiments, with the algorithm stopping whenever a (first) program failure was detected.
Accordingly, the \textit{F-measure} was the metric used to evaluate the failure-finding effectiveness of the three algorithms.

\subsection{External Validity}

\textit{External validity} refers to the extent to which our experiments are generally valid, and to what degree can the results and findings be generalized.
A potential external threat relates to the existing empirical research (e.g., \KDFC\/~\cite{Mao2019}).
Although the subject programs in our study cover a wide range of failure rates and dimensions, the programs are relatively small in size.
To address this threat, our future work will involve more subject programs, including both different types of subject programs (e.g., object-oriented software/systems), and larger-scale programs.

\subsection{Internal Validity}

\textit{Internal validity} refers to the accuracy and completeness of the experiments.
Each of our experiments was repeated 3000 times, allowing confidence in the calculated average data.
Different parameter settings for \VPP\/ will lead to different outcomes, which is also a potential threat.
For example, we recommended setting the parameter pair values as $\langle \varepsilon,\lambda \rangle =\langle 3,10 \rangle$ (Section~\ref{Determining-Parameters-of-VPPS-ART}):
\VPP\/ will produce different behavior and results when different parameter values are chosen.
As discussed in Section~\ref{Determining-Parameters-of-VPPS-ART}, a smaller $\lambda$ and a smaller $\varepsilon$ can enhance the VPP-ART performance.
However, it is not possible to enumerate every possible parameter value combination.

\section{Related Work
\label{SEC:relatedWork}}

Many techniques have been proposed to address the ART computational overhead problem, including:
\textit{forgetting}~\cite{Chan2006};
\textit{Distance-aware Forgetting for Fixed-Size-Candidate-Set} ART (DF-\FSCS\/)~\cite{Mao2017};
\textit{KD-tree-enhanced Fixed-Size-Candidate-Set ART} (\KDFC\/)~\cite{Mao2019}; and
\textit{Fixed-Size-Candidate-Set ART using SIMD instructions} (FSCS-SIMD)~\cite{Ashfaq2020}.

\subsection{Forgetting}

With ART, generation of the next test case depends on the $|E|$ test cases that were already executed, without revealing any failures.
As the number of test cases in $E$ grows, the computational overhead of generating test cases becomes greater.
Chan \textit{et al.}~\cite{Chan2006} proposed a \textit{forgetting} strategy that only examines up to a constant $M$ number of executed test cases when generating new test cases, typically with $M$ being $\ll |E|$.
The forgetting strategy thus makes generation of the next test case independent of the size of $E$.
\VPP\/ uses the concept of partitioning, with all $|E|$ test cases, but only $\lambda$ test cases are considered in the search process.
Both forgetting and \VPP\/ can reduce the searching scale of $E$, but there is a difference between the algorithms:
The forgetting strategy ignores some executed test cases without considering the relationship between them;
with \VPP\/, however,  the attribute (distance) information of each test case is indexed by vantage points,
thus implicitly including the information when reducing the search scale.

\subsection{DF-\FSCS\/}

DF-\FSCS\/~\cite{Mao2017} uses information about the distribution of executed test cases, ignoring those not in the \textit{sight} of a given candidate test case.
The input domain is first divided into $p \times p$ sub-domains (where $p$ is a predefined parameter).
During test case generation, \FSCS\/ is applied to the entire input domain, but for each candidate test case, only those executed test cases lying in adjacent sub-domains are included in the distance calculation process.
When DF-\FSCS\/ is dividing the input domain, as long as the division conditions are satisfied, a new round of division will be performed.
To avoid too many test cases within the sub-domains, DF-\FSCS\/ randomly filters them to keep the number below a threshold value $\lambda$.

DF-\FSCS\/ faces some challenges:
(1) Two parameters ($p$ and $\lambda$) need to be set before testing, with different values of these two parameters causing the performance of DF-\FSCS\/ to vary widely.
This is also a threat to our proposed \VPP\/, which requires that the parameter pair $\langle \varepsilon,\lambda \rangle$ be set before testing.
(2) As the number of executed test cases increases, more distance calculations need to be performed, increasing the time overheads.
Although \VPP\/ and DF-\FSCS\/ share the same goal of reducing the \FSCS\/ computational overheads, the dimensionality and number of executed test cases are significantly stronger limitations for DF-\FSCS\/:
As seen in this paper, the \VPP\/ time cost changes very little in response to increases in the executed test set size and input domain dimension.

\subsection{\KDFC\/}

To reduce the high computational overheads of \FSCS\/, Mao \textit{et al.}~\cite{Mao2019} proposed three \KDFC\/ algorithms, based on the \textit{K-dimensional tree} (KD-tree)~\cite{Bentley1975,Kubica2005,Adams2009,Foley2005} structure:
\textit{\Naive\/};
\textit{\SemiBal\/}; and
\textit{\LimBal\/}.
\textit{\Naive\/} sequentially divides the input space in each dimension to construct the KD-tree.
To improve the balance of the KD-tree, \textit{\SemiBal\/} prioritizes the partitioning using a splitting strategy~\cite{Mao2019}
Because the KD-tree structure has a drawback that, in the worst case (especially when the dimensionality increases), all nodes may need to be traversed, the third algorithm, \textit{\LimBal\/}, was proposed:
\textit{\LimBal\/} uses an upper limit to control the number of traversed nodes in the backtracking.
\textit{\Naive\/} and \textit{\SemiBal\/} are based on an exact \NN\/ search; while \textit{\LimBal\/} uses on approximate \NN\/ search
---
the same as \VPP\/.
As shown in this paper, both \VPP\/ and \KDFC\/ focus on reducing the high computational overheads of ART, and each has its own advantages and disadvantages.
\VPP\/ shows the improvement in high-dimensional efficiency due to its use of the VPP for distances among test cases:
\VPP\/ test case generation time changes little as the dimensionality increases.

\subsection{FSCS-SIMD}

Ashfaq \textit{et al.}~\cite{Ashfaq2020} proposed an efficient \FSCS\/ implementation, \textit{Fixed-Size-Candidate-Set} using \textit{Single-Instruction-Multiple-Data} (FSCS-SIMD), which uses the SIMD instruction architecture to calculate the distances among multiple test cases simultaneously, in a many-to-many manner.
FSCS-SIMD loads a batch of multiple test cases from the candidate and executed test case sets at once.
FSCS-SIMD accelerates the distance computation by increasing the CPU execution utilization, but is dependent on the available hardware resources~\cite{Ashfaq2020}.
\VPP\/, in contrast, improves the efficiency at the level of the algorithm and data structures.
In poorer hardware environments (such as with a low number of CPU cores), FSCS-SIMD has greater limitations than \VPP\/ because FSCS-SIMD depends more on the hardware resources.

\section{Conclusions and Future Work
\label{SEC:conclusion}}

\textit{Adaptive Random Testing} (ART)~\cite{Chen2004a} improves on the failure-detection effectiveness of \textit{Random Testing} (RT)~\cite{Arcuri2012}, by making the test cases more evenly distributed in the input domain.
The \textit{Fixed-Size-Candidate-Set} ART (\FSCS\/)~\cite{Chen2004a} was among the first ART implementations, and has remained among the most popular \cite{Huang2019}:
\FSCS\/ is widely used in practice, and has been shown to have better failure-detection effectiveness~\cite{Chen2004a,Chen2004c,Chen2006a,Chen2007}, test-case distribution~\cite{Chen2007b}, and code coverage~\cite{Chen2013} than RT.
However, \FSCS\/ still has a significant computational overhead problem~\cite{Huang2019}:
The time required to generate/select \FSCS\/ test cases is much greater than their execution time.
In this paper, we have proposed the \VPP\/ algorithm, which make use of \textit{Vantage Point Partitioning} (VPP) and an approximate nearest neighbor (\NN\/) search strategy.
\VPP\/  achieves a significant reduction in the time overheads for \FSCS\/ while maintaining its effectiveness.

We conducted a series of simulations and experimental studies to validate the \VPP\/ approach, using \FSCS\/ and three \KDFC\/ algorithms for comparison.
The simulations showed that \VPP\/ not only requires less time to detect software failures compared with \FSCS\/, but also retains comparable failure-detection effectiveness.
\VPP\/ also has similar, or better, failure-detection effectiveness compared with \KDFC\/~\cite{Mao2019}.
In terms of efficiency, \KDFC\/ has lower overheads in low dimensions, but \VPP\/ performs better in high dimensions.
Furthermore, the \VPP\/ time overheads only increase slightly for increases in the input domain dimensionality.
\VPP\/ effectively reduces the computational overhead problem of \FSCS\/, and is cost-effective.

The ART algorithms based on exact \NN\/ searches have better failure-detection effectiveness in lower dimensional input domains, while those based on approximate \NN\/ are better in higher dimensions.
This is because of the clustering of similar test cases, which is more pronounced as the dimensionality and number of test cases increases, resulting in more similar test cases clustering in the same sub-domain.
Moreover, when there is little difference between the approximate and the exact \NN\/s, the approximate \NN\/ can replace the exact \NN\/,
improving the search efficiency (test case generation efficiency) at the cost of a tolerable loss in accuracy.

In this paper, we introduced a modified VP-tree to support the dynamic nature of \FSCS\/.
Because many other ART approaches and implementations also suffer from the same drawbacks and overheads as \FSCS\/,
our future work will include exploration of application of the insert-able VP-tree to these other ART algorithms.
Furthermore, as mentioned in the Threats to Validity section (Section~\ref{SEC:Threat-to-Validity}), confirmation of the \VPP\/ parameter settings for $\lambda$ and $\varepsilon$ is also one of the future research directions.
The influence of the parameter pairs ($\lambda$ and $\varepsilon$) on the effectiveness of the experiments has been discussed, with the experimental parameter pair values determined from the simulations.
However, because there has not yet been much research into these two parameters, exploring the functional or dependency relationship between them, so that our proposed insert-able VP-tree structure can be better applied, will also form part of our future work.
\VPP\/ is an ART algorithm based on an approximate \NN\/ search (like \LimBal\/):
Our experimental analysis has shown that \VPP\/ and \LimBal\/ can effectively improve on the high dimensional failure-detection effectiveness of the original ART algorithm, and greatly reduce the test case generation time.
This gives us some inspiration to identify and create other algorithms or data structures that may be enhanced through an approximate, rather than exact, \NN\/ search.

\ifCLASSOPTIONcaptionsoff
    \newpage
\fi
{
\bibliographystyle{IEEEtran}
\bibliography{VPP-ART}
}

\end{document}